\documentclass[twocolumn,times]{aastex63} 

\usepackage{amssymb}
\usepackage{amsmath}
\usepackage{xspace}
\usepackage{epstopdf}
\usepackage[english]{babel}
\usepackage{mathrsfs}
\usepackage{mathtools}
\usepackage{graphicx}
\usepackage{listings}
\usepackage{xcolor}
\usepackage{array}
\usepackage{listings}

\definecolor{dkgreen}{rgb}{0,0.6,0}
\definecolor{gray}{rgb}{0.5,0.5,0.5}
\definecolor{mauve}{rgb}{0.58,0,0.82}
\definecolor{red}{rgb}{1,0,0}

\newcommand{\kepler}{\textit{Kepler}}

\newcommand{\feh}{$\rm{[Fe/H]}$}

\newcommand{\logg}{$\log g$}
\newcommand{\logrhk}{$\log R_{HK}^\prime$}

\newcommand{\mstar}{$M_{\star}$}

\newcommand{\prot}{$P_{\mathrm{rot}}$}

\newcommand{\rhostar}{$\bar{\rho}_{\star}$}
\newcommand{\rstar}{$R_{\star}$}
\newcommand{\teff}{$T_{\mathrm{eff}}$}
\newcommand{\vsini}{$v\sin i$}

\newcommand{\msun}{$M_{\odot}$}

\newcommand{\rsun}{$R_{\odot}$}

\newcommand{\ator}{$a/R_{\star}$}

\newcommand{\massp}{$M_{\mathrm{p}}$}

\newcommand{\radiusp}{$R_{\mathrm{p}}$}
\newcommand{\rhop}{$\bar{\rho}_{\mathrm{p}}$}
\newcommand{\rprs}{$R_{\mathrm{p}}/R_{\star}$}

\newcommand{\tc}{$T_{\mathrm{conj}}$}
\newcommand{\tcb}{${T_{\mathrm{conj}}}_{b}$}

\newcommand{\masse}{$M_{\oplus}$}
\newcommand{\radiuse}{$R_{\oplus}$}

\newcommand{\massj}{$M_{\mathrm{J}}$}

\newcommand{\sig}{$\sigma$}

\newcommand{\kms}{$\rm km\, s^{-1}$}
\newcommand{\ms}{$\rm m\, s^{-1}$}
\newcommand{\msd}{$\rm m\, s^{-1}\, d^{-1}$}

\newcommand{\gcc}{$\rm g\,cm^{-3}$}

\newcommand{\snr}{S/N}

\shorttitle{A Homogeneous Catalog of Subgiant Exoplanet Systems}
\shortauthors{Chontos, Huber et al.}

\begin{document}

\title{The TESS-Keck Survey XXI: \\ 13 New Planets and Homogeneous Properties for 21 Subgiant Systems}

\author[0000-0003-1125-2564]{Ashley~Chontos}
\altaffiliation{Henry Norris Russell Fellow}
\affiliation{Institute for Astronomy, University of Hawai`i, 2680 Woodlawn Drive, Honolulu, HI 96822, USA}
\affiliation{Department of Astrophysical Sciences, Princeton University, 4 Ivy Lane, Princeton, NJ 08544, USA}

\author[0000-0001-8832-4488]{Daniel~Huber}
\affiliation{Institute for Astronomy, University of Hawai`i, 2680 Woodlawn Drive, Honolulu, HI 96822, USA}
\affiliation{Sydney Institute for Astronomy (SIfA), School of Physics, University of Sydney, NSW 2006, Australia}

\author[0000-0003-4976-9980]{Samuel~K.~Grunblatt}
\affiliation{Department of Physics and Astronomy, Johns Hopkins University, 3400 N Charles St, Baltimore, MD 21218, USA}

\author[0000-0003-2657-3889]{Nicholas~Saunders}
\affiliation{Institute for Astronomy, University of Hawai`i, 2680 Woodlawn Drive, Honolulu, HI 96822, USA}
\altaffiliation{NSF Graduate Research Fellow}

\author[0000-0002-4265-047X]{Joshua~N.~Winn}
\affiliation{Department of Astrophysical Sciences, Princeton University, 4 Ivy Lane, Princeton, NJ 08544, USA}

\author[0000-0002-1463-9847]{Mason~McCormack}
\affiliation{Department of Astronomy \& Astrophysics, University of Chicago, Chicago, IL 60637, USA}

\author[0000-0001-7880-594X]{Emil Knudstrup}
\affiliation{Department of Space, Earth and Environment, Chalmers University of Technology, 412 93, Gothenburg, Sweden}
\affiliation{Stellar Astrophysics Centre, Department of Physics and Astronomy, Aarhus University, Ny Munkegade 120, DK-8000 Aarhus C, Denmark}

\author[0000-0003-1762-8235]{Simon~H.~Albrecht}
\affiliation{Stellar Astrophysics Centre, Department of Physics and Astronomy, Aarhus University, Ny Munkegade 120, DK-8000 Aarhus C, Denmark}

\author{Ian~J.~M.~Crossfield}
\affiliation{Department of Physics \& Astronomy, University of Kansas, 1082 Malott, 1251 Wescoe Hall Dr., Lawrence, KS 66045, USA}

\author[0000-0001-8812-0565]{Joseph~E.~Rodriguez}
\affiliation{Center for Data Intensive and Time Domain Astronomy, Department of Physics and Astronomy, Michigan State University, East Lansing, MI 48824, USA}

\author[0000-0002-5741-3047]{David~R.~Ciardi}
\affiliation{Caltech/IPAC-NASA Exoplanet Science Institute, 770 S. Wilson Avenue, Pasadena, CA 91106, USA}

\author[0000-0001-6588-9574]{Karen~A.~Collins}
\affiliation{Center for Astrophysics \textbar \ Harvard \& Smithsonian, 60 Garden Street, Cambridge, MA 02138, USA}

\author[0000-0002-4715-9460]{Jon~M.~Jenkins}
\affiliation{NASA Ames Research Center, Moffett Field, CA 94035, USA}

\author[0000-0001-6637-5401]{Allyson~Bieryla}
\affiliation{Center for Astrophysics \textbar \ Harvard \& Smithsonian, 60 Garden Street, Cambridge, MA 02138, USA}


\author[0000-0002-7030-9519]{Natalie~M.~Batalha}
\affiliation{Department of Astronomy and Astrophysics, University of California, Santa Cruz, CA 95060, USA}

\author[0000-0001-7708-2364]{Corey~Beard}
\altaffiliation{NASA FINESST Fellow}
\affiliation{Department of Physics \& Astronomy, University of California Irvine, Irvine, CA 92697, USA}

\author[0000-0002-8958-0683]{Fei~Dai} 
\affiliation{Division of Geological and Planetary Sciences,
1200 E California Blvd, Pasadena, CA, 91125, USA}
\affiliation{Department of Astronomy, California Institute of Technology, Pasadena, CA 91125, USA}
\altaffiliation{NASA Sagan Fellow}

\author[0000-0002-4297-5506]{Paul~A.~Dalba}
\affiliation{Department of Astronomy and Astrophysics, University of California, Santa Cruz, CA 95064, USA}

\author[0000-0002-3551-279X]{Tara Fetherolf}
\affiliation{Department of Physics, California State University, San Marcos, CA 92096, USA}
\affiliation{Department of Earth and Planetary Sciences, University of California, Riverside, CA 92521, USA}


\author[0000-0002-8965-3969]{Steven~Giacalone}
\altaffiliation{NSF Astronomy and Astrophysics Postdoctoral Fellow}
\affiliation{Department of Astronomy, California Institute of Technology, Pasadena, CA 91125, USA}

\author[0000-0002-0139-4756]{Michelle L. Hill}
\altaffiliation{NASA FINESST Fellow}
\affiliation{Department of Earth and Planetary Sciences, University of California, Riverside, CA 92521, USA}


\author[0000-0001-8638-0320]{Andrew~W.~Howard}
\affiliation{Department of Astronomy, California Institute of Technology, Pasadena, CA 91125, USA}

\author[0000-0002-0531-1073]{Howard~Isaacson}
\affiliation{{Department of Astronomy,  University of California Berkeley, Berkeley CA 94720, USA}}
\affiliation{Centre for Astrophysics, University of Southern Queensland, Toowoomba, QLD, Australia}

\author[0000-0002-7084-0529]{Stephen~R.~Kane}
\affiliation{Department of Earth and Planetary Sciences, University of California, Riverside, CA 92521, USA}

\author[0000-0001-8342-7736]{Jack Lubin}
\affiliation{Department of Physics \& Astronomy, University of California Los Angeles, Los Angeles, CA 90095, USA}
\affiliation{Department of Physics \& Astronomy, The University of California Irvine, Irvine, CA 92697, USA}

\author[0000-0003-2562-9043]{Mason~G.~MacDougall}
\affiliation{Department of Physics \& Astronomy, University of California Los Angeles, Los Angeles, CA 90095, USA}

\author[0000-0003-4603-556X]{Teo~Mo\v{c}nik}
\affiliation{Gemini Observatory/NSF's NOIRLab, 670 N. A'ohoku Place, Hilo, HI 96720, USA}

\author[0000-0001-8898-8284]{Joseph~M.~Akana~Murphy}
\altaffiliation{NSF Graduate Research Fellow}
\affiliation{Department of Astronomy and Astrophysics, University of California, Santa Cruz, CA 95064, USA}

\author[0000-0003-0967-2893]{Erik~A.~Petigura}
\affiliation{Department of Physics \& Astronomy, University of California Los Angeles, Los Angeles, CA 90095, USA}

\author[0000-0001-9771-7953]{Daria~Pidhorodetska} 
\affiliation{Department of Earth and Planetary Sciences, University of California, Riverside, CA 92521, USA}

\author[0000-0001-7047-8681]{Alex~S.~Polanski}
\affil{Department of Physics and Astronomy, University of Kansas, Lawrence, KS 66045, USA}

\author[0000-0003-0149-9678]{Paul~Robertson}
\affiliation{Department of Physics \& Astronomy, University of California Irvine, Irvine, CA 92697, USA}

\author[0000-0003-3856-3143]{Ryan~A.~Rubenzahl}
\altaffiliation{NSF Graduate Research Fellow}
\affiliation{Department of Astronomy, California Institute of Technology, Pasadena, CA 91125, USA}


\author[0000-0002-1845-2617]{Emma~V.~Turtelboom}
\affiliation{Department of Astronomy, 501 Campbell Hall, University of California, Berkeley, CA 94720, USA}

\author[0000-0002-3725-3058]{Lauren M. Weiss}
\affiliation{Department of Physics and Astronomy, University of Notre Dame, Notre Dame, IN 46556, USA}

\author[0000-0002-4290-6826]{Judah Van Zandt}
\affiliation{Department of Physics \& Astronomy, University of California Los Angeles, Los Angeles, CA 90095, USA}

%
%
%
%

\author[0000-0003-2058-6662]{George~R.~Ricker}
\affiliation{Department of Physics and Kavli Institute for Astrophysics and Space Research, Massachusetts Institute of Technology, 77 Massachusetts Avenue, Cambridge, MA 02139, USA}

\author[0000-0001-6763-6562]{Roland~Vanderspek}
\affiliation{Department of Physics and Kavli Institute for Astrophysics and Space Research, Massachusetts Institute of Technology, 77 Massachusetts Avenue, Cambridge, MA 02139, USA}

\author[0000-0001-9911-7388]{David~W.~Latham}
\affiliation{Center for Astrophysics \textbar \ Harvard \& Smithsonian, 60 Garden Street, Cambridge, MA 02138, USA}

\author[0000-0002-6892-6948]{Sara~Seager}
\affiliation{Department of Physics and Kavli Institute for Astrophysics and Space Research, Massachusetts Institute of Technology, 77 Massachusetts Avenue, Cambridge, MA 02139, USA}
\affiliation{Department of Earth, Atmospheric and Planetary Sciences, Massachusetts Institute of Technology, 77 Massachusetts Avenue, Cambridge, MA 02139, USA}
\affiliation{Department of Aeronautics and Astronautics, MIT, 77 Massachusetts Avenue, Cambridge, MA 02139, USA}

\author[0000-0002-8964-8377]{Samuel N. Quinn}
\affiliation{Center for Astrophysics \textbar \ Harvard \& Smithsonian, 60 Garden Street, Cambridge, MA 02138, USA}


\author[0000-0002-1836-3120]{Avi~Shporer}
\affiliation{Department of Physics and Kavli Institute for Astrophysics and Space Research, Massachusetts Institute of Technology, 77 Massachusetts Avenue, Cambridge, MA 02139, USA}

\author[0000-0002-0786-7307]{Nora L. Eisner}
\altaffiliation{Flatiron Research Fellow}
\affiliation{Center for Computational Astrophysics, Flatiron Institute, 162 Fifth Avenue, New York, NY 10010, USA}

\author{Robert~F.~Goeke}
\affiliation{Department of Physics and Kavli Institute for Astrophysics and Space Research, Massachusetts Institute of Technology, 77 Massachusetts Avenue, Cambridge, MA 02139, USA}

\author[0000-0001-8172-0453]{Alan~M.~Levine}
\affiliation{Department of Physics and Kavli Institute for Astrophysics and Space Research, Massachusetts Institute of Technology, 77 Massachusetts Avenue, Cambridge, MA 02139, USA}

\author[0000-0002-8219-9505]{Eric~B.~Ting}
\affiliation{NASA Ames Research Center, Moffett Field, CA 94035, USA}

\author{Steve~Howell}
\affiliation{NASA Ames Research Center, Moffett Field, CA 94035, USA}

\author{Joshua~E.~Schlieder}
\affiliation{NASA Goddard Space Flight Center, 8800 Greenbelt Rd., Greenbelt MD, 22071, USA}



\author[0000-0001-6981-8722]{Paul~Benni}
\affiliation{Acton Sky Portal private observatory, Acton, MA, USA}

\author[0000-0001-6037-2971]{Andrew~W.~Boyle}
\affiliation{Department of Astronomy, California Institute of Technology, Pasadena, CA 91125, USA}



\author[0000-0002-4503-9705]{Tianjun~Gan}
\affil{Department of Astronomy, Tsinghua University, Beijing 100084, People's Republic of China}

\author[0000-0002-5443-3640]{Eric~Girardin}
\affiliation{Grand-Pra Observatory, 1984 Les Hauderes, Switzerland}

\author[0000-0002-9329-2190]{Erica~Gonzalez}
\affiliation{University of California, Santa Cruz, 1156 High Street, Santa Cruz, CA 95065, USA}

\author[0000-0002-0145-5248]{Joao~Gregorio}
\affiliation{Crow Observatory, Portalegre, Portugal}

\author[0000-0003-1728-0304]{Keith~Horne}
\affiliation{SUPA Physics and Astronomy, University of St. Andrews, Fife, KY16 9SS Scotland, UK}


\author{John~Livingston}

\author[0000-0003-2527-1598]{Michael~B.~Lund}
\affiliation{Caltech/IPAC-NASA Exoplanet Science Institute, 770 S. Wilson Avenue, Pasadena, CA 91106, USA}

\author[0000-0002-9312-0073]{Christopher~R.~Mann}
\affiliation{National Research Council Canada, Herzberg Astronomy \& Astrophysics Research Centre, 5071 West Saanich Road, Victoria, BC V9E 2E7, Canada}

\author[0000-0001-8879-7138]{Bob~Massey}
\affiliation{Villa '39 Observatory, Landers, CA 92285, USA}

\author[0000-0003-0593-1560]{Elisabeth C. Matthews}
\affiliation{Max-Planck-Institut f\"ur Astronomie, K\"onigstuhl 17, 69117 Heidelberg, Germany}

\author[0000-0001-9504-1486]{Kim~K.~McLeod}
\affil{Department of Astronomy, Wellesley College, Wellesley, MA 02481, USA}


\author[0000-0003-0987-1593]{Enric~Palle}
\affiliation{Instituto de Astrof\'\i sica de Canarias (IAC), 38205 La Laguna, Tenerife, Spain}
\affiliation{Departamento de Astrof\'\i sica, Universidad de La Laguna (ULL), 38206, La Laguna, Tenerife, Spain}

\author[0000-0003-3184-5228]{Adam~Popowicz} 
\affiliation{Silesian University of Technology, Department of Electronics, Electrical Engineering and Microelectronics, Akademicka 16, 44-100 GLiwice, Poland}

\author[0009-0009-5132-9520]{Howard~M.~Relles}
\affiliation{Center for Astrophysics \textbar \ Harvard \& Smithsonian, 60 Garden Street, Cambridge, MA 02138, USA}


\author[0000-0001-8227-1020]{Richard~P.~Schwarz}
\affiliation{Center for Astrophysics \textbar \ Harvard \& Smithsonian, 60 Garden Street, Cambridge, MA 02138, USA}

\author[0000-0003-3904-6754]{Ramotholo~Sefako}  
\affiliation{South African Astronomical Observatory, P.O. Box 9, Observatory, Cape Town 7935, South Africa}

\author{Gregor~Srdoc}
\affiliation{Kotizarovci Observatory, Sarsoni 90, 51216 Viskovo, Croatia}

\author[0000-0001-5603-6895]{Thiam-Guan~Tan}
\affiliation{Perth Exoplanet Survey Telescope, Perth, Western Australia}

\author[0000-0003-3092-4418]{Gavin~Wang}
\affiliation{Tsinghua International School, Beijing 100084, China}

\author[0000-0002-0619-7639]{Carl~Ziegler}
\affiliation{Department of Physics, Engineering and Astronomy, Stephen F. Austin State University, TX 75962, USA}
\correspondingauthor{Ashley Chontos}
\email{ashleychontos@astro.princeton.edu}

\begin{abstract}
\noindent We present a dedicated transit and radial velocity survey of planets orbiting subgiant stars observed by the TESS Mission. Using $\sim$16 nights on Keck/HIRES, we confirm and characterize 12 new transiting planets -- TOI-329~b, HD 39688~b (TOI-480), TOI-603~b, TOI-1199~b, TOI-1294~b, TOI-1439~b, TOI-1605~b, TOI-1828~b, HD 148193~b (TOI-1836), TOI-1885~b, HD 83342~b (TOI-1898), TOI-2019~b -- and provide updated properties for 9 previously confirmed TESS subgiant systems (TOI-197, TOI-954, TOI-1181, TOI-1296, TOI-1298, TOI-1601, TOI-1736, TOI-1842, TOI-2145). We also report the discovery of an outer, non-transiting planet, TOI-1294~c ($P = 160.1 \pm 2.5$\,d, \massp\ $= 148.3^{+18.2}_{-16.4}$ \masse), and three additional stars with long-term RV trends. We find that at least 19$\pm$8\% of subgiants in our sample of 21 stars have outer companions, comparable to main-sequence stars. We perform a homogeneous analysis of the stars and planets in the sample, with median uncertainties of 3\%, 8\% and 15\% for planet radii, masses and ages, doubling the number of known planets orbiting subgiant stars with bulk densities measured to better than $10\%$.  We observe a dearth of giant planets around evolved stars with short orbital periods, consistent with tidal dissipation theories that predict the rapid inspiral of planets as their host stars leave the main sequence. We note the possible evidence for two distinct classes of hot Jupiter populations, indicating multiple formation channels to explain the observed distributions around evolved stars. Finally, continued RV monitoring of planets in this sample will provide a more comprehensive understanding of demographics for evolved planetary systems.

\end{abstract}

\section{Introduction} \label{sec:intro}

The existence of exoplanets orbiting stars in advanced evolutionary stages such as helium-core burning red giants \citep{hon2023}, white dwarfs \citep{vanderburg2020} and pulsars \citep{wolszczan1992} suggests that planets are able to withstand extreme environments present at the latest stages of stellar evolution. Most known exoplanets will face a similar fate as their host stars eventually evolve off the main-sequence \citep{kane2023}, yet the planet populations around post-main-sequence stars remain poorly understood. 


The subgiant branch is a rapid phase in stellar evolution, causing stars with small differences in mass and age to have significantly different temperatures and radii. Both temperature and radii can now be precisely measured for most exoplanet host stars by combining photometry, spectroscopy and Gaia parallaxes \citep[e.g.][]{berger2018}, allowing subgiants to be well characterized and thus used to explore post-main-sequence planet demographics. For example, precise planet densities (which require precise knowledge of stellar radii and masses) are important to infer both the bulk compositions and atmospheric properties of exoplanets \citep{batalha2019}. Indeed, while the number of planets orbiting subgiants is relatively small ($\sim$$15\%$), they are statistically over-represented ($\sim$$32\%$) when only considering planets with the most precisely measured densities.

Precise ages are valuable for placing observational constraints on the timescales of important processes affecting exoplanets. Previous studies suggested that the timescales for processes like tidal circularization and inward migration through tidal dissipation are strongly dependent on the orbital distance and stellar radius \citep{zahn1977,hut1981,zahn1989}. As the stellar radius rapidly changes for subgiant stars, stellar evolution is expected to affect tidal circularization timescales of close-in giant planets, producing a transient population of midly eccentric planets orbiting evolved stars \citep{villaver2009}. \textit{Kepler} and K2 data have yielded intriguing evidence which supports this theory, but were based on a small sample of planets \citep{vaneylen2016,grunblatt2018,chontos2019,vissapragada2022}.

\begin{table}
\caption{TESS Subgiant Planet Sample\label{tab:sample}} \vspace{-0.25cm}
\setlength{\tabcolsep}{7pt}
\centering
\begin{tabular}{cr rrrr}
\noalign{\smallskip}
\hline\hline
\noalign{\smallskip}
System & TOI & RA $[^{o}]$ & Dec $[^{o}]$ & $V_{\mathrm{mag}}$ & $T_{\mathrm{mag}}$ \\
\noalign{\smallskip}
\hline 
\noalign{\smallskip}
\multicolumn{6}{c}{New TESS systems} \\
\noalign{\smallskip}
\hline
\noalign{\smallskip}
TOI-329 & 329 & 351.32 & $-$15.63 & 11.3 & 10.7 \\ 
HD 39688 & 480 & 88.38 & $-$16.27 & 7.3 & 6.8 \\ 
TOI-603 & 603 & 141.11 & $+$5.77 & 10.3 & 9.7 \\ 
TOI-1199 & 1199 & 166.88 & $+$61.35 & 11.1 & 10.4 \\ 
TOI-1294 & 1294 & 223.09 & $+$70.48 & 11.3 & 10.9 \\ 
TOI-1439 & 1439 & 241.76 & $+$67.88 & 10.6 & 10.0 \\ 
TOI-1605 & 1605 & 54.48 & $+$33.08 & 10.2 & 9.6 \\ 
TOI-1828 & 1828 & 241.56 & $+$69.62 & 11.6 & 11.1 \\ 
HD 148193 & 1836 & 245.91 & $+$54.69 & 9.8 & 9.3 \\ 
TOI-1885 & 1885 & 304.67 & $+$66.16 & 12.7 & 12.0 \\ 
HD 83342 & 1898 & 144.56 & $+$23.55 & 7.9 & 7.4 \\ 
TOI-2019 & 2019 & 234.43 & $+$48.96 & 10.3 & 9.6 \\ 
\noalign{\smallskip}
\hline 
\noalign{\smallskip}
\multicolumn{6}{c}{Updated TESS systems} \\
\noalign{\smallskip}
\hline
\noalign{\smallskip}
HD 221416 & 197 & 353.03 & $-$21.80 & 8.2 & 7.4 \\ 
TOI-954 & 954 & 61.94 & $-$25.21 & 10.3 & 9.8 \\ 
TOI-1181 & 1181 & 297.22 & $+$64.35 & 10.6 & 10.1 \\ 
TOI-1296 & 1296 & 256.77 & $+$70.24 & 11.4 & 10.8 \\ 
TOI-1298 & 1298 & 241.32 & $+$70.19 & 11.9 & 11.0 \\ 
TOI-1601 & 1601 & 38.36 & $+$41.01 & 10.7 & 10.1 \\ 
TOI-1736 & 1736 & 43.44 & $+$69.10 & 8.9 & 8.3 \\ 
TOI-1842 & 1842 & 201.96 & $+$9.03 & 9.8 & 9.3 \\ 
TOI-2145 & 2145 & 263.76 & $+$40.70 & 9.1 & 8.6 \\ 
\noalign{\smallskip}
\hline
\end{tabular}
\end{table}

Early radial velocity (RV) surveys aimed to probe the sensitivity of gas-giant planet occurrence to stellar mass using subgiant host stars \citep{johnson2007,bowler2010,johnson2013,wolfhoff2022}. Additionally, \textit{Kepler} and TESS have identified individual planets that transit subgiants for which masses could be measured \citep{chontos2019,huber2019,addison2021,rodriguez2021,saunders2022,grunblatt2022,grunblatt2023a}. However, a systematic transit and radial velocity survey of subgiants has not yet been performed. In this paper we present a homogeneous population of TESS planets orbiting subgiant stars acquired as part of the TESS-Keck Survey \citep[TKS;][]{chontos2022a}. Further RV monitoring of planets in this sample will provide a more comprehensive understanding of planet demographics for evolved systems.

\section{Survey Description}

\subsection{Initial Target List} \label{sec:list}

In order to select the most promising targets in a programmatic way, we assembled a master target list starting with the catalog of TESS Objects of Interest \citep[TOI;][]{guerrero2021} and combined it with additional resources. The summary statistics for the TESS planet candidates were downloaded from the publicly-available database\footnote{\url{https://tev.mit.edu/}}. 

We queried the TESS Input Catalog \citep[version 8, TICv8;][]{stassun2019} for  stellar properties (e.g., radius, etc.) and available photometry (i.e. Johnson $V$ and 2-MASS $JHK$) for the current list of TOIs, as well as corresponding $Gaia$ \citep{gaia2016,gaia2018} source IDs. The source IDs were then used to gather additional astrometric (i.e. RUWE) and photometric information from $Gaia$ that were not directly accessible via TICv8. The California Planet Search (CPS) RV archive\footnote{\url{https://jump.caltech.edu/}} was queried for targets since many of the bright TESS stars already have archival HIRES data available. As a final query, TOI dispositions were updated with the TESS Follow-up Observing Program (TFOP) Working Group\footnote{\url{https://tess.mit.edu/followup/}} (WG) Sub Group 1 (hereafter referred to as SG1). 

The evolutionary state of the star was determined with \texttt{evolstate}\footnote{\url{https://github.com/danxhuber/evolstate/}}, given its effective temperature and surface gravity \citep{huber2017,berger2018}. Here, stellar parameters were first drawn from HIRES-derived spectroscopic parameters and then TICv8 if those were not available. In the event that parameters were not available from either of these sources, the parameters provided in the original TOI table were used. 

\begin{figure*}
\includegraphics[width=\textwidth]{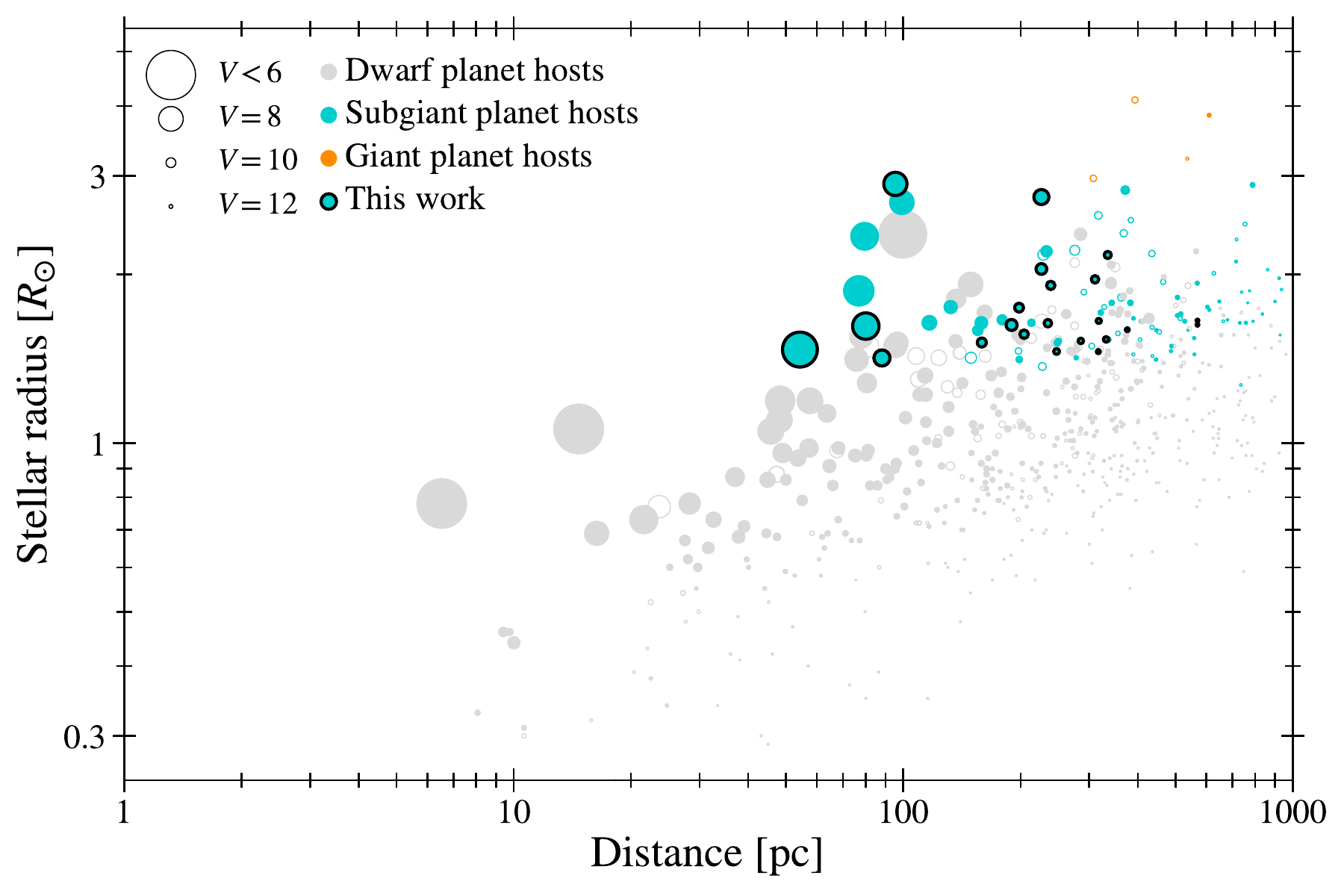} \vspace{-0.75cm}
\caption{Stellar size and distance for stars that host exoplanets with measured densities according to the NASA Exoplanet Archive\footnote{\url{https://exoplanetarchive.ipac.caltech.edu}}, colored by their approximate evolutionary state (i.e. Section \ref{sec:list}). Markers are inversely sized by their visual magnitudes (i.e. bigger is brighter) and filled in for stars that host planets with the most precise ($\geq4\sigma$) planet densities. The subgiant hosts in the present sample (highlighted with a thick, black outline) are, on average, closer and brighter than other known subgiant planet hosts.}
\label{fig:dist}
\end{figure*}

\subsection{Target Selection}

When the survey was initially constructed, the maximum TOI  number available was TOI-2145. From these candidates, we first removed TOIs with unfavorable or ambiguous SG1 dispositions (i.e. APC, BEB, FA, FP, NEB, PNEB, EB, SB1, SB2)\footnote{Ambiguous planet candidate (APC), blended eclipsing binary (BEB), false alarm (FA), false positive (FP), nearby eclipsing binary (NEB), probable nearby eclipsing binary (PNEB), eclipsing binary (EB), single-lined spectroscopic binary (SB1), double-lined spectroscopic binary (SB2)}. False positive and unfavorable dispositions accounted for $\sim25\%$ of the sample (514 TOIs). We excluded targets at low declinations ($\delta<-40^{o}$) that would not be accessible with Keck and also targets with a high $Gaia$ re-normalized unit weight error\footnote{\url{https://gea.esac.esa.int/archive/}} (RUWE $>2$), indicative of an unresolved companion \citep{belokurov2020,evans2018b}. 

Finally, we performed a search for close companions for the remaining targets using their $Gaia$ DR2 coordinates and the MAST (Mikulski Archive for Space Telescopes) Portal\footnote{\url{https://mast.stsci.edu/portal/Mashup/Clients/Mast/Portal.html}}. Any target with a relatively close ($<2"$), bright ($\Delta V<5$) companion or any companion within $1"$ were not considered for further follow-up, because of the danger that high-resolution spectroscopy would be compromised by flux contamination from the secondary star during sub-optimal observing conditions. The final cut required a subgiant (SG) evolutionary state and brought the final sample down to 93 TOIs.

\subsection{Target Vetting} \label{sec:vet}

Before being followed up with precise radial velocity (PRV) observations, the sample was subjected to one final round of vetting. The vetting described here was originally adapted from the TESS-Keck Survey (TKS) target selection work presented in \citet{chontos2022a} and is briefly summarized here.

Data validation (DV) reports from the SPOC \citep[Science Processing Operations Center;][]{jenkins2016,twicken2018,li2019} and QLP \citep[Quick Look Pipeline;][]{huang2020a,huang2020b} pipelines were downloaded and inspected. Careful consideration was given to standard diagnostics for threshold crossing events that indicate possible false positive scenarios such as significant odd-even differences or large centroid offsets. Full DV reports were downloaded to examine the best-fit model and model parameters to identify any possible inconsistencies or other concerning features.

After passing the photometric vetting, targets were queued for reconnaissance (or recon) spectra to check for spectroscopic false positives (FP). All recon spectra were processed by \texttt{ReaMatch} \citep{kolbl2015} to search for and identify faint stellar companions in double-lined spectroscopic binaries. Systemic radial velocities, computed according to \citet{chubak2012}, were compared to those from $Gaia$ to identify any large discrepancies which are indicative of single-lined spectroscopic binaries (SB1). Targets for which the recon and $Gaia$ RV differed by more than 5 \kms\ typically failed this vetting step. However for more ambiguous cases near this cutoff, a second recon spectrum was taken to test for significant linear trends which provide additional evidence of the SB1-like nature. 

\subsection{Final Survey Sample} 

Our survey goal was to maximize the number of precise mass and density measurements of evolved TESS planets that could be obtained within the given 16-night Keck/HIRES allocation. Thus brighter targets and targets with existing HIRES data, both which would require less telescope time to achieve this goal, were typically prioritized to be vetted and followed up first. The final sample is provided in Table \ref{tab:sample}, which consists of 21 subgiant stars that have at least one transiting TESS planet candidates. 

\begin{figure*}
\includegraphics[width=\linewidth]{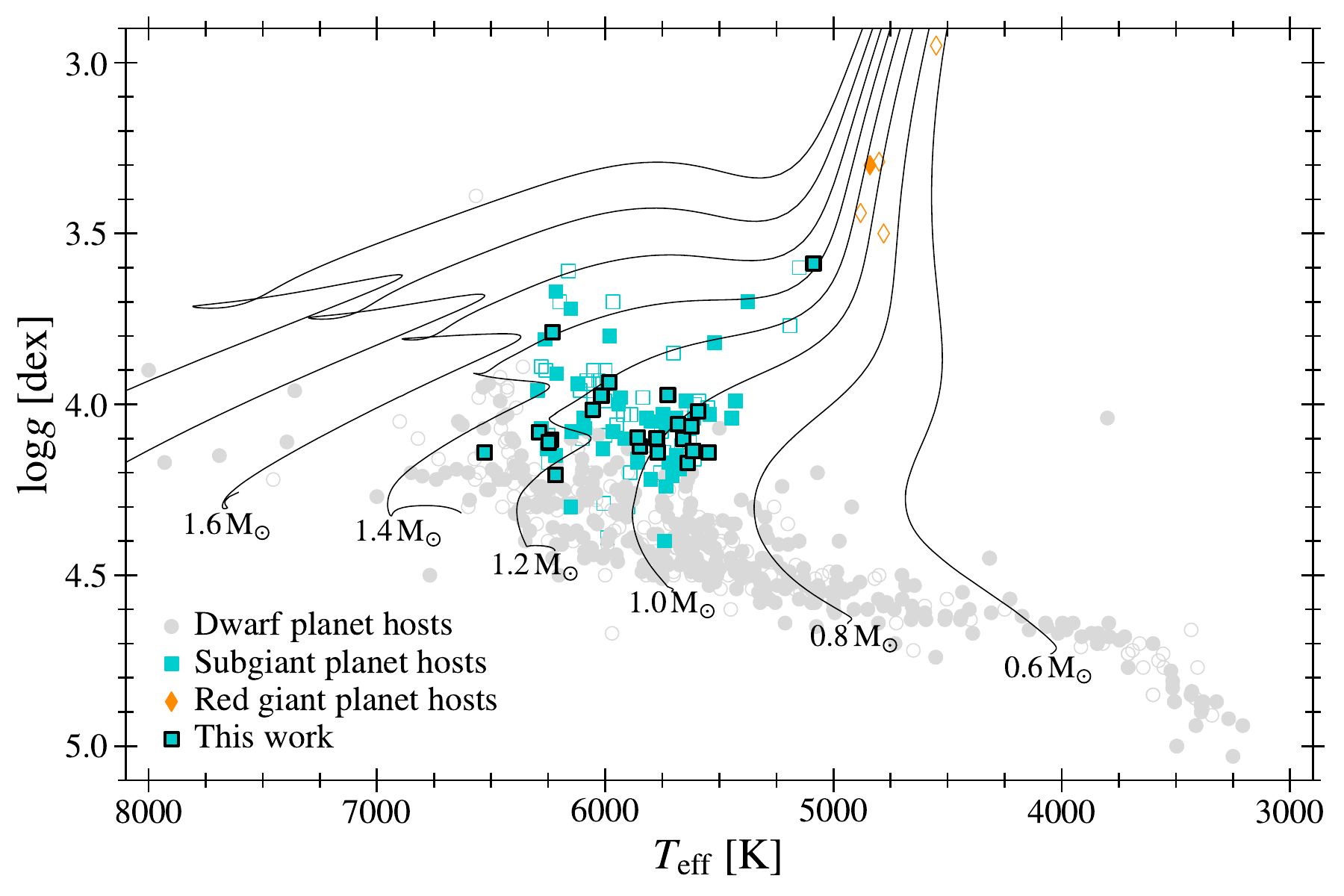} \vspace{-0.75cm}
\caption{Same as Figure \ref{fig:dist} but in a Kiel diagram, plotted with surface gravity and effective temperature. Here, markers also indicate stars of different evolutionary states. MIST solar-metallicity tracks are added for reference.}
\label{fig:hrdiagram}
\end{figure*}

Figures \ref{fig:dist} and \ref{fig:hrdiagram} contextualize the sample with the current population of single stars that host planets with measured bulk densities. As expected with TESS, the stars in this sample are on average closer and brighter than other subgiant stars known to be transiting planet hosts. The stars in the sample also span a wide range of masses and ages, as expected for subgiant stars. 

\section{Observations} \label{sec:obs}

\subsection{TESS Photometry} \label{sec:phot}

\begin{table}
\begin{center}
\caption{Number of sectors of available TESS data\label{tab:phot}} \vspace{-0.25cm}
\setlength{\tabcolsep}{8pt}
\begin{tabular}{lc cccc}
\hline\hline
\noalign{\smallskip}
System & TOI & 20s & 120s & 600s & 1800s \\
\noalign{\smallskip}
\hline
\noalign{\smallskip}
HD 221416 & 197 & 1 & 2$^{*}$ & 1 & 1 \\ 
TOI-329 & 329 & -- & 2$^{*}$ & 2 & 1 \\ 
HD 39688 & 480 & -- & 2$^{*}$ & 1 & 1 \\ 
TOI-603 & 603 & 1 & 4$^{*}$ & 3 & 1 \\ 
TOI-954 & 954 & -- & 1$^{*}$ & 1 & 2 \\ 
TOI-1181 & 1181 & -- & 19$^{*}$ & 10 & 12 \\ 
TOI-1199 & 1199 & -- & 2$^{*}$ & 2 & 2 \\ 
TOI-1294 & 1294 & -- & 5$^{*}$ & 5 & 6 \\ 
TOI-1296 & 1296 & 5 & 18$^{*}$ & 8 & 12 \\ 
TOI-1298 & 1298 & -- & 19$^{*}$ & 10 & 13 \\ 
TOI-1439 & 1439 & -- & 19$^{*}$ & 10 & 13 \\ 
TOI-1601 & 1601 & -- & -- & -- & 1$^{*}$ \\ 
TOI-1605 & 1605 & 1 & 1 & -- & 1$^{*}$ \\ 
TOI-1736 & 1736 & 3 & 6$^{*}$ & 1 & 3 \\ 
TOI-1828 & 1828 & -- & 20$^{*}$ & 10 & 13 \\ 
HD 148193 & 1836 & 1 & 7$^{*}$ & 4 & 4 \\ 
TOI-1842 & 1842 & -- & 2$^{*}$ & 1 & 1 \\ 
TOI-1885 & 1885 & -- & 10$^{*}$ & 3 & 8 \\ 
HD 83342 & 1898 & -- & 4$^{*}$ & 3 & 1 \\ 
TOI-2019 & 2019 & -- & 3$^{*}$ & 1 & 3  \\ 
TOI-2145 & 2145 & -- & 4$^{*}$ & 3 & 2 \\ 
\noalign{\smallskip}
\hline
\noalign{\smallskip}
\end{tabular}
\end{center}
\vspace{-0.25cm}
$^*$Indicates the data used for the joint fits (Section \ref{sec:global}). \\
\end{table}

We downloaded all available light curve products for each target by using the MAST\footnote{Mikulski Archive for Space Telescopes (MAST) \url{https://mast.stsci.edu/portal/Mashup/Clients/Mast/Portal.html}} Portal, which primarily came from the SPOC and QLP pipelines. In order to decide which light curves would be used in the joint transit and RV analysis (Section \ref{sec:global}), we inspected all available light curve data to look for systematics or other undesirable features as a result of differences in data reduction methods. Table \ref{tab:phot} summarizes the TESS data that was available for the sample during the analysis, including how many sectors a given target was observed for, at which cadence(s) and which data was used for further analyses. 

For the planet fits we typically used the specially curated TESS light curves, corresponding to the PDC-SAP \citep[pre-search data conditioning;][]{smith2012,stumpe2012,stumpe2014} flux from the SPOC pipeline and the KSP-SAP (for \kepler\ spline) flux from the QLP pipeline. Light curves were then processed following the methodology in \citet{chontos2019}. To summarize, light curve processing included rejecting points with poor quality flags, clipping outliers by using a sliding two-day filter at the 5\sig\ level, and flattening out-of-transit data by applying a median boxcar filter. As a final step, only data within $\pm 1.5$ times the transit duration (centered on the mid-transit times) was retained to speed up the planet-fitting routine.


\begin{table*}
\centering
\caption{Summary of RV observations taken of the sample. All RVs in this work are made publicly available in Table \ref{tab:RVs}.\label{tab:summary}}
\setlength{\tabcolsep}{6pt}
\begin{tabular}{lc ccccc}
\hline\hline
\noalign{\smallskip}
System & TOI & $N$ RVs (unbinned) & Instrument & First Observation [UT] & Last Observation [UT] & Baseline$\,\,\mathrm{[yr]}$ \\
\noalign{\smallskip}
\hline
\noalign{\smallskip}
HD~221416 & 197 & 57 (64) & HIRES & 2018 Nov 3 & 2021 Dec 25 & 3.1 \\ 
TOI-329 & 329 & 43 & HIRES & 2020 Jun 3 & 2021 Dec 25 & 1.6 \\ 
HD~39688 & 480 & 62 (181) & HIRES & 2019 Dec 27 & 2021 Dec 26 & 2.0 \\ 
HD~39688 & 480 & 20 (33) & HARPS-N & 2019 Sep 26 & 2023 Feb 20 & 3.4 \\ 
TOI-603 & 603 & 12 & HIRES & 2020 Oct 31 & 2022 May 13 & 1.5 \\ 
TOI-954 & 954 & 28 & HIRES & 2020 Aug 27 & 2021 Nov 26 & 1.2  \\ 
TOI-1181 & 1181 & 50 (87) & HIRES & 2019 Dec 2 & 2022 May 16 & 2.5 \\ 
TOI-1199 & 1199 & 16 & HIRES & 2019 Dec 10 & 2022 May 13 & 2.4 \\ 
TOI-1294 & 1294 & 32 & HIRES & 2020 Mar 10 & 2022 May 13 & 2.2 \\ 
TOI-1296 & 1296 & 24 & HIRES & 2019 Nov 28 & 2022 May 16 & 2.5 \\ 
TOI-1298 & 1298 & 22 & HIRES & 2020 Jun 3 & 2022 May 13 & 1.9 \\ 
TOI-1439 & 1439 & 53 & HIRES & 2020 Mar 10 & 2022 May 13 & 2.2 \\ 
TOI-1601 & 1601 & 37 & HIRES & 2020 Feb 25 & 2022 Dec 31 & 1.8 \\ 
TOI-1605 & 1605 & 44 & HIRES & 2020 Aug 26 & 2022 Sep 2 & 1.3 \\ 
TOI-1736 & 1736 & 77 (79) & HIRES & 2020 Aug 2 & 2022 Jan 19 & 2.0 \\ 
TOI-1736 & 1736 & 35 (37) & HARPS-N & 2020 Oct 10 & 2023 Feb 20 & 2.3 \\ 
TOI-1828 & 1828 & 19 & HIRES & 2020 Jun 12 & 2022 May 13 & 1.9 \\ 
HD~148193 & 1836 & 53 & HIRES & 2020 May 26 & 2022 May 13 & 2.0 \\ 
HD~148193 & 1836 & 45 & HARPS-N & 2021 Feb 13 & 2021 Jul 21 & 0.5 \\ 
TOI-1842 & 1842 & 38 & HIRES & 2020 May 26 & 2022 May 9 & 1.2 \\ 
TOI-1885 & 1885 & 15 & HIRES & 2020 Jun 10 & 2022 May 13 & 1.9 \\ 
HD~83342 & 1898 & 48 & HIRES & 2020 Jun 10 & 2022 May 13 & 2.1 \\ 
TOI-2019 & 2019 & 43 & HIRES & 2020 Jun 24 & 2022 May 13 & 1.9 \\ 
TOI-2145 & 2145 & 40 & HIRES & 2020 Aug 25 & 2022 May 13 & 1.7 \\
\noalign{\smallskip}
\hline
\noalign{\smallskip}
\end{tabular}
\end{table*}

\subsection{HIRES Radial Velocities} \label{sec:hires}

The survey results presented here were conducted over 16 nights in 2020 and 2021 using the High-Resolution Echelle Spectrometer \citep[HIRES;][]{vogt1994}, which is mounted on the 10-m Keck I telescope on Maunakea in Hawai`i. Observations were acquired as part of the TESS-Keck Survey (TKS), where baselines for program targets are on average $\sim$2 years. HIRES operates in the spectral range of $0.3-1.0$ microns and uses an iodine cell configuration that when placed in the light path imprints a dense forest of absorption lines and thus serves as a robust wavelength calibration from which Doppler shifts can be measured \citep{butler1996}. 

Every target presented in this sample has two iodine-free observations: 1) a low \snr\ recon spectrum and 2) a high \snr\ template. A recon spectrum is an iodine-free exposure with an \snr\ of $\approx40/\mathrm{pixel}$ to check for false positives as part of the vetting procedure discussed in Section \ref{sec:vet}. A ``template'' is also an iodine-free exposure but typically obtained at a very high \snr, since this is used to compute the rest of the RVs for a given target. The rest of the iodine-in observations were reduced according to the well-tested procedures of the California Planet Search documented in \citet{howard2010}. 

\subsection{HARPS-N Radial Velocities} \label{sec:harps}

We acquired high-resolution ($R=115000$) spectra of HD~148193 with the High Accuracy Radial velocity Planet Searcher for the Northern hemisphere \citep[HARPS-N;][]{cosentino2012} mounted on the 3.58~m Telescopio Nazionale Galileo (TNG) located on Roque de los Muchachos, La Palma, Spain. A total of 46 spectra were collected between UT 2021 February 13 and UT 2021 June 21. We set the exposure time to 900~s, which led to a median SNR of $\sim$75 per pixel at 550~nm. We used the second fibre of the instrument to monitor the sky background. The HARPS-N spectra were reduced and extracted using the dedicated Data Reduction Software \citep[DRS;][]{lovis2007} available at the telescope.

Ground-based RV observations for the sample are summarized in Table \ref{tab:summary}, including the number of observations per target, the first and last observation date and current baseline coverage. All radial velocity data used in the analysis are provided in Table \ref{tab:RVs}.

\begin{table}
\centering
\caption{All radial velocities used in this work. \label{tab:RVs}}\vspace{-0.5cm}
\begin{tabular}{lcccc}
\noalign{\smallskip}
\hline\hline
\noalign{\smallskip}
System & Time & RV & RV Unc. & Instrument \\
 & [BJD] & $\rm [m s^{-1}]$ & $\rm [m s^{-1}]$ & \\
\noalign{\smallskip}
\hline
\noalign{\smallskip}
$\cdots$ & $\cdots$ & $\cdots$ & $\cdots$ & $\cdots$ \\
TOI-329 & 2459028.108213 & $-9.74$ & 2.15 & HIRES \\
$\cdots$ & $\cdots$ & $\cdots$ & $\cdots$ & $\cdots$ \\
\noalign{\smallskip}
\hline 
\noalign{\smallskip}
\end{tabular}
(This table is available in machine-readable form.)
\end{table}

\begin{table*}
\caption{Summary of high-resolution imaging observations (Section \ref{sec:image}) for new TESS systems presented in this work\label{tab:imaging}.}\centering
\setlength{\tabcolsep}{10pt}
\begin{tabular}{lccccc}
\hline\hline
\noalign{\smallskip}
System & Instrument & Observation Date & Filter & Resolution & Contrast at $0.^{''}5$ \\
 & & [UT] & & [FWHM] & [$\Delta$ mag] \\
\noalign{\smallskip}
\hline
\noalign{\smallskip}
TOI-329 & Palomar/PHARO & 2019 Jul 13 & Br-$\gamma$ & $0.^{''}10$ & 7.0 \\ 
HD 39688 & Keck/NIRC2 & 2019 Mar 25 & $K$-cont & $0.^{''}05$ & 8.4 \\ 
TOI-603 & Keck/NIRC2 & 2019 Jun 09 & $K$-cont & $0.^{''}05$ & 7.5 \\ 
TOI-1199 & Palomar/PHARO & 2020 Jan 08 & Br-$\gamma$ & $0.^{''}13$ & 6.2 \\ 
TOI-1294 & Keck/NIRC2 & 2020 May 28 & $K$-cont & $0.^{''}05$ & 7.9 \\ 
TOI-1439 & Keck/NIRC2 & 2020 May 28 & Br-$\gamma$ & $0.^{''}05$ & 7.9 \\ 
TOI-1605 & Keck/NIRC2 & 2020 Sep 09 & Br-$\gamma$ & $0.^{''}05$ & 6.8 \\ 
TOI-1605 & Keck/NIRC2 & 2020 Sep 09 & $J$-cont & $0.^{''}03$ & 6.9 \\ 
HD 148193 & Palomar/PHARO & 2021 June 22 & Br-$\gamma$ & $0.^{''}09$ & 6.8 \\ 
HD 148193 & Palomar/PHARO & 2021 June 22 & $H$-cont & $0.^{''}08$ & 7.8 \\ 
HD 83342 & Palomar/PHARO & 2020 Dec 04 & Br-$\gamma$ & $0.^{''}14$ & 6.1 \\ 
TOI-2019 & Keck/NIRC2 & 2020 Sep 09 & $Ks$ & $0.^{''}05$ & 7.3 \\ 
\noalign{\smallskip}
\hline
\noalign{\smallskip}
\end{tabular}
\end{table*}

\subsection{High-resolution Imaging} \label{sec:image}

Table \ref{tab:imaging} summarizes the high-resolution imaging observations acquired for targets that are being published for the first time. Our targets HD 39688 and TOIs 603, 1294, 1439, 1605, and 2019 were all observed using Keck/NIRC2 natural guide star adaptive optics imaging in the 2.2$\mu$m Brackett-$\gamma$ (Br-$\gamma$) filter. We obtained the data in narrow-angle mode to provide a pixel scale of $\sim$0.01'' and a total field of view of $\sim$10''. We used NIRC2's standard three-point, 3''-step dither pattern to avoid the noisy, lower-left detector
quadrant. We acquired three sets of dithers for each target, with 0.5 positional offsets between each observation. Data were analyzed following \citet{furlan2017}.

\begin{figure*}
\centering
\includegraphics[width=\textwidth]{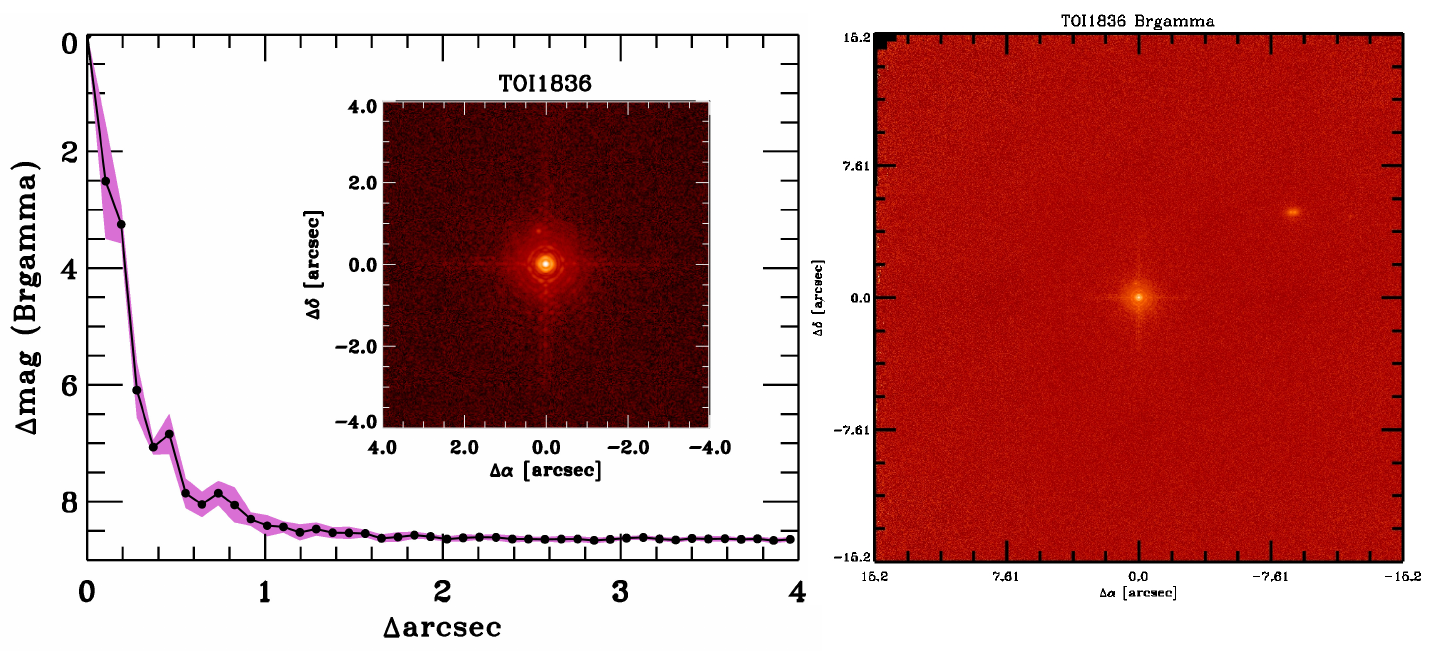}
\caption{High-resolution image of TOI-1836 taken by the PHARO instrument at Palomar on UT 2021-06-22 in the narrowband Br-$\gamma$ filter. Imaging revealed a multiple star system in an interesting heirarchy, the primary star which is orbited by a single star in addition to a well-separated binary at a projected separation of $10''$.}
\label{fig:imaging}
\end{figure*}

No companions were detected for our sample, with the exception of TOI-1836, which was observed using the PHARO (Palomar High Angular Resolution Observer) instrument on the 5-m Hale telescope at the Palomar Observatory in California. The observations of TOI-1836 were taken on (UT) 2021-06-22 with the narrowband Br-$\gamma$ filter. The sensitivity curve and image preview is shown in Figure \ref{fig:imaging}, revealing a companion at a separation of $\rho=0.819''$ with $\Delta m=5.722$. The zoomed out image on the right of Figure \ref{fig:imaging} shows a $0.2''$ binary (TIC 207468069) approximately 10'' away that is astrometrically associated with TOI-1836. We include the analysis and results from the joint fits discussed in Section \ref{sec:global} but later remove TOI-1836 b from the discussion in Section \ref{sec:disc}.

\section{Host Star Characterization} 

\subsection{High-resolution Spectroscopy} \label{sec:spec}

Spectroscopic stellar parameters (\feh, \teff, \vsini, \logg) were derived using \texttt{SpecMatch}. \texttt{SpecMatch} derives parameters either from synthetic  model atmospheres (i.e. \texttt{SpecMatch-Synth}; \citealt[]{petigura2015}) or empirically from  a library of HIRES templates (i.e. \texttt{SpecMatch-Emp}; \citealt[]{yee2017}), the latter of which was primarily developed for later-type cool dwarfs. 

Initial inspections of recon spectra for the sample reported effective temperatures in the range $5080\,$K $<$ \teff\ $<6600\,$K. Therefore, we adopted the derived spectroscopic values from \texttt{SpecMatch-Synth} for the entire sample. These values are subjected to uncertainties of $\sigma_{T_{\mathrm{eff}}}=100$ K in effective temperature, $\sigma_{\mathrm{log}g}=0.1$ dex in surface gravity, $\sigma_{\mathrm{[Fe/H]}}=0.06$ dex in metallicity, and $\sigma_{v\sin i}=1.0$ \kms\ in projected velocity. 


\subsection{Broadband Photometry and Gaia Parallax} \label{sec:gaia}

We derived fundamental stellar properties from MIST \citep[MESA Isochrones and Stellar Tracks; ][]{choi2016} model grids, as implemented in \texttt{isoclassify}\footnote{\url{https://github.com/danxhuber/isoclassify}} \citep{huber2017}. Given a set of observables (i.e., photometry, spectroscopy, etc.), \texttt{isoclassify} interpolates the stellar grids to derive properties like their masses, luminosities, radii, densities and ages.

\begin{table*}
\begin{center}
\caption{Summary of subgiant host star properties\label{tab:stars}}\vspace{-0.35cm}
\begin{tabular}{rc | cccc | cccc}
\hline\hline
\noalign{\smallskip}
\multicolumn{2}{r|}{ } & \multicolumn{4}{c|}{Spectroscopy$^{*}$} & \multicolumn{4}{c}{Derived parameters$^{\dagger}\,$} \\
TOI & TIC & \teff\ [K] & \logg\ [cgs] & \feh\ & \vsini\ & \rstar [\rsun] & \mstar\ [\msun] & \rhostar\ [g~cm$^{-3}$] & Age [Gyr] \\
\noalign{\smallskip}
\hline
\noalign{\smallskip}
197 & 441462736 & $5080\pm90$ & $3.584\pm0.010$ & $-0.08\pm0.08$ & $<$2.0 & $2.943\pm0.064$ & $1.212\pm0.074$ & $0.067\pm0.001$ & $4.9\pm1.1$ \\
329 & 169765334 & 5635 & 4.00 & $+$0.18 & $<$2.0 & $1.52^{+0.04}_{-0.04}$ & $1.07^{+0.06}_{-0.04}$ & $0.300\pm0.034$ & $8.3^{+1.4}_{-2.0}$ \\ 
480 & 317548889 & 6224 & 4.23 & $+$0.17 & 8.7 & $1.49^{+0.05}_{-0.03}$ & $1.28^{+0.03}_{-0.03}$ & $0.385\pm0.032$ & $2.6^{+0.6}_{-0.6}$ \\ 
603 & 262746281 & 5894 & 4.07 & $+$0.20 & 4.0 & $1.58^{+0.03}_{-0.03}$ & $1.23^{+0.04}_{-0.07}$ & $0.309\pm0.020$ & $4.2^{+1.6}_{-0.7}$ \\ 
954 & 44792534 & 5788 & 3.85 & $+$0.40 & 3.4 & $1.90^{+0.04}_{-0.04}$ & $1.37^{+0.03}_{-0.11}$ & $0.196\pm0.015$ & $3.4^{+1.6}_{-0.4}$ \\ 
1181 & 229510866 & 6107 & 4.02 & $+$0.42 & 10.6 & $1.93^{+0.04}_{-0.04}$ & $1.47^{+0.03}_{-0.03}$ & $0.204\pm0.014$ & $2.2^{+0.3}_{-0.3}$ \\ 
1199 & 99869022 & 5619 & 3.92 & $+$0.42 & $<$2.0 & $1.46^{+0.03}_{-0.03}$ & $1.15^{+0.05}_{-0.06}$ & $0.366\pm0.028$ & $6.2^{+2.0}_{-1.4}$ \\ 
1294 & 219015370 & 5690 & 4.02 & $+$0.27 & $<$2.0 & $1.55^{+0.03}_{-0.03}$ & $1.16^{+0.07}_{-0.06}$ & $0.310\pm0.025$ & $6.1^{+1.9}_{-1.5}$ \\ 
1296 & 219854185 & 5562 & 3.88 & $+$0.42 & $<$2.0 & $1.68^{+0.04}_{-0.03}$ & $1.16^{+0.12}_{-0.04}$ & $0.241\pm0.033$ & $6.9^{+1.1}_{-2.4}$ \\ 
1298 & 237104103 & 5783 & 4.01 & $+$0.41 & 2.6 & $1.45^{+0.03}_{-0.03}$ & $1.21^{+0.03}_{-0.06}$ & $0.395\pm0.032$ & $4.3^{+1.8}_{-0.8}$ \\ 
1439 & 232982558 & 5843 & 4.02 & $+$0.22 & 2.6 & $1.61^{+0.03}_{-0.03}$ & $1.23^{+0.04}_{-0.09}$ & $0.291\pm0.024$ & $4.5^{+1.8}_{-0.8}$ \\ 
1601 & 139375960 & 6020 & 3.94 & $+$0.36 & 5.7 & $2.20^{+0.06}_{-0.07}$ & $1.51^{+0.04}_{-0.05}$ & $0.140\pm0.013$ & $2.3^{+0.4}_{-0.3}$ \\ 
1605 & 101037590 & 5752 & 4.06 & $+$0.00 & $<$2.0 & $1.49^{+0.03}_{-0.03}$ & $1.04^{+0.05}_{-0.04}$ & $0.308\pm0.030$ & $8.3^{+1.4}_{-1.4}$ \\ 
1736 & 408618999 & 5724 & 4.21 & $+$0.16 & $<$2.0 & $1.41^{+0.03}_{-0.03}$ & $1.08^{+0.06}_{-0.05}$ & $0.382\pm0.033$ & $7.3^{+1.8}_{-1.9}$ \\ 
1828 & 232982938 & 5780 & 4.04 & $+$0.14 & $<$2.0 & $1.56^{+0.03}_{-0.03}$ & $1.17^{+0.06}_{-0.07}$ & $0.302\pm0.023$ & $5.5^{+1.8}_{-1.3}$ \\ 
1836 & 207468071 & 6198 & 4.18 & $-$0.13 & 5.9 & $1.66^{+0.03}_{-0.03}$ & $1.22^{+0.03}_{-0.08}$ & $0.265\pm0.018$ & $3.5^{+1.3}_{-0.5}$ \\ 
1842 & 404505029 & 6084 & 3.98 & $+$0.33 & 6.2 & $2.02^{+0.04}_{-0.04}$ & $1.47^{+0.03}_{-0.04}$ & $0.175\pm0.013$ & $2.4^{+0.4}_{-0.3}$ \\ 
1885 & 258510872 & 6617 & 4.31 & $+$0.03 & 7.8 & $1.62^{+0.04}_{-0.04}$ & $1.19^{+0.05}_{-0.08}$ & $0.272\pm0.024$ & $3.7^{+1.4}_{-0.6}$ \\ 
1898 & 91987762 & 6241 & $4.18$ & $-$0.10 & 7.0 & $1.62^{+0.03}_{-0.03}$ & $1.23^{+0.03}_{-0.06}$ & $0.283 \pm 0.021$ & $3.4^{+1.0}_{-0.5}$ \\
2019 & 159781361 & 5630 & 3.89 & $+$0.41 & 1.6 & $1.73^{+0.03}_{-0.03}$ & $1.20^{+0.11}_{-0.04}$ & $0.233\pm0.024$ & $6.0^{+1.0}_{-2.0}$ \\ 
2145 & 88992642 & 6232 & 4.00 & $+$0.29 & 17.8 & $2.69^{+0.06}_{-0.06}$ & $1.78^{+0.04}_{-0.05}$ & $0.091\pm0.006$ & $1.2^{+0.1}_{-0.1}$ \\ 
\noalign{\smallskip}
\hline
\noalign{\smallskip}
\end{tabular}
\end{center}\vspace{-0.25cm}
{\sc Note--}\\
Values for TOI-197 were adopted from the asteroseismic analysis in \citet{huber2019}.\\
$^{*}\,$Spectroscopic parameters were derived with \texttt{SpecMatch-Synth} \citep{petigura2017a} (see Section \ref{sec:spec}), which has the following systematic uncertainties: $\sigma_{T_{\mathrm{eff}}}=100$ K in effective temperature; $\sigma_{\mathrm{log}g}=0.1$ dex in surface gravity; $\sigma_{\mathrm{[Fe/H]}}=0.06$ dex in metallicity; $\sigma_{v\sin i}=1.0$ \kms\ in projected velocity. \\
$^{\dagger}\,$Derived stellar parameters were estimated from model grids using \texttt{isoclassify} \citep{huber2017}. Reported uncertainties are intrinsic precisions of the sample. When comparing these values to other methods, systematic errors as described in \citet{tayar22} should be considered. \\
\end{table*}

We used the values of \teff\ and \feh\ (discussed in the previous section) as spectroscopic inputs, along with parallaxes from Gaia Data Release 3 \citep[DR3; ][]{gaia2021,gaia2023}. In order to further constrain the results and fit for extinction directly, we also included Johnson $BV$ photometry, $JHK_s$ from 2MASS \citep{skrutskie2006} and $G B_p R_p$ from Gaia \citep{gaia2021,gaia2023}. For Gaia DR3 photometry, we adopted a systematic noise floor of 0.01 mag. A summary of derived stellar parameters are provided in Table \ref{tab:stars}. We note that the uncertainties on the reported masses, radii, densities and ages are intrinsic precisions of the sample. When comparing these values to other methods, systematic errors as described in \citet{tayar22} should be considered.

\section{Planet Characterization} \label{sec:planet}

\subsection{System Confirmation} \label{sec:confirm}

We fitted Keplerian orbits to all systems with significant RV variations that are in phase with reported transits seen by TESS. Keplerian RV orbits are described by the orbital period ($P$), argument of periastron ($\omega$), orbital eccentricity ($e$), Doppler velocity semi-amplitude ($K$) as well as the time of inferior conjunction (\tc). We also included instrument-specific ``jitter'' (\sig) and RV offset ($\gamma$) terms. RV models can be further described by long-term properties which include linear ($\dot{\gamma}$) and quadratic ($\ddot{\gamma}$) acceleration terms, referred to as trend and curvature respectively. The two parameters $\dot{\gamma}$ and $\ddot{\gamma}$ capture additional Keplerian signals present in a time series with periods longer than the timespan of the available data. We did not fit for curvature in any system and linear trends were only added in systems that required an additional parameter to reasonably fit the data, the latter which is discussed in more detail in the following section. 

Each system was fit with a single-planet model using \texttt{radvel} \citep{fulton2018r}, which performs rigorous model fitting and selection using well-known metrics (i.e. BIC, AIC, etc.). Since transits typically constrain periods well, we placed strong Gaussian priors on the periods using the values and uncertainties reported by TESS. For conjunction times we placed flat, uniform priors and enforced positive Doppler semi-amplitudes ($K>0$). Finally, to avoid implicit biases we used the default \texttt{radvel} fitting basis \{$P$, \tc, $\sqrt{e}\cos\omega$, $\sqrt{e}\sin\omega$, $K$\}.

\begin{table}
\begin{center}
\caption{Models used during joint transit and RV fits (Section \ref{sec:global}).\label{tab:models}}
\setlength{\tabcolsep}{8pt}
\begin{tabular}{ccc}
\hline\hline
\noalign{\smallskip}
System & TOI & Model \\
\noalign{\smallskip}
\hline
\noalign{\smallskip}
HD~221416 & 197 & 1-planet model \\ 
TOI-329 & 329 & 1-planet model \\ 
HD~39688 & 480 & 1-planet model \\ 
TOI-603 & 603 & 1-planet model \\ 
TOI-954 & 954 & 1-planet model  \\ 
TOI-1181 & 1181 & 1-planet model \\ 
TOI-1199 & 1199 & 1-planet model \\ 
TOI-1294 & 1294 & 2-planet model $+$ linear trend \\ 
TOI-1296 & 1296 & 1-planet model \\ 
TOI-1298 & 1298 & 1-planet model \\ 
TOI-1439 & 1439 & 1-planet model \\ 
TOI-1601 & 1601 & 1-planet model \\ 
TOI-1605 & 1605 & 1-planet model $+$ linear trend \\ 
TOI-1736 & 1736 & 2-planet model $+$ linear trend \\ 
TOI-1828 & 1828 & 1-planet model $+$ linear trend \\ 
HD~148193 & 1836 & 1-planet model \\ 
TOI-1842 & 1842 & 1-planet model \\ 
TOI-1885 & 1885 & 1-planet model \\ 
HD~83342 & 1898 & 1-planet model \\ 
TOI-2019 & 2019 & 1-planet model \\ 
TOI-2145 & 2145 & 1-planet model \\ 
\noalign{\smallskip}
\hline
\noalign{\smallskip}
\end{tabular}
\end{center}
\end{table}

For a given iteration, the system was initialized with all available  parameters and then \texttt{radvel} iterated through all combinations of free parameters. In other words, we set the initial number of possible parameters, which in the case of a single planet here is at most eight (i.e. $P_b$, $K_b$, $e_b$, $\omega_b$, \tcb, $\sigma_i$, $\gamma_i$, $\dot{\gamma}$). Each system required several iterations of this procedure in order to identify the best possible solution and ultimately, confirm that we were recovering the same signal identified by TESS. 

\subsection{Searching for Additional Companions}

With continued RV follow up, some systems exhibited significant scatter that could not be explained by a single-planet model alone. Since it has been shown that RV data can be systematically contaminated by stellar variability \citep{lubin2021b}, we first estimated what the expected contribution would be from the host star given its physical properties. This step was especially important with our sample of evolved host stars, which are notoriously more noisy than their main-sequence counterparts. We calculated the stellar ``jitter'' using Equation 3 from \citet{chontos2022a}, which includes noise contributions from granulation and p-mode oscillations \citep{yu2018}, as well as rotation \citep{chontos2022a} and activity \citep{isaacson2010}. We then use four categories to understand their long-term characteristics:

\begin{table*}
\begin{center}
\caption{Priors used for the joint transit and RV fitting analysis discussed in Section \ref{sec:global}. \label{tab:prior}}
\setlength{\tabcolsep}{25pt}
\begin{tabular}{lccc}
\hline\hline
\noalign{\smallskip}
Parameter & Symbol & Units & Prior \\
\noalign{\smallskip}
\hline
\noalign{\smallskip}
\multicolumn{4}{c}{\textit{Default model parameters}} \\
\textit{Instrument-specific terms} & & & \\
TESS zero-point offset & $z$ & ppm & $\mathcal{U}[-\infty;+\infty]$ \\
Keck/HIRES RV offset & $\gamma$ & \ms\ & $\mathcal{U}[-\infty;+\infty]$ \\
\textit{Transiting planet parameters} & & & \\
Mean stellar density & {$\bar{\rho}_{\star}$} & g cm$^{-3}$ & $\mathcal{U}[0;+\infty]$ \\
Limb-darkening coefficient 1 & $u_1$ & & $\mathcal{N}(x;0.6)[0,2]$ \\
Limb-darkening coefficient 2 & $u_2$ & & $\mathcal{N}(x;0.6)[-1,1]$\\
Orbital period & $P$ & days & $\mathcal{U}[0;+\infty]$ \\
Time of inferior conjunction & $T_{\mathrm{c}}$ & days & $\mathcal{U}[-\infty;+\infty]$ \\
Impact parameter & $b$ & & $\mathcal{U}[0;1+R_{\mathrm{p}}/R_{\star}]$ \\
Planet-to-star radius ratio & \rprs\ & & $\mathcal{U}[0;+\infty]$ \\
Eccentricity & $e$ & & 1/$e$ \\
& $\sqrt{e}\sin\omega$ & & $\mathcal{U}[-1;1]$ \\
& $\sqrt{e}\cos\omega$ & & $\mathcal{U}[-1;1]$ \\
RV semiamplitude & $K$ & \ms\ & $\mathcal{U}[0;+\infty]$ \\
\noalign{\smallskip}
\hline
\noalign{\smallskip}
\multicolumn{4}{c}{\textit{Optional model parameters}} \\
\textit{RV trends} & & & \\
Linear RV trend$^{*}$ & $\Dot{\gamma}$ & \ms $\rm \,d^{-1}$ & $\mathcal{U}[-\infty;+\infty]$ \\
\textit{Nontransiting planet parameters} & & & \\
Orbital period & $P$ & days & $\mathcal{U}[0;0.5\times \mathrm{baseline}]$ \\
Time of inferior conjunction & $T_{\mathrm{c}}$ & days & $\mathcal{U}[-\infty;+\infty]$ \\
Eccentricity & $e$ & & 1/$e$ \\
& $\sqrt{e}\sin\omega$ & & $\mathcal{U}[-1;1]$ \\
& $\sqrt{e}\cos\omega$ & & $\mathcal{U}[-1;1]$ \\
RV semiamplitude & $K$ & \ms\ & $\mathcal{U}[0;+\infty]$ \\
\noalign{\smallskip}
\hline
\end{tabular}\vspace{-0.15cm}
\end{center}
$^{*}$Time reference was set to the RV baseline midpoint, which is provided in the tables for each system that included a trend. 
\end{table*}

\begin{figure*}
\centering
\includegraphics[width=\textwidth]{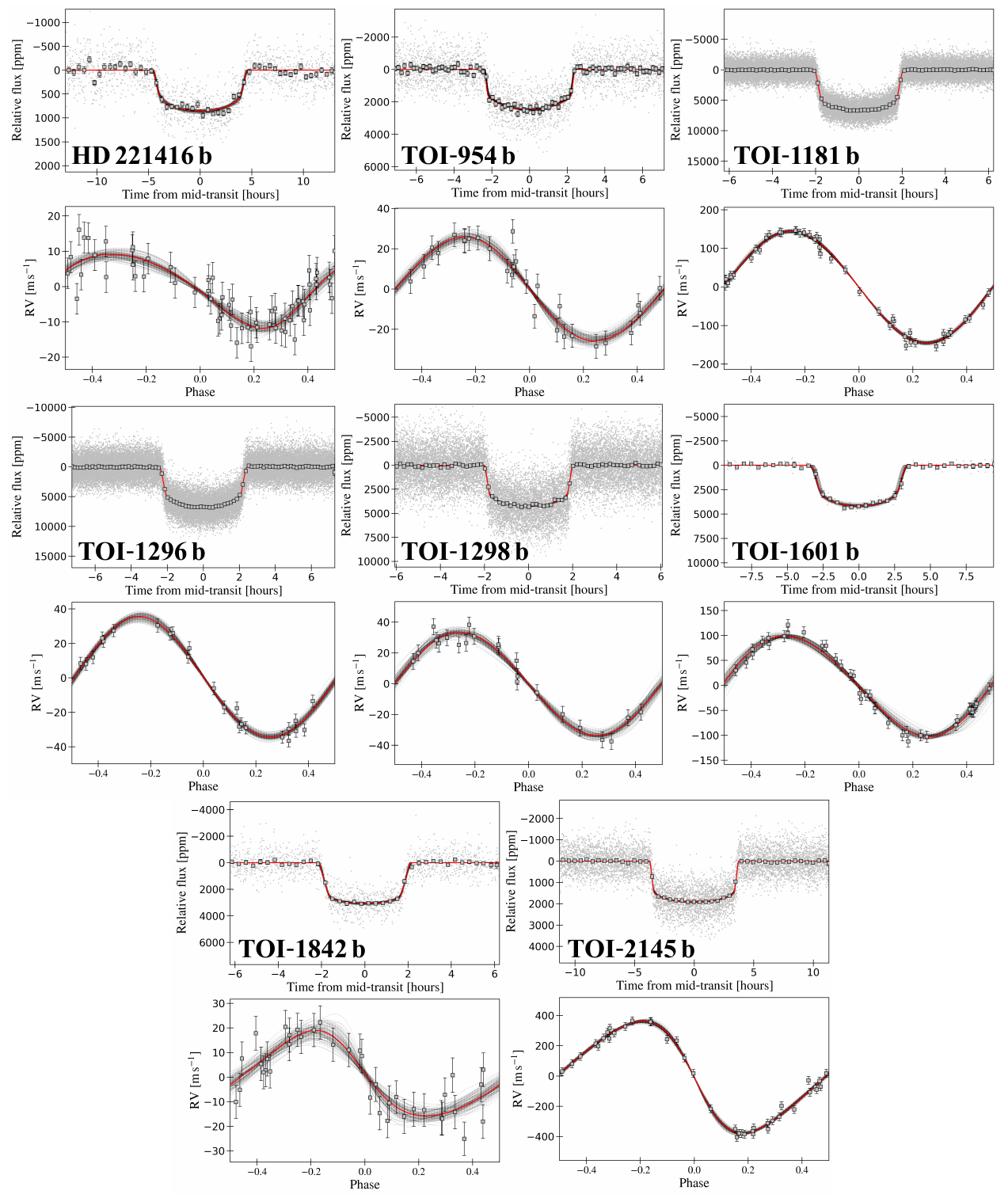}
\caption{Phase-folded transit and radial velocity data for previously confirmed single-planet TESS systems: HD~221416~b (TOI-197; top left), TOI-954~b (top middle), TOI-1181~b (top right), TOI-1296~b (middle left), TOI-1298~b (middle middle), TOI-1601~b (middle right), TOI-1842~b (bottom left), and TOI-2145~b (bottom right). Best-fit joint models are overplotted in red, including 100 samples randomly drawn from posterior distributions and plotted in gray (with transparency) to show model uncertainties.}\label{fig:known}
\end{figure*}

\begin{enumerate}
    \item Systems with observations that behaved as expected (observed scatter $\approx$ expected scatter). This included many of the known TESS systems in Figure \ref{fig:known}, including HD 221416, TOI-954, TOI-1296, TOI-1298, TOI-1601, and TOI-1842. TOI-603, a new confirmation, also falls into this category but has a limited number of RVs.
    
    \item Systems that show evidence for additional companions through significant linear trends. Figure \ref{fig:trends} shows 3 newly confirmed TESS systems, TOI-1294 (left), TOI-1605 (middle) and TOI-1828 (right), all which showed significant RV offsets between observing seasons that could only be explained by an additional companion. The other system is TOI-1736, which also exhibits an astrometric acceleration and therefore, may be affected by a distant but bound stellar companion. For the systems that included a linear trend component, time references were defined as the midpoint of the RV time series and are included in the supplementary table(s). 
    
    \item Systems with scatter higher than expected due to the presence of new non-transiting companions detected in the RV observations. Here, we define ``non-transiting'' as a planet with no reported detection of a transit in the TESS photometry. For systems in this category, we implemented a blind search approach by adding parameters for a second planet to the model, whose only constraint was the orbital period, which was restricted to one half of the total observation baseline. To facilitate the interpretation of various models, the analysis was complemented by \texttt{rvsearch}\footnote{\url{https://github.com/California-Planet-Search/rvsearch}}, which is a tool to search for Keplerian signals in RV time series data \citep{rosenthal2021}. For example, we typically ranked a potential planet relatively high when \texttt{radvel} converged on a fairly precise period which was then matched by a significant period peak in the \texttt{rvsearch} results. In other words, \texttt{rvsearch} was not used to enforce more strict priors but used as a second interpretation of the system. Our three new, non-transiting planetary detections are discussed in Section \ref{sec:new} in more detail. 
    
    \item Systems with higher residual RV scatter than expected but no straightforward model to best explain the current set of observations. For systems like TOI-1181, TOI-1885 and TOI-2145, residual scatter exceeded 50 \ms. The residuals were also high for TOI-329, TOI-1199, TOI-1439, HD~148193 (TOI-1836), HD~83342 (TOI-1898) and TOI-2019 but did not exceed 50 \ms. In none of these systems was there an obvious way to reduce the scatter with an additional companion and consequently, more RV observations are needed to reveal the true nature of the systems. 

\end{enumerate}

\subsection{Joint Transit \& RV Fitting} \label{sec:global}

Table \ref{tab:models} defines the models used for each system. We simultaneously fit TESS photometry and RV data for each target, as indicated in Table \ref{tab:phot} and Table \ref{tab:RVs}, respectively. For fitting we followed \citet{chontos2019}, who used the \texttt{ktransit}\footnote{\url{https://github.com/mrtommyb/ktransit}} package, an implementation of the analytical model by \citet{mandel2002}. The analysis uses the following parameters as model input: orbital period ($P$), mid-transit time ($T_0$), quadratic limb-darkening coefficients ($u_1,u_2$), mean stellar density (\rhostar), impact parameter ($b$), ratio of the planet radius to the stellar radius (\rprs) and Doppler amplitude ($K_{\star}$). Other free parameters include the photometric ($z$) and spectroscopic ($\gamma$) zero-point offsets, the eccentricity of the orbit as well as the argument of periapsis, where $e$ and $\omega$ are re-parameterized to avoid parameter biases.

Table \ref{tab:prior} defines the standard priors assigned in the joint transit and RV fitting, which were generally flat, uninformative priors. We used weak priors for quadratic limb-darkening coefficients (LDCs) $\{u_1,u_2\}$ by setting Gaussian priors with wide widths ($\sigma=0.6$) and centers set by theoretical TESS band values from \citet{claret2016}, which are interpolated in \teff, \logg\ and \feh. Moreover, additional constraints on $u_1$ and $u_2$ were implemented to prevent the limb-darkening properties from taking nonphysical values \citep{burke2008,barclay2015}. We also implemented the $\{q_1,q_2\}$ parametrization from \citet{kipping2013} to more efficiently sample the two-dimensional LDC parameter space. We sampled uniformly in $e\cos\omega$ and $e\sin\omega$ and therefore also added an additional Jeffreys prior (1/$e$) to effectively set a uniform prior on eccentricity \citep{eastman2013}.

To sample the multi-dimensional parameter space, we used the affine-invariant ensemble Markov Chain Monte Carlo (MCMC) sampler, \texttt{emcee} \citep{fm2013}. For joint global fits, we assumed a linear ephemeris and quadratic limb darkening law.  Every joint transit and RV fit initialized 50 walkers, i.e., at least twice the maximum number of free parameters for a given system. Each walker took $10^4$ steps and then implemented a burn-in phase of $10^3$ steps before concatenating samples to obtain the final posterior distributions for the parameters. 

Table \ref{tab:planets} summarizes the primary properties of interest for the full subgiant planet sample, which uses HD designations where available and remaining targets are assigned by their TOI number. More than half are newly confirmed TESS systems, comprising TOI-329, HD~39688 (TOI-480), TOI-603, TOI-1199, TOI-1294, TOI-1439, TOI-1605, TOI-1828, HD~148193 (TOI-1836), TOI-1885, HD~83342 (TOI-1898) and TOI-2019. We also provide updated, homogeneous parameters for the 9 other known TESS systems in Table \ref{tab:sample} \citep{huber2019,rodriguez2021,sha2021,kabath2022,wittenmyer2022,rodriguez2023,murphy2023}. 


\subsection{Previously known TESS Systems}

Our homogeneous sample of precise RVs combined with considerably longer time baselines than previous studies provides the opportunity to better characterize and understand evolved TESS systems. Figure \ref{fig:known} shows transit and RV data for previously confirmed systems plotted with our global results, as well as updated properties for each system in the supplementary tables. Here we summarize any major differences we find in the updated properties for known TESS systems.

Our most evolved system, HD 221416 (TOI-197), was the first asteroseismic host discovered by TESS and sits at the base of the red giant branch. Our derived mass of \massp\ $=44.6\pm4.0$ \masse\ is nearly one third smaller than the value of \massp\ $=60.5\pm5.7$ \masse\ reported in \citet{huber2019}, highlighting the importance for long-term RV monitoring. We also find a mild eccentricity ($e=0.167\pm0.050$) that is consistent with their reported value of $e=0.115\pm0.032$. Combining the eccentricity with the asteroseismic age, \citet{huber2019} constrained the planetary tidal quality to be near or larger than a lower limit of $Q_{\mathrm{p}}\approx3.2\times10^4$, because otherwise the orbit would have already circularized within its lifetime of $\sim$5 Gyr.

First identified by \citet{sha2021}, TOI-954 b is a short-period ($P=3.68$ days) Saturn-sized planet. \citet{sha2021} reported a value of \massp\ $=55.3\pm5.6$ \massp\ while we found a mass that is $4\sigma$ higher (\massp\ $=76.8\pm5.6$ \masse). Our median eccentricity of $e=0.045$ is smaller than the value of $e=0.14$ reported by \citet{sha2021}, but both estimates are consistent with zero. Finally, our $10\sigma$ mass measurement improves the bulk planet density precision by a factor of two, with \rhop\ $=0.50^{+0.06}_{-0.05}$ \gcc.

\citet{rodriguez2021} reported a bimodal posterior probability distributions for the mass and age of the TESS planet host TOI 1601, which they attributed to the star's relatively ambiguous location on an HR diagram given the precision of their observations. Our analysis agrees with their younger, more massive solution, with an age of $2.3 \pm 0.4$ Gyr and mass of $1.51 \pm 0.05$ $M_{*}$. 

\section{New TESS Systems} \label{sec:new}

Phase-folded data and global fits are shown for all new, single-planet TESS systems in Figure \ref{fig:singles}. New systems with significant RV trends, including TOI-1294, TOI-1605 and TOI-1828, are shown in Figure \ref{fig:trends}. We discuss each of the new TESS confirmations in the following subsections below. Planet properties derived from the global fits are summarized in Table \ref{tab:planets}.

\subsection{TOI-329} 

The $V$$=$$11.3$ subgiant star is a cool (\teff\ $=5660\pm75$ K), low mass (\mstar\ $=1.08\pm1.03$ \msun) star with a moderate metallicity of \feh\ $=+0.16\pm0.05$. The TOI-329.01 $\sim$1 parts per thousand (ppt) transit-like events were also detected twice by TFOP SG1 ground-based transit observations in Sloan $i'$ band using 7.8'' photometric apertures that excluded the flux of all known Gaia DR3 stars. The SPOC difference image analysis for TOI-329.01 constrained the host star to within 1.2$\pm$3.3'' of the location of the transit source.

\begin{figure*}
\centering
\includegraphics[width=\textwidth]{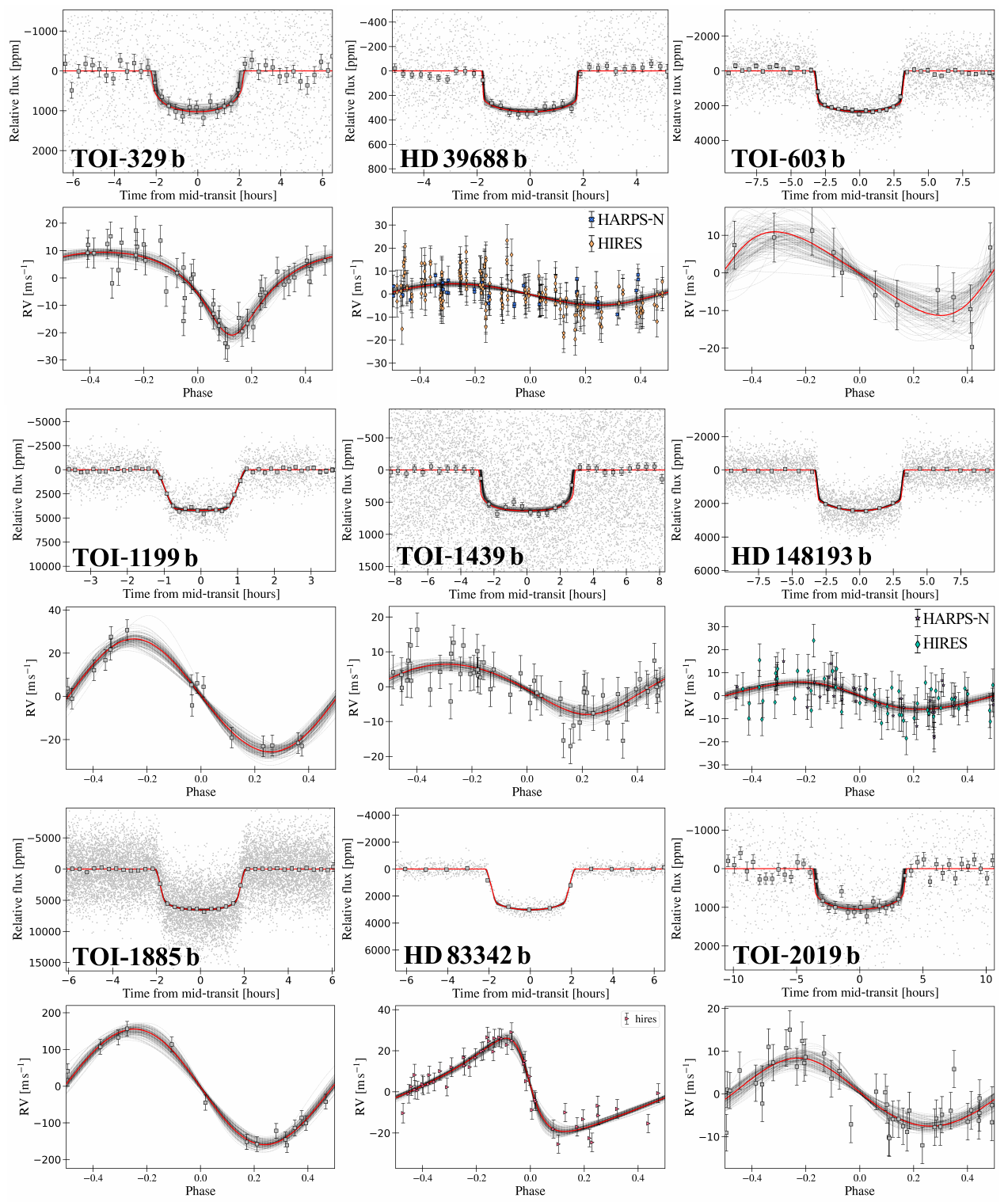}
\caption{Same as Figure \ref{fig:known} but for new TESS single-planet systems: TOI-329~b (top left), HD~39688~b (TOI-480, top middle), TOI-603~b (top right), TOI-1199~b (middle left), TOI-1439~b (middle middle), HD~148193~b (TOI-1836; middle right), TOI-1885~b (bottom left), HD~83342~b (TOI-1898; bottom middle), and TOI-2019 (bottom right). Best-fit joint models are overplotted in red with 100 samples randomly drawn from posterior distributions and plotted in gray (with transparency) to demonstrate the model uncertainties.}\label{fig:singles}
\end{figure*}

Follow-up RV measurements resolved the 5.7 day orbit, where the phase-folded HIRES data in Figure \ref{fig:singles} (top left) clearly demonstrate that the planet has a significantly ($\sim$$13\sigma$) non-zero eccentricity. The planet is slightly larger than Neptune at \radiusp $\sim4.7$ \radiuse, which is a rare planet type in the context of the observed planet radius distribution (e.g., \citealt[]{fulton2017}{}). We report a mass of \massp\ $=40.8\pm3.6$ \masse\ and a bulk density of \rhop\ $=2.13\pm0.54$ \gcc. The combination of the relatively small separation (\ator$\leq10$) and older age ($8.3\pm1.0$ Gyr) would suggest that the orbit has had sufficient time to circularize via tidal interactions of the planet with the host star. Yet the eccentricity is found to be moderate ($e=0.39\pm0.03$), suggesting some other dynamical process as a potential source for the non-zero eccentricity.

\subsection{HD~39688 (TOI-480)}

TFOP SG1 ground-based lightcurves ruled out nearby eclipsing binaries as the source of the TOI-480.01 transit-like event detected in the TESS data. However, HD~39688 ($V$$=$$7.3$) is a challenging system since the host star has a rotational velocity of $\sim$8 \kms\ while the estimated Doppler amplitude of the transiting $\sim$2.8 \radiuse\ candidate was 2.8 \ms. The $\sigma_{\mathrm{RV}}-$\vsini\ relation of \citet{chontos2022a} predicted an additional scatter of $\sim$2.6 \ms\ in the time series due to the rotation. Therefore, during each visit to HD~39688 we obtained three back-to-back observations to improve the per-visit uncertainty by a factor of $\sqrt{3}$. The transit and RV observations indeed confirm the planetary nature of HD~39688~b (\massp$\,=15.7\pm2.2$ \masse), which has a bulk density consistent with a rocky, terrestrial planet ($\rho_{\mathrm{p}}=4.4\pm0.8$ gcc). 

Due to the high \vsini, we checked the raw TESS photometry (SAP-FLUX) for evidence of stellar rotation via star spot modulation. We identified a significant peak at a period of \prot\ $\sim6.92$ days, which if due to the host star rotation, suggests that the system is either near or at synchronous rotation with the transiting planet, which has an orbital period, $P\sim6.86$ days. The stellar inclination to the line of sight measured from the observations (by combining \vsini, \prot, \rstar) is $\sim52^{o}$, which suggests a probable misalignment of the orbital plane of the planets with respect to the rotation axis of the star. In the event that the misalignment is confirmed then synchronization would not be expected.


\subsection{TOI-603}

The effective temperature and age of TOI-603 are roughly consistent with those of the Sun, with \teff\ $=5850\pm70$ K and an age of $4.9\pm1.2$ Gyr. The subgiant star is larger and more massive though, with a size of \rstar\ $=1.56\pm0.03$ \rsun\ and a mass of \mstar\ $=1.20\pm0.05$ \msun. With only 12 HIRES RVs, the mass measurement of \massp\ $=47.5\pm13.7$ \masse\ has the lowest statistical significance of all new confirmations, but the $\sim$$3\sigma$ detection is sufficient to confirm the planetary nature of the transiting planet, TOI-603~b.

TOI-603.01 is one example from the TESS Primary Mission where an alias of the true period was reported; it was only observed for a single sector during the nominal mission, and it was initially assumed that a transit occurred during the data gap. Further TFOP efforts consistently confirmed that the period is close to twice that of the previously reported period, which is supported by the phase-folded HIRES RVs shown in Figure \ref{fig:singles}. Fortunately the target was re-observed in 2-minute cadence in the first Extended Mission in sectors 35, 45 and 46, which provided more transits to better constrain the observed planet properties (i.e. \radiusp\ $=7.93\pm0.20$ \radiuse, $P=16.17989\pm0.00006$ days). 

\subsection{TOI-1199}

TOI-1199 is one the smallest host stars in the sample, with \rstar\ $=1.46\pm0.02$ \rsun, \teff\ $=5630\pm54$ K and an age of $\sim6.3$ Gyr. The transiting planet, TOI-1199~b, is a giant planet (\radiusp\ $=10.6\pm0.3$ \radiuse) on an approximately circular orbit with a period of $P\sim3.7$ days. Using 16 HIRES RVs, we report a precise mass of \massp\ $=69.4\pm5.7$ \masse\ and bulk density of \rhop\ $=0.32\pm0.04$ g$\,$cm$^{-3}$. 

Intriguingly, a total of 7 different threshold crossing events (TCEs) with similar periods were all assigned to the same TIC, which we speculate is evidence for transit timing variations (TTVs). We also found additional power near the reported period in the residual RV periodogram after subtracting out the single-planet model, which could be evidence for another planet in the system. However given the small number of RVs, more observations are needed to confirm the existence of any additional companions. 

\subsection{TOI-1294} \label{sec:1294}

TOI-1294 is a massive (\mstar\ $=1.20\pm0.05$ \msun) and metal-rich (\feh\ $=+0.28\pm0.06$) host star, with an effective temperature of \teff\ $\sim5770\pm80$ K and age of $\sim5.2^{+1.4}_{-0.9}$ Gyr. Achromatic $\sim$3 ppt transit-like events were also detected on target by TFOP SG1 ground-based transit observations in Sloan $g'$ and Sloan $i'$ bands using 5.8'' or smaller photometric apertures that excluded the flux of all known Gaia DR3 stars. Initial follow up RVs immediately confirmed the planetary nature of the transiting, Saturn-sized (\radiusp$\,=9.2\pm0.3$) TOI-1294~b, with a mass of \massp$\,=62.2\pm5.0$ \masse\ on a $\sim3.9$-day orbit. 

\begin{figure*}
\centering
\includegraphics[width=\textwidth]{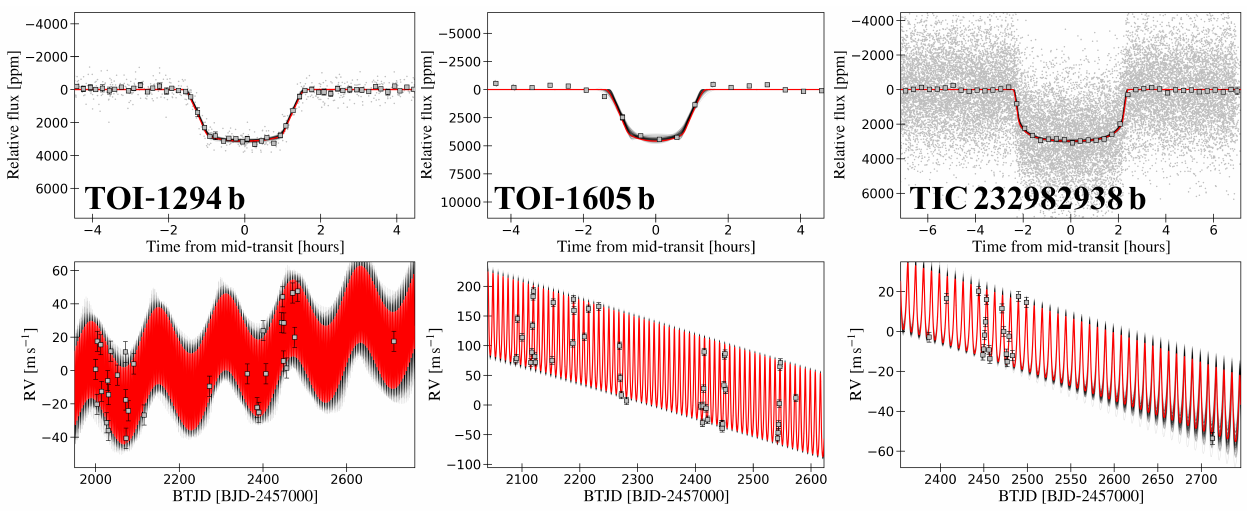}
\caption{Phase-folded TESS photometry (top) and Keck/HIRES RV time series (bottom) with the joint transit and RV fit overplotted in red. The RV observations presented here enabled the discovery of non-transiting planetary companion, TOI-1294~c (Section \ref{sec:1294}), as well as three additional outer companions observed via significant trends in TOI-1294 (left), TOI-1605 (middle) and TOI-1828 (right).}\label{fig:trends}
\end{figure*}

A significant offset was observed during the second observing season near the 2400 time stamp, seen in the bottom left panel of Figure \ref{fig:trends}, which could only be explained by at least one additional companion. A blind RV search unambiguously recovered a Keplerian signal at a period of $P=160.9\pm2.5$ days, with a mass corresponding to \massp$\,=148.3\pm17.3$ \masse, hereafter referred to as TOI-1294~c. The best-fit model also exhibits an RV trend of $\dot{\gamma}= 0.052\pm 0.007$ \msd\ ($T_{\mathrm{ref}}=2459198.114897$), where further RV observations will enable a more comprehensive understanding of the evolved, multi-planet TOI-1294 system.

\subsection{TOI-1439}

TOI-1439 ($V$$=$$10.6$) has an effective temperature \teff\ $=5857\pm60$ K and metallicity of \feh\ $=+0.22\pm0.06$. The radius and mass of the subgiant host are \rstar\ $=1.64$ \rsun\ and \mstar\ $=1.24$ \msun, respectively, and it has an intermediate age of $\sim4.5^{+1.6}_{-0.6}$ Gyr. The observed activity indicator from the RV time series, \logrhk\ $=-5.26$, suggests that the star is relatively inactive. This is further supported by back-to-back HIRES observations obtained on UT 2020-07-27, which were separated by one hour and showed very little jitter.

TOI-1439~b is a Neptune-sized (\radiusp\ $=4.1\pm0.2$ \radiuse) planet on an orbit with a period of $P\sim27.6$ days, which is slightly longer than a single TESS sector. We measured a mass and bulk density corresponding to \massp\ $=38.4\pm5.5$ \masse\ and \rhop\ $=2.9\pm0.6$ \gcc, respectively. After subtracting the best-fit model from the RV time series, however, the residual RVs still showed high scatter and given the relatively inactive host star, could indicate the presence of additional planets in the system.

\subsection{TOI-1605}

TOI-1605 ($V$$=$$10.2$) has an effective temperature and metallicity which are fully consistent with those of the Sun, with \teff\ $=5774\pm75$ K and \feh\ $=+0.03\pm0.06$. The host star also has a radius of \rstar\ $=1.51\pm0.03$ \rsun, a mass of \mstar\ $=1.05\pm0.03$ \msun, and an age of $\sim7.9\pm1.0$ Gyr. Achromatic $\sim$5 ppt transit-like events of TOI-1605.01 were detected by TFOP SG1 ground-based transit observations in multiple bands spanning Sloan $g'$ through Pan-STARRS $z$-short bands. 

TOI-1605~b is currently challenging to characterize, given the limited availability of TESS photometry and the high impact parameter that is implied by the v-shaped transit shown in Figure \ref{fig:trends} (middle). The Jupiter-sized planet (\radiusp\ $=10.34\pm0.64$ \radiuse) has a mass of \massp\ $=221.3\pm7.7$ \masse\ and is on an eccentric ($e=0.28\pm0.01$) orbit, which is interesting given the short orbital period ($<10$ days). As seen in the bottom of Figure \ref{fig:trends}, the RV observations also indicate the presence of an outer companion ($\dot{\gamma}= -0.298 \pm 0.006$ \msd\ using a time reference of $T_{\mathrm{ref}}=2459497.065739$) that could be responsible for the observed high eccentricity but more RVs are needed to characterize the outer companion.

\subsection{TOI-1828}

The host star TOI-1828 ($V$$=$$11.6$) is a subgiant with an effective temperature of \teff\ $=5680$ K, a radius of \rstar\ $=1.60$ \rsun, and a mass of \mstar\ $=1.06$ \msun. TFOP SG1 ground-based transit observations in Sloan $g'$ and Sloan $i'$ bands also detected achromatic $\sim$3 ppt transit-like events of TOI-1828.01 using 7.8'' or smaller photometric apertures that excluded the flux of all known Gaia DR3 stars. The SPOC difference image analysis also constrained the host star to within 0.12$\pm$2.4'' of the location of the transit source.

The close-in ($P=9.094$ days), eccentric ($e=0.31\pm0.03$) HJ is strikingly similar to TOI-1605~b (discussed above), with the exception of the lower planet mass of \massp\ $=58.5\pm3.8$ (and bulk density of \rhop\ $=0.50\pm0.05$ \gcc). The panels on the right side in Figure \ref{fig:trends} show the phase-folded TESS photometry (top) and RV time series (bottom). The last RV observation clearly indicates an additional companion in the system. As expected though, the single, most recent RV does not constrain the slope well but the measured trend is significant nevertheless, with $\dot{\gamma}= -0.147\pm 0.012$ \msd\ ($T_{\mathrm{ref}}=2459445.570399$).

\begin{table*}[ht!]
\begin{center}
\caption{Summary of derived planet properties from the joint transit and RV fits.\label{tab:planets}}
\begin{tabular}{ccccccc}
\noalign{\smallskip}
\hline\hline
\noalign{\smallskip}
Planet & $P \,\, \mathrm{[days]}$ & Epoch [BTJD] & $R_p \,\, [R_{\oplus}]$ & $M_p \,\, [M_{\oplus}]$ & $\rm \bar{\rho}_{_p} \,\, [g\, cm^{-3}]$ & $e$ \\
\noalign{\smallskip}
\hline
\noalign{\smallskip}
HD~221416~b & $14.2790\pm0.0027$ & $1357.0127\pm0.0022$ & $9.36 \pm 0.33$ & $44.6 \pm 4.0$ & $0.30 \pm 0.04$ & $0.17 \pm 0.05$ \\ 
TOI-329~b & $5.7044\pm0.0001$ & $2090.7935\pm0.0049$ & $4.72 \pm 0.36$ & $40.8 \pm 3.6$ & $2.13 \pm 0.54$ & $0.39 \pm 0.03$ \\ 
HD~39688~b & $6.86588\pm0.00001$ & $1469.5660\pm0.0016$ & $2.69 \pm 0.10$ & $15.7 \pm 2.2$ & $4.42 \pm 0.80$ & $0.10 \pm 0.07$ \\ 
TOI-603~b & $16.17989\pm0.00006$ & $2268.0635\pm0.0015$ & $7.93 \pm 0.20$ & $47.5 \pm 13.7$ & $0.52 \pm 0.16$ & $0.22 \pm 0.19$ \\ 
TOI-954~b & $3.6843\pm0.0002$ & $2145.2170\pm0.0008$ & $9.47 \pm 0.25$ & $76.8 \pm 5.6$ & $0.50 \pm 0.06$ & $0.05 \pm 0.04$ \\ 
TOI-1181~b & $2.1031937\pm0.0000005$ & $1957.8213\pm0.0001$ & $16.14 \pm 0.34$ & $374.7 \pm 8.9$ & $0.49 \pm 0.03$ & $0.01 \pm 0.01$ \\ 
TOI-1199~b & $3.67147\pm0.00001$ & $2611.4540\pm0.0005$ & $10.60 \pm 0.33$ & $69.4 \pm 5.7$ & $0.32 \pm 0.04$ & $0.05 \pm 0.04$ \\ 
TOI-1294~b & $3.91529\pm0.00001$ & $2393.0110\pm0.0033$ & $9.19 \pm 0.31$ & $62.2 \pm 5.0$ & $0.44 \pm 0.06$ & $0.07 \pm 0.04$ \\ 
TOI-1294~c & $160.1 \pm 2.5$ & $2349.0106\pm5.7665$ & -- & $148.3 \pm 17.0$ & -- & $0.10 \pm 0.09$ \\ 
TOI-1296~b & $3.944373\pm0.000001$ & $1930.7553\pm0.0002$ & $13.93 \pm 0.30$ & $95.4 \pm 5.35$ & $0.19 \pm 0.02$ & $0.02 \pm 0.02$ \\ 
TOI-1298~b & $4.537143\pm0.000003$ & $1929.5853\pm0.0003$ & $9.64 \pm 0.21$ & $98.3 \pm 5.2$ & $0.60 \pm 0.05$ & $0.03 \pm 0.03$ \\ 
TOI-1439~b & $27.643927\pm0.000090$ & $1703.4752\pm0.0023$ & $4.15 \pm 0.18$ & $38.4 \pm 5.5$ & $2.94 \pm 0.57$ & $0.15 \pm 0.07$ \\ 
TOI-1601~b & $5.33206\pm0.00034$ & $1793.2741\pm0.0026$ & $14.07 \pm 0.49$ & $361.4 \pm 7.6$ & $0.71 \pm 0.08$ & $0.07 \pm 0.04$ \\ 
TOI-1605~b & $8.7099\pm0.0006$ & $2887.4759\pm0.0766$ & $10.34 \pm 0.64$ & $221.3 \pm 7.7$ & $1.10 \pm 0.21$ & $0.28 \pm 0.01$ \\ 
TOI-1736~b & $7.073091\pm0.000008$ & $2740.5891\pm0.0007$ & $3.05 \pm 0.19$ & $11.9 \pm 1.6$ & $2.30 \pm 0.50$ & $0.04 \pm 0.04$ \\ 
TOI-1736~c & $571.3 \pm 0.5$ & $2273.1\pm0.4$ & -- & $2477 \pm 118$ & -- & $0.37 \pm 0.01$ \\ 
TOI-1828~b & $9.0941045\pm0.0000083$ & $1936.2770\pm0.0007$ & $8.62 \pm 0.19$ & $58.5 \pm 3.8$ & $0.50 \pm 0.05$ & $0.31 \pm 0.03$ \\ 
HD~148193~b & $20.38085\pm0.000025$ & $1933.1655\pm0.0008$ & $8.38 \pm 0.19$ & $28.4 \pm 4.3$ & $0.27 \pm 0.04$ & $0.14 \pm 0.11$  \\ 
TOI-1842~b & $9.573922\pm0.000013$ & $1933.3360\pm0.0008$ & $12.29 \pm 0.32$ & $73.4 \pm 7.5$ & $0.22 \pm 0.03$ & $0.18 \pm 0.08$ \\ 
TOI-1885~b & $6.544060\pm0.000006$ & $1958.2553\pm0.0008$ & $13.66 \pm 0.41$ & $516.4 \pm 32.7$ & $1.12 \pm 0.12$ & $0.04 \pm 0.03$ \\ 
HD~83342~b & $45.522149\pm0.000039$ & $1894.2540\pm0.0005$ & $9.74 \pm 0.22$ & $127.5 \pm 6.6$ & $0.76 \pm 0.07$ & $0.48 \pm 0.04$ \\ 
TOI-2019~b & $15.3444\pm0.0055$ & $1958.2895\pm0.0040$ & $5.66 \pm 0.28$ & $34.6 \pm 4.2$ & $1.05 \pm 0.21$ & $0.09 \pm 0.07$ \\ 
TOI-2145~b & $10.26111\pm0.00001$ & $2013.2807\pm0.0006$ & $12.25 \pm 0.30$ & $1810.2 \pm 42.5$ & $5.42 \pm 0.42$ & $0.21 \pm 0.02$ \\ 
\noalign{\smallskip}
\hline
\end{tabular}
\end{center} \vspace{-0.25cm}
\end{table*}

\subsection{HD~148193 (TOI-1836)}

HD~148193 ($V$$=$$\,9.6$) is a young ($\sim$$3.3$ Gyr) and hot ($\sim$$6235$ K) star, and is one of the more metal-poor subgiants in the sample, with \feh\ $=-0.16$. As discussed in Section \ref{sec:image}, HD~148193 is the only host star in the sample with any known stellar companions. High-resolution imaging revealed a close ($<1''$, $\Delta m=5.7$) companion and a $0.2''$ binary (TIC~207468069) that is $\sim$$10''$ away. TFOP SG1 ground-based transit observations in Pan-STARRS $z$-short band and Cousins I band detected $\sim$2.5 ppt transit-like events on target. The transiting $\sim$20-day planet HD~148193~b has a size of \radiusp\ $=8.38\pm0.19$ \radiuse\ and a mass of \massp\ $=28.4\pm4.3$ \masse. 

The SPOC pipeline more recently identified another candidate signal using data from the extended mission. The candidate is interior to the confirmed planet, with a radius of \radiuse\ $=2.5$ \radiuse\ and a period of $P=$ 1.77 days, but ultimately its confirmation is beyond the scope of this paper.

\subsection{TOI-1885}

TOI-1885 is a young ($2.0\pm0.3$ Gyr) and hot (\teff\ $=6527\pm98$ K) host star with a solar-like metallicity (\feh\ $=0.02\pm$0.05). Gaia DR3 also reported an effective temperature that is fully consistent with the value of  derived from the spectrum. The effective temperature of the star places it on the hot side of the Kraft break \citep{kraft1967}, suggesting that the host star lacks a significant convective envelope. 

Our analysis estimates a planet size of \radiusp\ $=13.66\pm0.41$ \radiuse\ and mass of \massp\ $=516.4\pm32.7$ \masse\ ($\sim1.6$ \massj), which together constrain the bulk density to $\sim$$10\%$, with \rhop$=1.12\pm0.12$ \gcc. The close-in giant planet has an orbital period of $P=6.544$ days and appears to be on a nearly circular orbit, but more RVs could better constrain the eccentricity of the orbit. 

\begin{figure*}
\includegraphics[width=\linewidth]{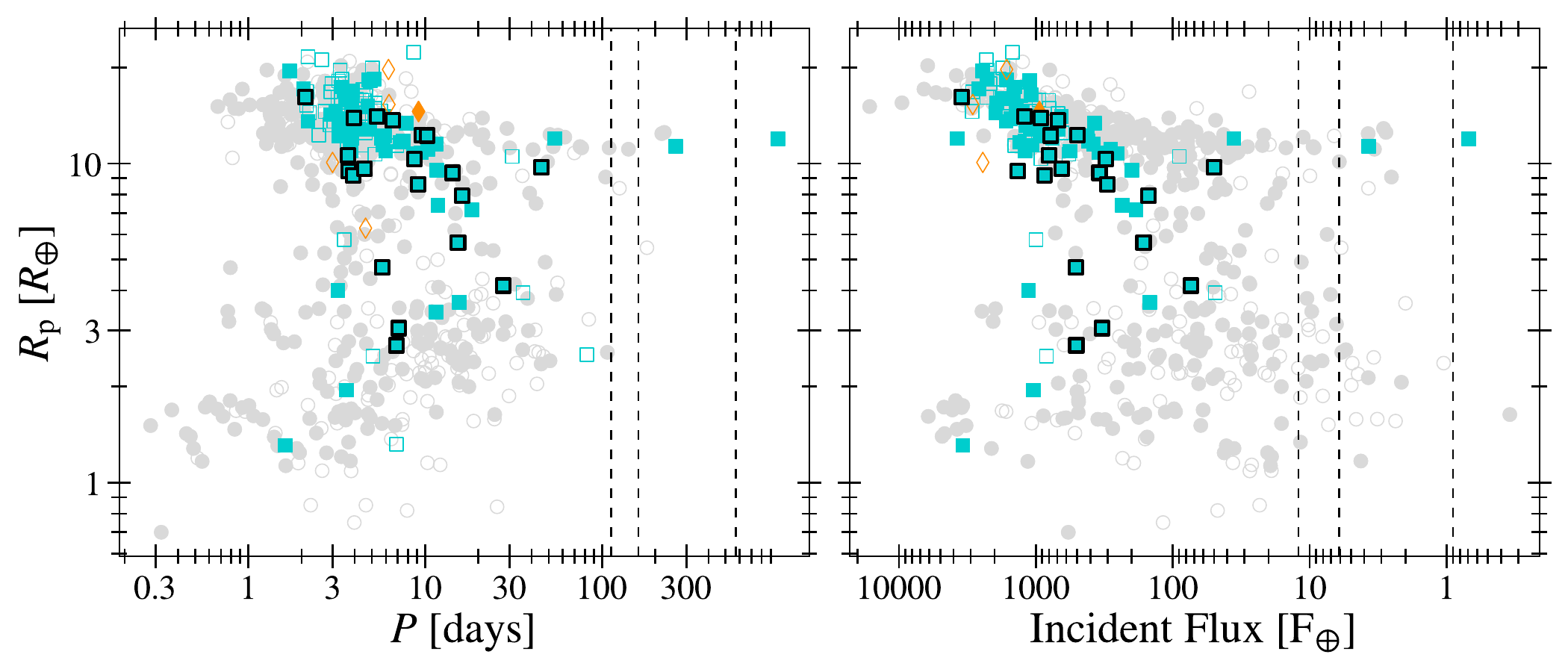}
\caption{Planet size versus orbital period (left) and incident flux (right) for all planets with measured radii and masses. Evolutionary states of the host stars are indicated by different colors: main-sequence (grey), subgiants (cyan) and red giants (orange). Filled markers show planets with the most precise densities ($\geq4\sigma$). Our subgiant sample is highlighted with thick black outlines, and positions for new non-transiting planets (i.e. with no measured sizes) are shown by vertical, dashed lines.}\label{fig:combo1}
\end{figure*}

\subsection{HD~83342 (TOI-1898)}

HD~83342 ($V$$=$$7.9$) was originally a community TESS object of interest due to the single transit that occurred during the nominal mission. However the target was reobserved for two additional sectors in the extended mission, hence providing two additional transits to better constrain the orbital period of the transiting candidate TOI-1898.01. The planet HD~83342~b is a warm Jupiter ($P=45.5$ days) with a mass of \massp\ $=127.5\pm6.6$ \masse\ and is on a highly eccentric orbit with $e=0.48\pm0.04$, as shown in the phase-folded RVs in the bottom left of Figure \ref{fig:singles}. The moderate \vsini\ of the host star also makes the system an ideal target to measure the sky-projected obliquity via Rossiter-McLaughlin or Doppler tomography observations, and especially because the highly eccentric, warm Jupiter could help elucidate gas giant formation and migration pathways.

\subsection{TOI-2019}

TOI-2019 is a cool ($\sim$$5590$ K), metal-rich (\feh\ $=+0.40\pm0.06$) star with an age of $6.7\pm0.6$ Gyr. Comparatively, the subgiant host has a more moderate size, with \rstar\ $=1.75\pm0.03$ \rsun\ and a mass of \mstar\ $=1.17\pm0.04$ \msun. The transiting planet, TOI-2019~b, is a rare planet type in context with the current distribution of planetary radii, with \radiusp\ $=5.7\pm0.3$ \radiuse, and is also one of the longer period planets in our sample, with $P=15.35$ days. Our measured mass is \massp\ $=34.60\pm4.26$ \masse, which translates to a bulk planet density of \rhop\ $=1.05\pm0.20$ \gcc.

\section{Discussion} \label{sec:disc}

\subsection{Planet Sizes and Multiplicity}

Figure \ref{fig:combo1} shows planet sizes versus orbital period and incident flux for all planets with measured radii and masses, separated according to the evolutionary state of the host star. Literature values were taken from the NASA Exoplanet Archive\footnote{\url{https://exoplanetarchive.ipac.caltech.edu}}. Our sample increases the number of known planets with measured radii and masses orbiting subgiant stars by 25\%, bringing the total population to $106$ planets. Further, our sample contribution increases to 50\% when only considering planets with the most precisely measured ($\leq10\%$) densities. Based on radius and incident flux, our sample consists mainly of warm and hot Jupiters and sub-Saturns, and a handful of sub-Neptune sized planets.

Figure \ref{fig:combo1} shows that planets orbiting subgiants show the same broad demographic features as planets orbiting main-sequence stars, including radius inflation with incident flux for gas-giant planets \citep{burrows2000,lopez2015} and a dearth of planets with high-incident fluxes and radii between $\approx 3-8$\,\radiuse\ \citep[the sub-Neptune desert,][]{lundkvist2016}. Quantitatively, planets orbiting subgiants are on average hotter and exclude the shortest orbital period planets. The former is consistent with selection bias, while the latter 
is consistent with tidal orbital period decay becoming more efficient as a star evolves off the main-sequence, causing short period planets to be engulfed \citep{schlaufman2013}.

Our sample includes 4 subgiants with outer non-transiting companions. Sample targets have observations that span two years on average and therefore assuming each target has equal sensitivity to such companions, corresponds to an approximate occurrence rate of outer Jovian planets in subgiant systems of 19$\pm$8\%. This fraction is consistent with the occurrence rate of Jovian planets beyond the ice line from long-term RV surveys \citep{fernandes2019,fulton2021}.
Since subgiants are on average more massive than main-sequence stars, this may imply that the correlation of stellar mass and gas-giant planet occurrence does not extend to long orbital periods. We caution that the above numbers are not based on a detailed occurrence rate study (using injection/recovery tests), which is beyond the scope of this paper. 


\begin{figure*}
\includegraphics[width=\linewidth]{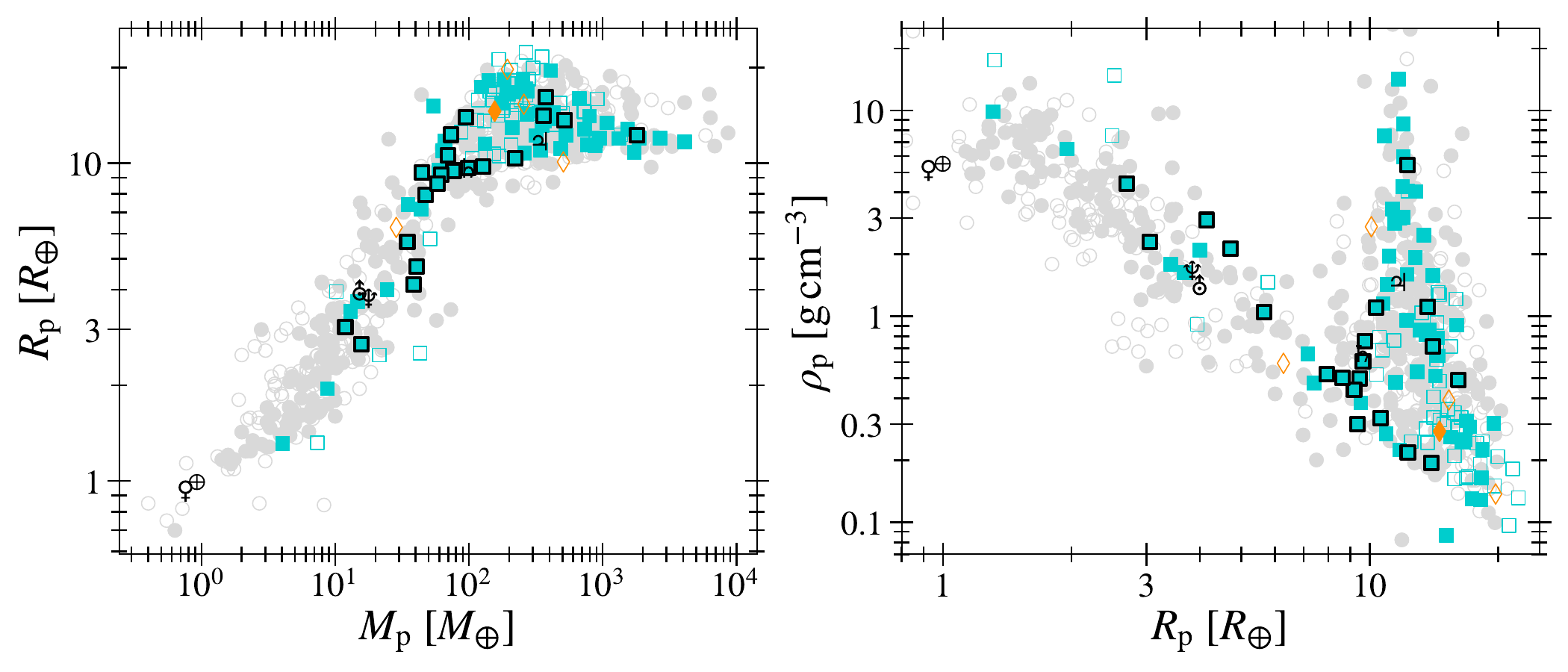}
\caption{Same as Figure \ref{fig:combo1} but now showing planet radius versus planet mass (left) and bulk density versus planet radius (right). Solar system planets are shown by the black symbols.}
\label{fig:combo2}
\end{figure*}

\subsection{Mass-Radius Relation}

Figure \ref{fig:combo2} compares the mass-radius relation and density versus radius for our sample with  values for other planets drawn from the literature. For giant planets ($>10$ \radiuse) we observe that planets orbiting subgiant stars follow largely the same distribution as planets orbiting main-sequence stars. We observe a tentative lack of inflated, massive ($>1000$\masse) planets around subgiant stars. However, a K-S test shows that this is not significant, likely due to the small sample size. 


A large fraction of our newly confirmed planets fall in the region between Neptune and sub-Saturn sized planets, for which radii increase as $R_{P} \approx M_{P}^{0.6}$, and Jovian planets, for which radius is nearly constant with mass \citep{weiss2013,chen2017}. Sub-Saturn sized planets (4--8\,\radiuse) around main-sequence stars show a wide variety of masses, (6--60\,\masse), even at roughly fixed radius \citep{petigura2017b,vaneylen2018b}. Conversely, our sample of planets orbiting subgiants appears to show a tighter mass range (35-50\,\masse), and the smallest ($<4$ \radiuse) planets found around subgiants are all consistently more massive. We speculate that this could signify that the atmospheres of low density planets are preferentially eroded away as the host star evolves into a subgiant. An alternative explanation is that low-mass planets are more difficult to confirm around subgiant stars, which exhibit increased RV jitter due to granulation.

\subsection{Eccentricity and Age}

Orbital eccentricities of exoplanets provide valuable clues for how planetary systems might have formed. For example, planet-planet scattering followed by high eccentricity migration is one of the favored formation scenarios for gas-giant planets on short orbits \citep{dawson2018}. Our sample is particularly useful for investigations of such theories due to the precise age constraints for subgiant host stars. Indeed, recent studies have found that hot Jupiter occurrence decreases with stellar age \citep{hamer2019,miyazaski2023,chen2023}.

\begin{figure*}
\includegraphics[width=\linewidth]{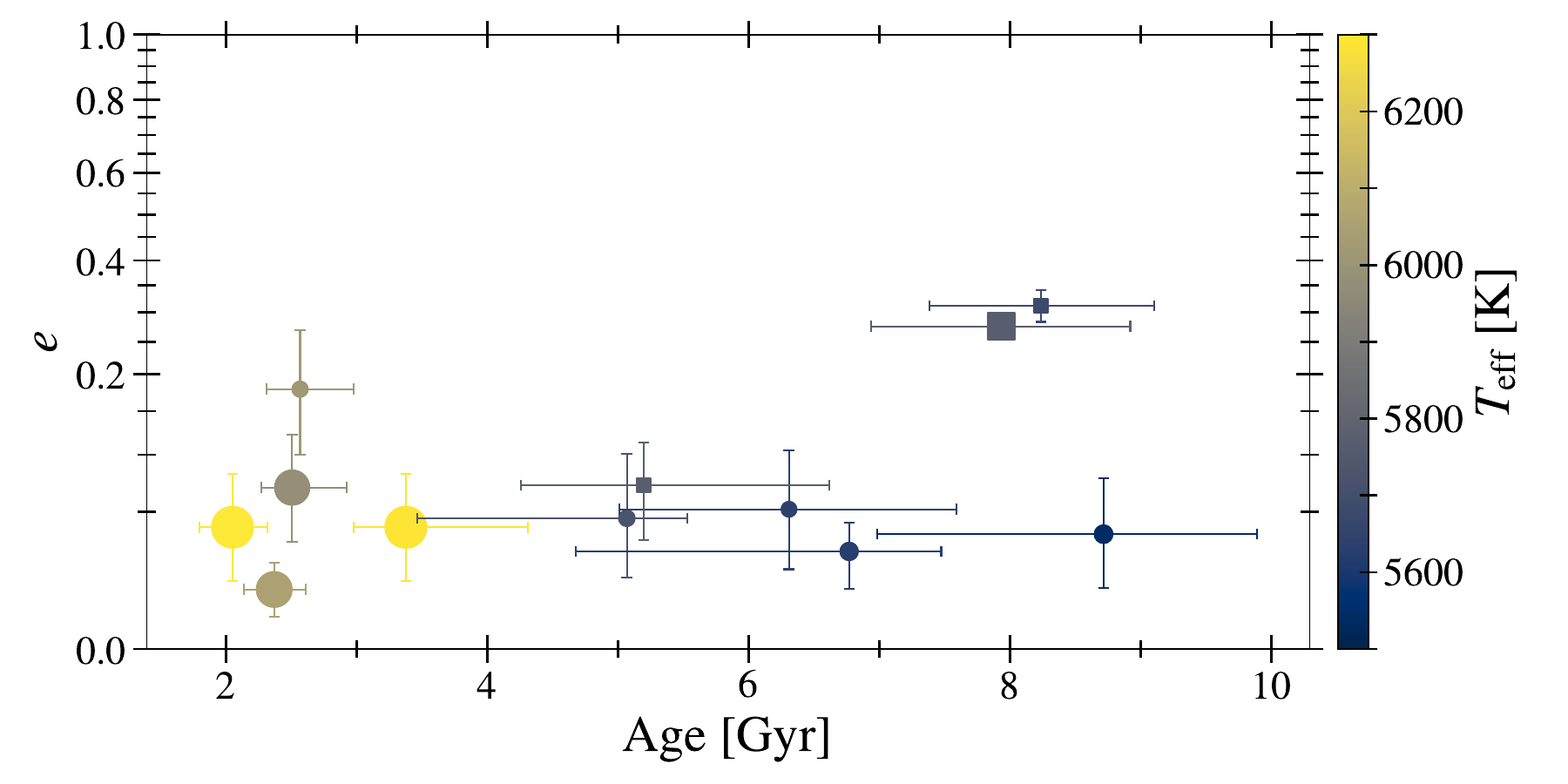}
\caption{Orbital eccentricity versus age for all hot Jupiters (i.e., \radiusp\ $>8$ \radiuse\ and $P\leq10$ days) in our sample. The $y$-axis is scaled non-linearly to emphasize more moderate eccentricities. The effective temperature of the host star is color coded and symbol size scales with planet mass. Planets with non-transiting outer companions are shown as squares.}
\label{fig:age_period}
\end{figure*}

Figure \ref{fig:age_period} shows orbital eccentricity versus age for all hot Jupiters (\radiusp\ $>8$ \radiuse, $P<10$ days) in our sample. Intriguingly, some of the most eccentric planets in our sample are found orbiting around some of the oldest stars. This suggests that planet-planet scattering, which can lead to high eccentricities, may occur on $>$ Gyr timescales. Alternatively, it suggests that tidal circularization can occur on significantly slower timescales than orbital period decay, consistent with a diversity of dissipation efficiencies. 

Another clue for the dynamical architectures of these systems is the presence of outer, non-transiting companions. Figure \ref{fig:age_period} shows that planets without outer companions are preferentially young and on circular orbits. Conversely, planets with outer companions are preferentially older and on eccentric orbits. If confirmed, this would imply possible distinct formation pathways for different hot Jupiter populations in which some undergo high-eccentricity migration and circularize quickly, whereas those with outer companions possibly gain their eccentricities through Kozai-Lidov oscillations \citep{fabrycky2007}. Additional systems will be required to confirm the significance of this trend, given the small number statistics in this sample.

\section{Conclusions}

We presented a dedicated transit and radial velocity survey of planets orbiting subgiants observed by TESS. Our main conclusions can be summarized as follows:

\begin{itemize}
\item We measured radii and masses for the 23 planets orbiting 21 subgiant stars, 13 of which are new planet discoveries. Of the 21 transiting planets with measured sizes, 19 planets have bulk densities measured to better than 25\%, including 6 with densities measured to better than 10\%.

\item Our sample includes 4 subgiants with outer non-transiting companions, corresponding to an approximate outer Jovian planet fraction of 19$\pm$8\%. This fraction is consistent with the occurrence rate of outer Jovian planets orbiting main-sequence stars.

\item The mass-radius diagram for giant planets orbiting subgiant stars roughly matches the observed distribution for main-sequence stars, with the exception of a tentative lack of inflated, massive planets.
We also observe that sub-Saturn sized planets orbiting subgiants are systematically more massive than similar sized planets around main-sequence stars, which may be related to detection bias or preferential atmosphere stripping in low-density planets.

\item We find tentative evidence for two populations among close-in ($P<10$ days) giant planets (\radiusp\ $>8\,$\radiuse) orbiting our subgiant sample: 1) ``lonely'' young ($\sim2$ Gyr), circular HJs orbiting hot (\teff\ $>6200$ K) stars, and 2) older ($\sim8-10$ Gyr), eccentric ($e\sim0.3$) HJs orbiting stars with \teff\ between $5700-5900$ K where there is evidence for outer companions. Thus the observed properties of hot Jupiters orbiting evolved stars are the likely consequence of multiple formation channels.
\end{itemize}

Future characterization of planets in this sample will provide a more comprehensive understanding of planet demographics for evolved systems. For example, many of the host stars in this sample are rapidly rotating (\vsini\ $>5$ \kms), facilitating the measurement of projected obliquities via Rossiter-McLaughlin or Doppler tomography observations (Saunders et al., in prep; Knudstrup et al., in prep). Continued monitoring by TESS and RV surveys will furthermore provide insights into additional planets in these systems.

%
%

\section*{Acknowledgements}

The authors wish to recognize and acknowledge the very significant cultural role and reverence that the summit of Maunakea has always had within the Native Hawaiian community. We are most fortunate to have the opportunity to conduct observations from this mountain.

We thank all the observers who spent time collecting data over the many years at Keck/HIRES. We gratefully acknowledge the efforts and dedication of the Keck Observatory staff for support of HIRES and remote observing. We thank the University of California and Google for supporting Lick Observatory and the UCO staff for their dedicated work scheduling and operating the telescopes of Lick Observatory. 

A.C. and D.H. acknowledge support from the National Aeronautics and Space Administration (80NSSC21K0652). D.H. also acknowledges support from the Alfred P. Sloan Foundation and the Australian Research Council (FT200100871). N.S. acknowledges support by the National Science Foundation Graduate Research Fellowship Program under Grant Number 2236415. E.K. and S.H.A. acknowledge support from the Danish Council for Independent Research through Grant Number 2032-00230B. M.L.H. would like to acknowledge NASA support via the FINESST Planetary Science Division, NASA award number 80NSSC21K1536. K.K.M. acknowledges support from the New York Community Trust Fund for Astrophysical Research). E.P. acknowledges financial support from the Agencia Estatal de Investigaci\'on of the Ministerio de Ciencia e Innovaci\'on MCIN/AEI/10.13039/501100011033 and the ERDF ``A way of making Europe'' through project PID2021-125627OB-C32, and from the Centre of Excellence ``Severo Ochoa'' award to the Instituto de Astrofisica de Canarias.\ldots

This paper made use of data collected by the TESS mission and are publicly available from the Mikulski Archive for Space Telescopes operated by the Space Telescope Science Institute. Funding for the TESS mission is provided by NASA's Science Mission Directorate. We acknowledge the use of public TESS data from pipelines at the TESS Science Office and at the TESS Science Processing Operations Center. Resources supporting this work were provided by the NASA High-End Computing Program through the NASA Advanced Supercomputing Division at Ames Research Center for the production of the SPOC data products. This research has made use of the Exoplanet Follow-up Observation Program website, which is operated by the California Institute of Technology, under contract with the National Aeronautics and Space Administration under the Exoplanet Exploration Program. This work has made use of observations (programme IDs: A42/TAC22 and A43/TAC11) from the Italian Telescopio Nazionale Galileo, which is operated on the island of La Palma by the Fundación Galileo Galilei of the Istituto Nazionale di Astrofisica at the Spanish Observatorio del Roque de los Muchachos of the Instituto de Astrofisica de Canarias.

\facilities{APF, Hale (PHARO), HARPS, Keck:I (HIRES), Keck:II (NIRC2), \href{https://archive.stsci.edu/index.html}{MAST}, TESS}

\software{We made use of the following publicly-available Python modules: \texttt{astroquery}, \texttt{emcee} \citep{fm2013}, \texttt{ktransit}, \texttt{evolstate} \citep{huber2017,berger2018}, \texttt{pySYD} \citep{chontos2022b,chontos2021} and \texttt{tesspoint}.}

\vspace{1cm}
\small
\bibliography{main.bib}

\newcommand{\SortNoop}[1]{}
\begin{thebibliography}{}
\expandafter\ifx\csname natexlab\endcsname\relax\def\natexlab#1{#1}\fi
\providecommand{\url}[1]{\href{#1}{#1}}

\bibitem[{{Addison} {et~al.}(2021){Addison}, {Wright}, {Nicholson}, {Cale},
  {Mocnik}, {Huber}, {Plavchan}, {Wittenmyer}, {Vanderburg}, {Chaplin},
  {Chontos}, {Clark}, {Eastman}, {Ziegler}, {Brahm}, {Carter}, {Clerte},
  {Espinoza}, {Horner}, {Bentley}, {Jord{\'a}n}, {Kane}, {Kielkopf},
  {Laychock}, {Mengel}, {Okumura}, {Stassun}, {Bedding}, {Bowler}, {Burnelis},
  {Blanco-Cuaresma}, {Collins}, {Crossfield}, {Davis}, {Evensberget},
  {Heitzmann}, {Howell}, {Law}, {Mann}, {Marsden}, {Matson}, {O'Connor},
  {Shporer}, {Stevens}, {Tinney}, {Tylor}, {Wang}, {Zhang}, {Henning},
  {Kossakowski}, {Ricker}, {Sarkis}, {Schlecker}, {Torres}, {Vanderspek},
  {Latham}, {Seager}, {Winn}, {Jenkins}, {Mireles}, {Rowden}, {Pepper},
  {Daylan}, {Schlieder}, {Collins}, {Collins}, {Tan}, {Ball}, {Basu}, {Buzasi},
  {Campante}, {Corsaro}, {Gonz{\'a}lez-Cuesta}, {Davies}, {de Almeida}, {do
  Nascimento}, {Garc{\'\i}a}, {Guo}, {Handberg}, {Hekker}, {Hey}, {Kallinger},
  {Kawaler}, {Kayhan}, {Kuszlewicz}, {Lund}, {Lyttle}, {Mathur}, {Miglio},
  {Mosser}, {Nielsen}, {Serenelli}, {Aguirre}, \& {Theme{\ss}l}}]{addison2021}
{Addison}, B.~C., {Wright}, D.~J., {Nicholson}, B.~A., {et~al.} 2021, \mnras,
  502, 3704

\bibitem[{{Akana Murphy} {et~al.}(2023){Akana Murphy}, {Batalha}, {Scarsdale},
  {Isaacson}, {Ciardi}, {Gonzales}, {Giacalone}, {Twicken}, {Dattilo},
  {Fetherolf}, {Rubenzahl}, {Crossfield}, {Dressing}, {Fulton}, {Howard},
  {Huber}, {Kane}, {Petigura}, {Robertson}, {Roy}, {Weiss}, {Beard}, {Chontos},
  {Dai}, {Rice}, {Van Zandt}, {Lubin}, {Blunt}, {Polanski}, {Behmard}, {Dalba},
  {Hill}, {Rosenthal}, {Brinkman}, {Mayo}, {Turtelboom}, {Angelo},
  {Mo{\v{c}}nik}, {MacDougall}, {Pidhorodetska}, {Tyler}, {Kosiarek},
  {Holcomb}, {Louden}, {Hirsch}, {Gilbert}, {Anderson}, \&
  {Valenti}}]{murphy2023}
{Akana Murphy}, J.~M., {Batalha}, N.~M., {Scarsdale}, N., {et~al.} 2023, \aj,
  166, 153

\bibitem[{{Barclay} {et~al.}(2015){Barclay}, {Endl}, {Huber}, {Foreman-Mackey},
  {Cochran}, {MacQueen}, {Rowe}, \& {Quintana}}]{barclay2015}
{Barclay}, T., {Endl}, M., {Huber}, D., {et~al.} 2015, \apj, 800, 46

\bibitem[{{Batalha} {et~al.}(2019){Batalha}, {Lewis}, {Fortney}, {Batalha},
  {Kempton}, {Lewis}, \& {Line}}]{batalha2019}
{Batalha}, N.~E., {Lewis}, T., {Fortney}, J.~J., {et~al.} 2019, \apjl, 885, L25

\bibitem[{{Belokurov} {et~al.}(2020){Belokurov}, {Penoyre}, {Oh}, {Iorio},
  {Hodgkin}, {Evans}, {Everall}, {Koposov}, {Tout}, {Izzard}, {Clarke}, \&
  {Brown}}]{belokurov2020}
{Belokurov}, V., {Penoyre}, Z., {Oh}, S., {et~al.} 2020, \mnras, 496, 1922

\bibitem[{{Berger} {et~al.}(2018){Berger}, {Huber}, {Gaidos}, \& {van
  Saders}}]{berger2018}
{Berger}, T.~A., {Huber}, D., {Gaidos}, E., \& {van Saders}, J.~L. 2018, ArXiv
  e-prints, arXiv:1805.00231

\bibitem[{{Bowler} {et~al.}(2010){Bowler}, {Johnson}, {Marcy}, {Henry}, {Peek},
  {Fischer}, {Clubb}, {Liu}, {Reffert}, {Schwab}, \& {Lowe}}]{bowler2010}
{Bowler}, B.~P., {Johnson}, J.~A., {Marcy}, G.~W., {et~al.} 2010, \apj, 709,
  396

\bibitem[{{Burke} {et~al.}(2008){Burke}, {McCullough}, {Valenti}, {Long},
  {Johns-Krull}, {Machalek}, {Janes}, {Taylor}, {Fleenor}, {Foote}, {Gary},
  {Garc{\'{\i}}a-Melendo}, {Gregorio}, \& {Vanmunster}}]{burke2008}
{Burke}, C.~J., {McCullough}, P.~R., {Valenti}, J.~A., {et~al.} 2008, \apj,
  686, 1331

\bibitem[{{Burrows} {et~al.}(2000){Burrows}, {Guillot}, {Hubbard}, {Marley},
  {Saumon}, {Lunine}, \& {Sudarsky}}]{burrows2000}
{Burrows}, A., {Guillot}, T., {Hubbard}, W.~B., {et~al.} 2000, \apjl, 534, L97

\bibitem[{{Butler} {et~al.}(1996){Butler}, {Marcy}, {Williams}, {McCarthy},
  {Dosanjh}, \& {Vogt}}]{butler1996}
{Butler}, R.~P., {Marcy}, G.~W., {Williams}, E., {et~al.} 1996, \pasp, 108, 500

\bibitem[{{Chen} {et~al.}(2023){Chen}, {Xie}, {Zhou}, {Dong}, {Yang}, {Zhu},
  {Liu}, {Huang}, {Xiang}, {Wang}, {Zheng}, {Luo}, {Zhang}, \&
  {Zhu}}]{chen2023}
{Chen}, D.-C., {Xie}, J.-W., {Zhou}, J.-L., {et~al.} 2023, Proceedings of the
  National Academy of Science, 120, e2304179120

\bibitem[{{Chen} \& {Kipping}(2017)}]{chen2017}
{Chen}, J., \& {Kipping}, D. 2017, \apj, 834, 17

\bibitem[{{Choi} {et~al.}(2016){Choi}, {Dotter}, {Conroy}, {Cantiello},
  {Paxton}, \& {Johnson}}]{choi2016}
{Choi}, J., {Dotter}, A., {Conroy}, C., {et~al.} 2016, \apj, 823, 102

\bibitem[{{Chontos} {et~al.}(2021{\natexlab{a}}){Chontos}, {Huber}, {Sayeed},
  \& {Yamsiri}}]{chontos2022b}
{Chontos}, A., {Huber}, D., {Sayeed}, M., \& {Yamsiri}, P. 2021{\natexlab{a}},
  arXiv e-prints, arXiv:2108.00582

\bibitem[{{Chontos} {et~al.}(2019){Chontos}, {Huber}, {Latham}, {Bieryla}, {Van
  Eylen}, {Bedding}, {Berger}, {Buchhave}, {Campante}, {Chaplin}, {Colman},
  {Coughlin}, {Davies}, {Hirano}, {Howard}, \& {Isaacson}}]{chontos2019}
{Chontos}, A., {Huber}, D., {Latham}, D.~W., {et~al.} 2019, \aj, 157, 192

\bibitem[{{Chontos} {et~al.}(2021{\natexlab{b}}){Chontos}, {Huber}, {Berger},
  {Kjeldsen}, {Serenelli}, {Silva Aguirre}, {Ball}, {Basu}, {Bedding},
  {Chaplin}, {Claytor}, {Corsaro}, {Garcia}, {Howell}, {Lundkvist}, {Mathur},
  {Metcalfe}, {Nielsen}, {Mian Joel Ong}, {{\c{C}}elik Orhan}, {{\"O}rtel},
  {Salama}, {Stassun}, {Townsend}, {van Saders}, {Winther}, {Yildiz}, {Butler},
  {Tinney}, \& {Wittenmyer}}]{chontos2021}
{Chontos}, A., {Huber}, D., {Berger}, T.~A., {et~al.} 2021{\natexlab{b}}, \apj,
  922, 229

\bibitem[{{Chontos} {et~al.}(2022){Chontos}, {Murphy}, {MacDougall},
  {Fetherolf}, {Van Zandt}, {Rubenzahl}, {Beard}, {Huber}, {Batalha},
  {Crossfield}, {Dressing}, {Fulton}, {Howard}, {Isaacson}, {Kane}, {Petigura},
  {Robertson}, {Roy}, {Weiss}, {Behmard}, {Dai}, {Dalba}, {Giacalone}, {Hill},
  {Lubin}, {Mayo}, {Mo{\v{c}}nik}, {Polanski}, {Rosenthal}, {Scarsdale},
  {Turtelboom}, {Ricker}, {Vanderspek}, {Latham}, {Seager}, {Winn}, {Jenkins},
  {Quinn}, {Guerrero}, {Collins}, {Ciardi}, {Shporer}, {Goeke}, {Levine},
  {Ting}, {Bieryla}, {Collins}, {Kielkopf}, {Barkaoui}, {Benni},
  {Esparza-Borges}, {Conti}, {Hooton}, {Kagetani}, {Laloum}, {Marino},
  {Massey}, {Murgas}, {Papini}, {Schwarz}, {Srdoc}, {Stockdale}, {Wang},
  {Wittrock}, \& {Zou}}]{chontos2022a}
{Chontos}, A., {Murphy}, J. M.~A., {MacDougall}, M.~G., {et~al.} 2022, \aj,
  163, 297

\bibitem[{{Chubak} {et~al.}(2012){Chubak}, {Marcy}, {Fischer}, {Howard},
  {Isaacson}, {Johnson}, \& {Wright}}]{chubak2012}
{Chubak}, C., {Marcy}, G., {Fischer}, D.~A., {et~al.} 2012, arXiv e-prints,
  arXiv:1207.6212

\bibitem[{{Claret}(2016)}]{claret2016}
{Claret}, A. 2016, VizieR Online Data Catalog, J/A+A/600/A30

\bibitem[{{Cosentino} {et~al.}(2012){Cosentino}, {Lovis}, {Pepe}, {Collier
  Cameron}, {Latham}, {Molinari}, {Udry}, {Bezawada}, {Black}, {Born},
  {Buchschacher}, {Charbonneau}, {Figueira}, {Fleury}, {Galli}, {Gallie},
  {Gao}, {Ghedina}, {Gonzalez}, {Gonzalez}, {Guerra}, {Henry}, {Horne},
  {Hughes}, {Kelly}, {Lodi}, {Lunney}, {Maire}, {Mayor}, {Micela}, {Ordway},
  {Peacock}, {Phillips}, {Piotto}, {Pollacco}, {Queloz}, {Rice}, {Riverol},
  {Riverol}, {San Juan}, {Sasselov}, {Segransan}, {Sozzetti}, {Sosnowska},
  {Stobie}, {Szentgyorgyi}, {Vick}, \& {Weber}}]{cosentino2012}
{Cosentino}, R., {Lovis}, C., {Pepe}, F., {et~al.} 2012, in Society of
  Photo-Optical Instrumentation Engineers (SPIE) Conference Series, Vol. 8446,
  Ground-based and Airborne Instrumentation for Astronomy IV, ed. I.~S.
  {McLean}, S.~K. {Ramsay}, \& H.~{Takami}, 84461V

\bibitem[{{Dawson} \& {Johnson}(2018)}]{dawson2018}
{Dawson}, R.~I., \& {Johnson}, J.~A. 2018, \araa, 56, 175

\bibitem[{{Eastman} {et~al.}(2013){Eastman}, {Gaudi}, \& {Agol}}]{eastman2013}
{Eastman}, J., {Gaudi}, B.~S., \& {Agol}, E. 2013, \pasp, 125, 83

\bibitem[{{Evans}(2018)}]{evans2018b}
{Evans}, D.~F. 2018, Research Notes of the American Astronomical Society, 2, 20

\bibitem[{{Fabrycky} \& {Tremaine}(2007)}]{fabrycky2007}
{Fabrycky}, D., \& {Tremaine}, S. 2007, \apj, 669, 1298

\bibitem[{{Fernandes} {et~al.}(2019){Fernandes}, {Mulders}, {Pascucci},
  {Mordasini}, \& {Emsenhuber}}]{fernandes2019}
{Fernandes}, R.~B., {Mulders}, G.~D., {Pascucci}, I., {Mordasini}, C., \&
  {Emsenhuber}, A. 2019, \apj, 874, 81

\bibitem[{{Foreman-Mackey} {et~al.}(2013){Foreman-Mackey}, {Hogg}, {Lang}, \&
  {Goodman}}]{fm2013}
{Foreman-Mackey}, D., {Hogg}, D.~W., {Lang}, D., \& {Goodman}, J. 2013, \pasp,
  125, 306

\bibitem[{{Fulton} {et~al.}(2018){Fulton}, {Petigura}, {Blunt}, \&
  {Sinukoff}}]{fulton2018r}
{Fulton}, B.~J., {Petigura}, E.~A., {Blunt}, S., \& {Sinukoff}, E. 2018, \pasp,
  130, 044504

\bibitem[{{Fulton} {et~al.}(2017){Fulton}, {Petigura}, {Howard}, {Isaacson},
  {Marcy}, {Cargile}, {Hebb}, {Weiss}, {Johnson}, {Morton}, {Sinukoff},
  {Crossfield}, \& {Hirsch}}]{fulton2017}
{Fulton}, B.~J., {Petigura}, E.~A., {Howard}, A.~W., {et~al.} 2017, \aj, 154,
  109

\bibitem[{{Fulton} {et~al.}(2021){Fulton}, {Rosenthal}, {Hirsch}, {Isaacson},
  {Howard}, {Dedrick}, {Sherstyuk}, {Blunt}, {Petigura}, {Knutson}, {Behmard},
  {Chontos}, {Crepp}, {Crossfield}, {Dalba}, {Fischer}, {Henry}, {Kane},
  {Kosiarek}, {Marcy}, {Rubenzahl}, {Weiss}, \& {Wright}}]{fulton2021}
{Fulton}, B.~J., {Rosenthal}, L.~J., {Hirsch}, L.~A., {et~al.} 2021, \apjs,
  255, 14

\bibitem[{{Furlan} {et~al.}(2017){Furlan}, {Ciardi}, {Everett}, {Saylors},
  {Teske}, {Horch}, {Howell}, {van Belle}, {Hirsch}, {Gautier}, {Adams},
  {Barrado}, {Cartier}, {Dressing}, {Dupree}, {Gilliland}, {Lillo-Box},
  {Lucas}, \& {Wang}}]{furlan2017}
{Furlan}, E., {Ciardi}, D.~R., {Everett}, M.~E., {et~al.} 2017, \aj, 153, 71

\bibitem[{{Gaia Collaboration} {et~al.}(2016){Gaia Collaboration}, {Prusti},
  {de Bruijne}, {Brown}, {Vallenari}, {Babusiaux}, {Bailer-Jones}, {Bastian},
  {Biermann}, {Evans}, {Eyer}, {Jansen}, {Jordi}, {Klioner}, {Lammers},
  {Lindegren}, {Luri}, {Mignard}, {Milligan}, {Panem}, {Poinsignon},
  {Pourbaix}, {Randich}, {Sarri}, {Sartoretti}, {Siddiqui}, {Soubiran},
  {Valette}, {van Leeuwen}, {Walton}, {Aerts}, {Arenou}, {Cropper}, {Drimmel},
  {H{\o}g}, {Katz}, {Lattanzi}, {O'Mullane}, {Grebel}, {Holland}, {Huc},
  {Passot}, {Bramante}, {Cacciari}, {Casta{\~n}eda}, {Chaoul}, {Cheek}, {De
  Angeli}, {Fabricius}, {Guerra}, {Hern{\'a}ndez}, {Jean-Antoine-Piccolo},
  {Masana}, {Messineo}, {Mowlavi}, {Nienartowicz}, {Ord{\'o}{\~n}ez-Blanco},
  {Panuzzo}, {Portell}, {Richards}, {Riello}, {Seabroke}, {Tanga},
  {Th{\'e}venin}, {Torra}, {Els}, {Gracia-Abril}, {Comoretto},
  {Garcia-Reinaldos}, {Lock}, {Mercier}, {Altmann}, {Andrae}, {Astraatmadja},
  {Bellas-Velidis}, {Benson}, {Berthier}, {Blomme}, {Busso}, {Carry},
  {Cellino}, {Clementini}, {Cowell}, {Creevey}, {Cuypers}, {Davidson}, {De
  Ridder}, {de Torres}, {Delchambre}, {Dell'Oro}, {Ducourant}, {Fr{\'e}mat},
  {Garc{\'\i}a-Torres}, {Gosset}, {Halbwachs}, {Hambly}, {Harrison}, {Hauser},
  {Hestroffer}, {Hodgkin}, {Huckle}, {Hutton}, {Jasniewicz}, {Jordan},
  {Kontizas}, {Korn}, {Lanzafame}, {Manteiga}, {Moitinho}, {Muinonen},
  {Osinde}, {Pancino}, {Pauwels}, {Petit}, {Recio-Blanco}, {Robin}, {Sarro},
  {Siopis}, {Smith}, {Smith}, {Sozzetti}, {Thuillot}, {van Reeven}, {Viala},
  {Abbas}, {Abreu Aramburu}, {Accart}, {Aguado}, {Allan}, {Allasia},
  {Altavilla}, {{\'A}lvarez}, {Alves}, {Anderson}, {Andrei}, {Anglada Varela},
  {Antiche}, {Antoja}, {Ant{\'o}n}, {Arcay}, {Atzei}, {Ayache}, {Bach},
  {Baker}, {Balaguer-N{\'u}{\~n}ez}, {Barache}, {Barata}, {Barbier}, {Barblan},
  {Baroni}, {Barrado y Navascu{\'e}s}, {Barros}, {Barstow}, {Becciani},
  {Bellazzini}, {Bellei}, {Bello Garc{\'\i}a}, {Belokurov}, {Bendjoya},
  {Berihuete}, {Bianchi}, {Bienaym{\'e}}, {Billebaud}, {Blagorodnova},
  {Blanco-Cuaresma}, {Boch}, {Bombrun}, {Borrachero}, {Bouquillon}, {Bourda},
  {Bouy}, {Bragaglia}, {Breddels}, {Brouillet}, {Br{\"u}semeister},
  {Bucciarelli}, {Budnik}, {Burgess}, {Burgon}, {Burlacu}, {Busonero}, {Buzzi},
  {Caffau}, {Cambras}, {Campbell}, {Cancelliere}, {Cantat-Gaudin}, {Carlucci},
  {Carrasco}, {Castellani}, {Charlot}, {Charnas}, {Charvet}, {Chassat},
  {Chiavassa}, {Clotet}, {Cocozza}, {Collins}, {Collins}, {Costigan}, {Crifo},
  {Cross}, {Crosta}, {Crowley}, {Dafonte}, {Damerdji}, {Dapergolas}, {David},
  {David}, {De Cat}, {de Felice}, {de Laverny}, {De Luise}, {De March}, {de
  Martino}, {de Souza}, {Debosscher}, {del Pozo}, {Delbo}, {Delgado},
  {Delgado}, {di Marco}, {Di Matteo}, {Diakite}, {Distefano}, {Dolding}, {Dos
  Anjos}, {Drazinos}, {Dur{\'a}n}, {Dzigan}, {Ecale}, {Edvardsson}, {Enke},
  {Erdmann}, {Escolar}, {Espina}, {Evans}, {Eynard Bontemps}, {Fabre},
  {Fabrizio}, {Faigler}, {Falc{\~a}o}, {Farr{\`a}s Casas}, {Faye}, {Federici},
  {Fedorets}, {Fern{\'a}ndez-Hern{\'a}ndez}, {Fernique}, {Fienga}, {Figueras},
  {Filippi}, {Findeisen}, {Fonti}, {Fouesneau}, {Fraile}, {Fraser}, {Fuchs},
  {Furnell}, {Gai}, {Galleti}, {Galluccio}, {Garabato}, {Garc{\'\i}a-Sedano},
  {Gar{\'e}}, {Garofalo}, {Garralda}, {Gavras}, {Gerssen}, {Geyer}, {Gilmore},
  {Girona}, {Giuffrida}, {Gomes}, {Gonz{\'a}lez-Marcos},
  {Gonz{\'a}lez-N{\'u}{\~n}ez}, {Gonz{\'a}lez-Vidal}, {Granvik}, {Guerrier},
  {Guillout}, {Guiraud}, {G{\'u}rpide}, {Guti{\'e}rrez-S{\'a}nchez}, {Guy},
  {Haigron}, {Hatzidimitriou}, {Haywood}, {Heiter}, {Helmi}, {Hobbs},
  {Hofmann}, {Holl}, {Holland }, {Hunt}, {Hypki}, {Icardi}, {Irwin}, {Jevardat
  de Fombelle}, {Jofr{\'e}}, {Jonker}, {Jorissen}, {Julbe}, {Karampelas},
  {Kochoska}, {Kohley}, {Kolenberg}, {Kontizas}, {Koposov}, {Kordopatis},
  {Koubsky}, {Kowalczyk}, {Krone-Martins}, {Kudryashova}, {Kull}, {Bachchan},
  {Lacoste-Seris}, {Lanza}, {Lavigne}, {Le Poncin-Lafitte}, {Lebreton},
  {Lebzelter}, {Leccia}, {Leclerc}, {Lecoeur-Taibi}, {Lemaitre}, {Lenhardt},
  {Leroux}, {Liao}, {Licata}, {Lindstr{\o}m}, {Lister}, {Livanou}, {Lobel},
  {L{\"o}ffler}, {L{\'o}pez}, {Lopez-Lozano}, {Lorenz}, {Loureiro},
  {MacDonald}, {Magalh{\~a}es Fernandes}, {Managau}, {Mann}, {Mantelet},
  {Marchal}, {Marchant}, {Marconi}, {Marie}, {Marinoni}, {Marrese},
  {Marschalk{\'o}}, {Marshall}, {Mart{\'\i}n-Fleitas}, {Martino}, {Mary},
  {Matijevi{\v{c}}}, {Mazeh}, {McMillan}, {Messina}, {Mestre}, {Michalik},
  {Millar}, {Miranda}, {Molina}, {Molinaro}, {Molinaro}, {Moln{\'a}r},
  {Moniez}, {Montegriffo}, {Monteiro}, {Mor}, {Mora}, {Morbidelli}, {Morel},
  {Morgenthaler}, {Morley}, {Morris}, {Mulone}, {Muraveva}, {Musella},
  {Narbonne}, {Nelemans}, {Nicastro}, {Noval}, {Ord{\'e}novic},
  {Ordieres-Mer{\'e}}, {Osborne}, {Pagani}, {Pagano}, {Pailler}, {Palacin},
  {Palaversa}, {Parsons}, {Paulsen}, {Pecoraro}, {Pedrosa}, {Pentik{\"a}inen},
  {Pereira}, {Pichon}, {Piersimoni}, {Pineau}, {Plachy}, {Plum}, {Poujoulet},
  {Pr{\v{s}}a}, {Pulone}, {Ragaini}, {Rago}, {Rambaux}, {Ramos-Lerate},
  {Ranalli}, {Rauw}, {Read}, {Regibo}, {Renk}, {Reyl{\'e}}, {Ribeiro},
  {Rimoldini}, {Ripepi}, {Riva}, {Rixon}, {Roelens}, {Romero-G{\'o}mez},
  {Rowell}, {Royer}, {Rudolph}, {Ruiz-Dern}, {Sadowski}, {Sagrist{\`a}
  Sell{\'e}s}, {Sahlmann}, {Salgado}, {Salguero}, {Sarasso}, {Savietto},
  {Schnorhk}, {Schultheis}, {Sciacca}, {Segol}, {Segovia}, {Segransan},
  {Serpell}, {Shih}, {Smareglia}, {Smart}, {Smith}, {Solano}, {Solitro},
  {Sordo}, {Soria Nieto}, {Souchay}, {Spagna}, {Spoto}, {Stampa}, {Steele},
  {Steidelm{\"u}ller}, {Stephenson}, {Stoev}, {Suess}, {S{\"u}veges}, {Surdej},
  {Szabados}, {Szegedi-Elek}, {Tapiador}, {Taris}, {Tauran}, {Taylor},
  {Teixeira}, {Terrett}, {Tingley}, {Trager}, {Turon}, {Ulla}, {Utrilla},
  {Valentini}, {van Elteren}, {Van Hemelryck}, {van Leeuwen}, {Varadi},
  {Vecchiato}, {Veljanoski}, {Via}, {Vicente}, {Vogt}, {Voss}, {Votruba},
  {Voutsinas}, {Walmsley}, {Weiler}, {Weingrill}, {Werner}, {Wevers},
  {Whitehead}, {Wyrzykowski}, {Yoldas}, {{\v{Z}}erjal}, {Zucker}, {Zurbach},
  {Zwitter}, {Alecu}, {Allen}, {Allende Prieto}, {Amorim},
  {Anglada-Escud{\'e}}, {Arsenijevic}, {Azaz}, {Balm}, {Beck}, {Bernstein},
  {Bigot}, {Bijaoui}, {Blasco}, {Bonfigli}, {Bono}, {Boudreault}, {Bressan},
  {Brown}, {Brunet}, {Bunclark}, {Buonanno}, {Butkevich}, {Carret}, {Carrion},
  {Chemin}, {Ch{\'e}reau}, {Corcione}, {Darmigny}, {de Boer}, {de Teodoro}, {de
  Zeeuw}, {Delle Luche}, {Domingues}, {Dubath}, {Fodor}, {Fr{\'e}zouls},
  {Fries}, {Fustes}, {Fyfe}, {Gallardo}, {Gallegos}, {Gardiol}, {Gebran},
  {Gomboc}, {G{\'o}mez}, {Grux}, {Gueguen}, {Heyrovsky}, {Hoar}, {Iannicola},
  {Isasi Parache}, {Janotto}, {Joliet}, {Jonckheere}, {Keil}, {Kim},
  {Klagyivik}, {Klar}, {Knude}, {Kochukhov}, {Kolka}, {Kos}, {Kutka}, {Lainey},
  {LeBouquin}, {Liu}, {Loreggia}, {Makarov}, {Marseille}, {Martayan},
  {Martinez-Rubi}, {Massart}, {Meynadier}, {Mignot}, {Munari}, {Nguyen},
  {Nordlander}, {Ocvirk}, {O'Flaherty}, {Olias Sanz}, {Ortiz}, {Osorio},
  {Oszkiewicz}, {Ouzounis}, {Palmer}, {Park}, {Pasquato}, {Peltzer}, {Peralta},
  {P{\'e}turaud}, {Pieniluoma}, {Pigozzi}, {Poels}, {Prat}, {Prod'homme},
  {Raison}, {Rebordao}, {Risquez}, {Rocca-Volmerange}, {Rosen}, {Ruiz-Fuertes},
  {Russo}, {Sembay}, {Serraller Vizcaino}, {Short}, {Siebert}, {Silva},
  {Sinachopoulos}, {Slezak}, {Soffel}, {Sosnowska}, {Strai{\v{z}}ys}, {ter
  Linden}, {Terrell}, {Theil}, {Tiede}, {Troisi}, {Tsalmantza}, {Tur},
  {Vaccari}, {Vachier}, {Valles}, {Van Hamme}, {Veltz}, {Virtanen}, {Wallut},
  {Wichmann}, {Wilkinson}, {Ziaeepour}, \& {Zschocke}}]{gaia2016}
{Gaia Collaboration}, {Prusti}, T., {de Bruijne}, J.~H.~J., {et~al.} 2016,
  \aap, 595, A1

\bibitem[{{Gaia Collaboration} {et~al.}(2018){Gaia Collaboration}, {Brown},
  {Vallenari}, {Prusti}, {de Bruijne}, {Babusiaux}, {Bailer-Jones}, {Biermann},
  {Evans}, {Eyer}, {Jansen}, {Jordi}, {Klioner}, {Lammers}, {Lindegren},
  {Luri}, {Mignard}, {Panem}, {Pourbaix}, {Randich}, {Sartoretti}, {Siddiqui},
  {Soubiran}, {van Leeuwen}, {Walton}, {Arenou}, {Bastian}, {Cropper},
  {Drimmel}, {Katz}, {Lattanzi}, {Bakker}, {Cacciari}, {Casta{\~n}eda},
  {Chaoul}, {Cheek}, {De Angeli}, {Fabricius}, {Guerra}, {Holl}, {Masana},
  {Messineo}, {Mowlavi}, {Nienartowicz}, {Panuzzo}, {Portell}, {Riello},
  {Seabroke}, {Tanga}, {Th{\'e}venin}, {Gracia-Abril}, {Comoretto},
  {Garcia-Reinaldos}, {Teyssier}, {Altmann}, {Andrae}, {Audard},
  {Bellas-Velidis}, {Benson}, {Berthier}, {Blomme}, {Burgess}, {Busso},
  {Carry}, {Cellino}, {Clementini}, {Clotet}, {Creevey}, {Davidson}, {De
  Ridder}, {Delchambre}, {Dell'Oro}, {Ducourant},
  {Fern{\'a}ndez-Hern{\'a}ndez}, {Fouesneau}, {Fr{\'e}mat}, {Galluccio},
  {Garc{\'\i}a-Torres}, {Gonz{\'a}lez-N{\'u}{\~n}ez}, {Gonz{\'a}lez-Vidal},
  {Gosset}, {Guy}, {Halbwachs}, {Hambly}, {Harrison}, {Hern{\'a}ndez},
  {Hestroffer}, {Hodgkin}, {Hutton}, {Jasniewicz}, {Jean-Antoine-Piccolo},
  {Jordan}, {Korn}, {Krone-Martins}, {Lanzafame}, {Lebzelter}, {L{\"o}ffler},
  {Manteiga}, {Marrese}, {Mart{\'\i}n-Fleitas}, {Moitinho}, {Mora}, {Muinonen},
  {Osinde}, {Pancino}, {Pauwels}, {Petit}, {Recio-Blanco}, {Richards},
  {Rimoldini}, {Robin}, {Sarro}, {Siopis}, {Smith}, {Sozzetti}, {S{\"u}veges},
  {Torra}, {van Reeven}, {Abbas}, {Abreu Aramburu}, {Accart}, {Aerts},
  {Altavilla}, {{\'A}lvarez}, {Alvarez}, {Alves}, {Anderson}, {Andrei},
  {Anglada Varela}, {Antiche}, {Antoja}, {Arcay}, {Astraatmadja}, {Bach},
  {Baker}, {Balaguer-N{\'u}{\~n}ez}, {Balm}, {Barache}, {Barata}, {Barbato},
  {Barblan}, {Barklem}, {Barrado}, {Barros}, {Barstow}, {Bartholom{\'e}
  Mu{\~n}oz}, {Bassilana}, {Becciani}, {Bellazzini}, {Berihuete}, {Bertone},
  {Bianchi}, {Bienaym{\'e}}, {Blanco-Cuaresma}, {Boch}, {Boeche}, {Bombrun},
  {Borrachero}, {Bossini}, {Bouquillon}, {Bourda}, {Bragaglia}, {Bramante},
  {Breddels}, {Bressan}, {Brouillet}, {Br{\"u}semeister}, {Brugaletta},
  {Bucciarelli}, {Burlacu}, {Busonero}, {Butkevich}, {Buzzi}, {Caffau},
  {Cancelliere}, {Cannizzaro}, {Cantat-Gaudin}, {Carballo}, {Carlucci},
  {Carrasco}, {Casamiquela}, {Castellani}, {Castro-Ginard}, {Charlot},
  {Chemin}, {Chiavassa}, {Cocozza}, {Costigan}, {Cowell}, {Crifo}, {Crosta},
  {Crowley}, {Cuypers}, {Dafonte}, {Damerdji}, {Dapergolas}, {David}, {David},
  {de Laverny}, {De Luise}, {De March}, {de Martino}, {de Souza}, {de Torres},
  {Debosscher}, {del Pozo}, {Delbo}, {Delgado}, {Delgado}, {Di Matteo},
  {Diakite}, {Diener}, {Distefano}, {Dolding}, {Drazinos}, {Dur{\'a}n},
  {Edvardsson}, {Enke}, {Eriksson}, {Esquej}, {Eynard Bontemps}, {Fabre},
  {Fabrizio}, {Faigler}, {Falc{\~a}o}, {Farr{\`a}s Casas}, {Federici},
  {Fedorets}, {Fernique}, {Figueras}, {Filippi}, {Findeisen}, {Fonti},
  {Fraile}, {Fraser}, {Fr{\'e}zouls}, {Gai}, {Galleti}, {Garabato},
  {Garc{\'\i}a-Sedano}, {Garofalo}, {Garralda}, {Gavel}, {Gavras}, {Gerssen},
  {Geyer}, {Giacobbe}, {Gilmore}, {Girona}, {Giuffrida}, {Glass}, {Gomes},
  {Granvik}, {Gueguen}, {Guerrier}, {Guiraud}, {Guti{\'e}rrez-S{\'a}nchez},
  {Haigron}, {Hatzidimitriou}, {Hauser}, {Haywood}, {Heiter}, {Helmi}, {Heu},
  {Hilger}, {Hobbs}, {Hofmann}, {Holland}, {Huckle}, {Hypki}, {Icardi},
  {Jan{\ss}en}, {Jevardat de Fombelle}, {Jonker}, {Juh{\'a}sz}, {Julbe},
  {Karampelas}, {Kewley}, {Klar}, {Kochoska}, {Kohley}, {Kolenberg},
  {Kontizas}, {Kontizas}, {Koposov}, {Kordopatis}, {Kostrzewa-Rutkowska},
  {Koubsky}, {Lambert}, {Lanza}, {Lasne}, {Lavigne}, {Le Fustec}, {Le
  Poncin-Lafitte}, {Lebreton}, {Leccia}, {Leclerc}, {Lecoeur-Taibi},
  {Lenhardt}, {Leroux}, {Liao}, {Licata}, {Lindstr{\o}m}, {Lister}, {Livanou},
  {Lobel}, {L{\'o}pez}, {Managau}, {Mann}, {Mantelet}, {Marchal}, {Marchant},
  {Marconi}, {Marinoni}, {Marschalk{\'o}}, {Marshall}, {Martino}, {Marton},
  {Mary}, {Massari}, {Matijevi{\v{c}}}, {Mazeh}, {McMillan}, {Messina},
  {Michalik}, {Millar}, {Molina}, {Molinaro}, {Moln{\'a}r}, {Montegriffo},
  {Mor}, {Morbidelli}, {Morel}, {Morris}, {Mulone}, {Muraveva}, {Musella},
  {Nelemans}, {Nicastro}, {Noval}, {O'Mullane}, {Ord{\'e}novic},
  {Ord{\'o}{\~n}ez-Blanco}, {Osborne}, {Pagani}, {Pagano}, {Pailler},
  {Palacin}, {Palaversa}, {Panahi}, {Pawlak}, {Piersimoni}, {Pineau}, {Plachy},
  {Plum}, {Poggio}, {Poujoulet}, {Pr{\v{s}}a}, {Pulone}, {Racero}, {Ragaini},
  {Rambaux}, {Ramos-Lerate}, {Regibo}, {Reyl{\'e}}, {Riclet}, {Ripepi}, {Riva},
  {Rivard}, {Rixon}, {Roegiers}, {Roelens}, {Romero-G{\'o}mez}, {Rowell},
  {Royer}, {Ruiz-Dern}, {Sadowski}, {Sagrist{\`a} Sell{\'e}s}, {Sahlmann},
  {Salgado}, {Salguero}, {Sanna}, {Santana-Ros}, {Sarasso}, {Savietto},
  {Schultheis}, {Sciacca}, {Segol}, {Segovia}, {S{\'e}gransan}, {Shih},
  {Siltala}, {Silva}, {Smart}, {Smith}, {Solano}, {Solitro}, {Sordo}, {Soria
  Nieto}, {Souchay}, {Spagna}, {Spoto}, {Stampa}, {Steele},
  {Steidelm{\"u}ller}, {Stephenson}, {Stoev}, {Suess}, {Surdej}, {Szabados},
  {Szegedi-Elek}, {Tapiador}, {Taris}, {Tauran}, {Taylor}, {Teixeira},
  {Terrett}, {Teyssand ier}, {Thuillot}, {Titarenko}, {Torra Clotet}, {Turon},
  {Ulla}, {Utrilla}, {Uzzi}, {Vaillant}, {Valentini}, {Valette}, {van Elteren},
  {Van Hemelryck}, {van Leeuwen}, {Vaschetto}, {Vecchiato}, {Veljanoski},
  {Viala}, {Vicente}, {Vogt}, {von Essen}, {Voss}, {Votruba}, {Voutsinas},
  {Walmsley}, {Weiler}, {Wertz}, {Wevers}, {Wyrzykowski}, {Yoldas},
  {{\v{Z}}erjal}, {Ziaeepour}, {Zorec}, {Zschocke}, {Zucker}, {Zurbach}, \&
  {Zwitter}}]{gaia2018}
{Gaia Collaboration}, {Brown}, A.~G.~A., {Vallenari}, A., {et~al.} 2018, \aap,
  616, A1

\bibitem[{{Gaia Collaboration} {et~al.}(2021){Gaia Collaboration}, {Brown},
  {Vallenari}, {Prusti}, {de Bruijne}, {Babusiaux}, {Biermann}, {Creevey},
  {Evans}, {Eyer}, {Hutton}, {Jansen}, {Jordi}, {Klioner}, {Lammers},
  {Lindegren}, {Luri}, {Mignard}, {Panem}, {Pourbaix}, {Randich}, {Sartoretti},
  {Soubiran}, {Walton}, {Arenou}, {Bailer-Jones}, {Bastian}, {Cropper},
  {Drimmel}, {Katz}, {Lattanzi}, {van Leeuwen}, {Bakker}, {Cacciari},
  {Casta{\~n}eda}, {De Angeli}, {Ducourant}, {Fabricius}, {Fouesneau},
  {Fr{\'e}mat}, {Guerra}, {Guerrier}, {Guiraud}, {Jean-Antoine Piccolo},
  {Masana}, {Messineo}, {Mowlavi}, {Nicolas}, {Nienartowicz}, {Pailler},
  {Panuzzo}, {Riclet}, {Roux}, {Seabroke}, {Sordo}, {Tanga}, {Th{\'e}venin},
  {Gracia-Abril}, {Portell}, {Teyssier}, {Altmann}, {Andrae}, {Bellas-Velidis},
  {Benson}, {Berthier}, {Blomme}, {Brugaletta}, {Burgess}, {Busso}, {Carry},
  {Cellino}, {Cheek}, {Clementini}, {Damerdji}, {Davidson}, {Delchambre},
  {Dell'Oro}, {Fern{\'a}ndez-Hern{\'a}ndez}, {Galluccio}, {Garc{\'\i}a-Lario},
  {Garcia-Reinaldos}, {Gonz{\'a}lez-N{\'u}{\~n}ez}, {Gosset}, {Haigron},
  {Halbwachs}, {Hambly}, {Harrison}, {Hatzidimitriou}, {Heiter},
  {Hern{\'a}ndez}, {Hestroffer}, {Hodgkin}, {Holl}, {Jan{\ss}en}, {Jevardat de
  Fombelle}, {Jordan}, {Krone-Martins}, {Lanzafame}, {L{\"o}ffler}, {Lorca},
  {Manteiga}, {Marchal}, {Marrese}, {Moitinho}, {Mora}, {Muinonen}, {Osborne},
  {Pancino}, {Pauwels}, {Petit}, {Recio-Blanco}, {Richards}, {Riello},
  {Rimoldini}, {Robin}, {Roegiers}, {Rybizki}, {Sarro}, {Siopis}, {Smith},
  {Sozzetti}, {Ulla}, {Utrilla}, {van Leeuwen}, {van Reeven}, {Abbas}, {Abreu
  Aramburu}, {Accart}, {Aerts}, {Aguado}, {Ajaj}, {Altavilla}, {{\'A}lvarez},
  {{\'A}lvarez Cid-Fuentes}, {Alves}, {Anderson}, {Anglada Varela}, {Antoja},
  {Audard}, {Baines}, {Baker}, {Balaguer-N{\'u}{\~n}ez}, {Balbinot}, {Balog},
  {Barache}, {Barbato}, {Barros}, {Barstow}, {Bartolom{\'e}}, {Bassilana},
  {Bauchet}, {Baudesson-Stella}, {Becciani}, {Bellazzini}, {Bernet}, {Bertone},
  {Bianchi}, {Blanco-Cuaresma}, {Boch}, {Bombrun}, {Bossini}, {Bouquillon},
  {Bragaglia}, {Bramante}, {Breedt}, {Bressan}, {Brouillet}, {Bucciarelli},
  {Burlacu}, {Busonero}, {Butkevich}, {Buzzi}, {Caffau}, {Cancelliere},
  {C{\'a}novas}, {Cantat-Gaudin}, {Carballo}, {Carlucci}, {Carnerero},
  {Carrasco}, {Casamiquela}, {Castellani}, {Castro-Ginard}, {Castro Sampol},
  {Chaoul}, {Charlot}, {Chemin}, {Chiavassa}, {Cioni}, {Comoretto}, {Cooper},
  {Cornez}, {Cowell}, {Crifo}, {Crosta}, {Crowley}, {Dafonte}, {Dapergolas},
  {David}, {David}, {de Laverny}, {De Luise}, {De March}, {De Ridder}, {de
  Souza}, {de Teodoro}, {de Torres}, {del Peloso}, {del Pozo}, {Delbo},
  {Delgado}, {Delgado}, {Delisle}, {Di Matteo}, {Diakite}, {Diener},
  {Distefano}, {Dolding}, {Eappachen}, {Edvardsson}, {Enke}, {Esquej}, {Fabre},
  {Fabrizio}, {Faigler}, {Fedorets}, {Fernique}, {Fienga}, {Figueras},
  {Fouron}, {Fragkoudi}, {Fraile}, {Franke}, {Gai}, {Garabato},
  {Garcia-Gutierrez}, {Garc{\'\i}a-Torres}, {Garofalo}, {Gavras}, {Gerlach},
  {Geyer}, {Giacobbe}, {Gilmore}, {Girona}, {Giuffrida}, {Gomel}, {Gomez},
  {Gonzalez-Santamaria}, {Gonz{\'a}lez-Vidal}, {Granvik},
  {Guti{\'e}rrez-S{\'a}nchez}, {Guy}, {Hauser}, {Haywood}, {Helmi}, {Hidalgo},
  {Hilger}, {H{\l}adczuk}, {Hobbs}, {Holland}, {Huckle}, {Jasniewicz},
  {Jonker}, {Juaristi Campillo}, {Julbe}, {Karbevska}, {Kervella}, {Khanna},
  {Kochoska}, {Kontizas}, {Kordopatis}, {Korn}, {Kostrzewa-Rutkowska},
  {Kruszy{\'n}ska}, {Lambert}, {Lanza}, {Lasne}, {Le Campion}, {Le Fustec},
  {Lebreton}, {Lebzelter}, {Leccia}, {Leclerc}, {Lecoeur-Taibi}, {Liao},
  {Licata}, {Lindstr{\o}m}, {Lister}, {Livanou}, {Lobel}, {Madrero Pardo},
  {Managau}, {Mann}, {Marchant}, {Marconi}, {Marcos Santos}, {Marinoni},
  {Marocco}, {Marshall}, {Martin Polo}, {Mart{\'\i}n-Fleitas}, {Masip},
  {Massari}, {Mastrobuono-Battisti}, {Mazeh}, {McMillan}, {Messina},
  {Michalik}, {Millar}, {Mints}, {Molina}, {Molinaro}, {Moln{\'a}r},
  {Montegriffo}, {Mor}, {Morbidelli}, {Morel}, {Morris}, {Mulone}, {Munoz},
  {Muraveva}, {Murphy}, {Musella}, {Noval}, {Ord{\'e}novic}, {Orr{\`u}},
  {Osinde}, {Pagani}, {Pagano}, {Palaversa}, {Palicio}, {Panahi}, {Pawlak},
  {Pe{\~n}alosa Esteller}, {Penttil{\"a}}, {Piersimoni}, {Pineau}, {Plachy},
  {Plum}, {Poggio}, {Poretti}, {Poujoulet}, {Pr{\v{s}}a}, {Pulone}, {Racero},
  {Ragaini}, {Rainer}, {Raiteri}, {Rambaux}, {Ramos}, {Ramos-Lerate}, {Re
  Fiorentin}, {Regibo}, {Reyl{\'e}}, {Ripepi}, {Riva}, {Rixon}, {Robichon},
  {Robin}, {Roelens}, {Rohrbasser}, {Romero-G{\'o}mez}, {Rowell}, {Royer},
  {Rybicki}, {Sadowski}, {Sagrist{\`a} Sell{\'e}s}, {Sahlmann}, {Salgado},
  {Salguero}, {Samaras}, {Sanchez Gimenez}, {Sanna}, {Santove{\~n}a},
  {Sarasso}, {Schultheis}, {Sciacca}, {Segol}, {Segovia}, {S{\'e}gransan},
  {Semeux}, {Shahaf}, {Siddiqui}, {Siebert}, {Siltala}, {Slezak}, {Smart},
  {Solano}, {Solitro}, {Souami}, {Souchay}, {Spagna}, {Spoto}, {Steele},
  {Steidelm{\"u}ller}, {Stephenson}, {S{\"u}veges}, {Szabados}, {Szegedi-Elek},
  {Taris}, {Tauran}, {Taylor}, {Teixeira}, {Thuillot}, {Tonello}, {Torra},
  {Torra}, {Turon}, {Unger}, {Vaillant}, {van Dillen}, {Vanel}, {Vecchiato},
  {Viala}, {Vicente}, {Voutsinas}, {Weiler}, {Wevers}, {Wyrzykowski}, {Yoldas},
  {Yvard}, {Zhao}, {Zorec}, {Zucker}, {Zurbach}, \& {Zwitter}}]{gaia2021}
---. 2021, \aap, 649, A1

\bibitem[{{Gaia Collaboration} {et~al.}(2023){Gaia Collaboration}, {Vallenari},
  {Brown}, {Prusti}, {de Bruijne}, {Arenou}, {Babusiaux}, {Biermann},
  {Creevey}, {Ducourant}, {Evans}, {Eyer}, {Guerra}, {Hutton}, {Jordi},
  {Klioner}, {Lammers}, {Lindegren}, {Luri}, {Mignard}, {Panem}, {Pourbaix},
  {Randich}, {Sartoretti}, {Soubiran}, {Tanga}, {Walton}, {Bailer-Jones},
  {Bastian}, {Drimmel}, {Jansen}, {Katz}, {Lattanzi}, {van Leeuwen}, {Bakker},
  {Cacciari}, {Casta{\~n}eda}, {De Angeli}, {Fabricius}, {Fouesneau},
  {Fr{\'e}mat}, {Galluccio}, {Guerrier}, {Heiter}, {Masana}, {Messineo},
  {Mowlavi}, {Nicolas}, {Nienartowicz}, {Pailler}, {Panuzzo}, {Riclet}, {Roux},
  {Seabroke}, {Sordo}, {Th{\'e}venin}, {Gracia-Abril}, {Portell}, {Teyssier},
  {Altmann}, {Andrae}, {Audard}, {Bellas-Velidis}, {Benson}, {Berthier},
  {Blomme}, {Burgess}, {Busonero}, {Busso}, {C{\'a}novas}, {Carry}, {Cellino},
  {Cheek}, {Clementini}, {Damerdji}, {Davidson}, {de Teodoro}, {Nu{\~n}ez
  Campos}, {Delchambre}, {Dell'Oro}, {Esquej}, {Fern{\'a}ndez-Hern{\'a}ndez},
  {Fraile}, {Garabato}, {Garc{\'\i}a-Lario}, {Gosset}, {Haigron}, {Halbwachs},
  {Hambly}, {Harrison}, {Hern{\'a}ndez}, {Hestroffer}, {Hodgkin}, {Holl},
  {Jan{\ss}en}, {Jevardat de Fombelle}, {Jordan}, {Krone-Martins}, {Lanzafame},
  {L{\"o}ffler}, {Marchal}, {Marrese}, {Moitinho}, {Muinonen}, {Osborne},
  {Pancino}, {Pauwels}, {Recio-Blanco}, {Reyl{\'e}}, {Riello}, {Rimoldini},
  {Roegiers}, {Rybizki}, {Sarro}, {Siopis}, {Smith}, {Sozzetti}, {Utrilla},
  {van Leeuwen}, {Abbas}, {{\'A}brah{\'a}m}, {Abreu Aramburu}, {Aerts},
  {Aguado}, {Ajaj}, {Aldea-Montero}, {Altavilla}, {{\'A}lvarez}, {Alves},
  {Anders}, {Anderson}, {Anglada Varela}, {Antoja}, {Baines}, {Baker},
  {Balaguer-N{\'u}{\~n}ez}, {Balbinot}, {Balog}, {Barache}, {Barbato},
  {Barros}, {Barstow}, {Bartolom{\'e}}, {Bassilana}, {Bauchet}, {Becciani},
  {Bellazzini}, {Berihuete}, {Bernet}, {Bertone}, {Bianchi}, {Binnenfeld},
  {Blanco-Cuaresma}, {Blazere}, {Boch}, {Bombrun}, {Bossini}, {Bouquillon},
  {Bragaglia}, {Bramante}, {Breedt}, {Bressan}, {Brouillet}, {Brugaletta},
  {Bucciarelli}, {Burlacu}, {Butkevich}, {Buzzi}, {Caffau}, {Cancelliere},
  {Cantat-Gaudin}, {Carballo}, {Carlucci}, {Carnerero}, {Carrasco},
  {Casamiquela}, {Castellani}, {Castro-Ginard}, {Chaoul}, {Charlot}, {Chemin},
  {Chiaramida}, {Chiavassa}, {Chornay}, {Comoretto}, {Contursi}, {Cooper},
  {Cornez}, {Cowell}, {Crifo}, {Cropper}, {Crosta}, {Crowley}, {Dafonte},
  {Dapergolas}, {David}, {David}, {de Laverny}, {De Luise}, {De March}, {De
  Ridder}, {de Souza}, {de Torres}, {del Peloso}, {del Pozo}, {Delbo},
  {Delgado}, {Delisle}, {Demouchy}, {Dharmawardena}, {Di Matteo}, {Diakite},
  {Diener}, {Distefano}, {Dolding}, {Edvardsson}, {Enke}, {Fabre}, {Fabrizio},
  {Faigler}, {Fedorets}, {Fernique}, {Fienga}, {Figueras}, {Fournier},
  {Fouron}, {Fragkoudi}, {Gai}, {Garcia-Gutierrez}, {Garcia-Reinaldos},
  {Garc{\'\i}a-Torres}, {Garofalo}, {Gavel}, {Gavras}, {Gerlach}, {Geyer},
  {Giacobbe}, {Gilmore}, {Girona}, {Giuffrida}, {Gomel}, {Gomez},
  {Gonz{\'a}lez-N{\'u}{\~n}ez}, {Gonz{\'a}lez-Santamar{\'\i}a},
  {Gonz{\'a}lez-Vidal}, {Granvik}, {Guillout}, {Guiraud},
  {Guti{\'e}rrez-S{\'a}nchez}, {Guy}, {Hatzidimitriou}, {Hauser}, {Haywood},
  {Helmer}, {Helmi}, {Sarmiento}, {Hidalgo}, {Hilger}, {H{\l}adczuk}, {Hobbs},
  {Holland}, {Huckle}, {Jardine}, {Jasniewicz}, {Jean-Antoine Piccolo},
  {Jim{\'e}nez-Arranz}, {Jorissen}, {Juaristi Campillo}, {Julbe}, {Karbevska},
  {Kervella}, {Khanna}, {Kontizas}, {Kordopatis}, {Korn}, {K{\'o}sp{\'a}l},
  {Kostrzewa-Rutkowska}, {Kruszy{\'n}ska}, {Kun}, {Laizeau}, {Lambert},
  {Lanza}, {Lasne}, {Le Campion}, {Lebreton}, {Lebzelter}, {Leccia}, {Leclerc},
  {Lecoeur-Taibi}, {Liao}, {Licata}, {Lindstr{\o}m}, {Lister}, {Livanou},
  {Lobel}, {Lorca}, {Loup}, {Madrero Pardo}, {Magdaleno Romeo}, {Managau},
  {Mann}, {Manteiga}, {Marchant}, {Marconi}, {Marcos}, {Marcos Santos},
  {Mar{\'\i}n Pina}, {Marinoni}, {Marocco}, {Marshall}, {Martin Polo},
  {Mart{\'\i}n-Fleitas}, {Marton}, {Mary}, {Masip}, {Massari},
  {Mastrobuono-Battisti}, {Mazeh}, {McMillan}, {Messina}, {Michalik}, {Millar},
  {Mints}, {Molina}, {Molinaro}, {Moln{\'a}r}, {Monari}, {Mongui{\'o}},
  {Montegriffo}, {Montero}, {Mor}, {Mora}, {Morbidelli}, {Morel}, {Morris},
  {Muraveva}, {Murphy}, {Musella}, {Nagy}, {Noval}, {Oca{\~n}a}, {Ogden},
  {Ordenovic}, {Osinde}, {Pagani}, {Pagano}, {Palaversa}, {Palicio},
  {Pallas-Quintela}, {Panahi}, {Payne-Wardenaar}, {Pe{\~n}alosa Esteller},
  {Penttil{\"a}}, {Pichon}, {Piersimoni}, {Pineau}, {Plachy}, {Plum}, {Poggio},
  {Pr{\v{s}}a}, {Pulone}, {Racero}, {Ragaini}, {Rainer}, {Raiteri}, {Rambaux},
  {Ramos}, {Ramos-Lerate}, {Re Fiorentin}, {Regibo}, {Richards}, {Rios Diaz},
  {Ripepi}, {Riva}, {Rix}, {Rixon}, {Robichon}, {Robin}, {Robin}, {Roelens},
  {Rogues}, {Rohrbasser}, {Romero-G{\'o}mez}, {Rowell}, {Royer}, {Ruz Mieres},
  {Rybicki}, {Sadowski}, {S{\'a}ez N{\'u}{\~n}ez}, {Sagrist{\`a} Sell{\'e}s},
  {Sahlmann}, {Salguero}, {Samaras}, {Sanchez Gimenez}, {Sanna},
  {Santove{\~n}a}, {Sarasso}, {Schultheis}, {Sciacca}, {Segol}, {Segovia},
  {S{\'e}gransan}, {Semeux}, {Shahaf}, {Siddiqui}, {Siebert}, {Siltala},
  {Silvelo}, {Slezak}, {Slezak}, {Smart}, {Snaith}, {Solano}, {Solitro},
  {Souami}, {Souchay}, {Spagna}, {Spina}, {Spoto}, {Steele},
  {Steidelm{\"u}ller}, {Stephenson}, {S{\"u}veges}, {Surdej}, {Szabados},
  {Szegedi-Elek}, {Taris}, {Taylor}, {Teixeira}, {Tolomei}, {Tonello}, {Torra},
  {Torra}, {Torralba Elipe}, {Trabucchi}, {Tsounis}, {Turon}, {Ulla}, {Unger},
  {Vaillant}, {van Dillen}, {van Reeven}, {Vanel}, {Vecchiato}, {Viala},
  {Vicente}, {Voutsinas}, {Weiler}, {Wevers}, {Wyrzykowski}, {Yoldas}, {Yvard},
  {Zhao}, {Zorec}, {Zucker}, \& {Zwitter}}]{gaia2023}
{Gaia Collaboration}, {Vallenari}, A., {Brown}, A.~G.~A., {et~al.} 2023, \aap,
  674, A1

\bibitem[{{Grunblatt} {et~al.}(2018){Grunblatt}, {Huber}, {Gaidos}, {Lopez},
  {Barclay}, {Chontos}, {Sinukoff}, {Van Eylen}, {Howard}, \&
  {Isaacson}}]{grunblatt2018}
{Grunblatt}, S.~K., {Huber}, D., {Gaidos}, E., {et~al.} 2018, \apjl, 861, L5

\bibitem[{{Grunblatt} {et~al.}(2022){Grunblatt}, {Saunders}, {Sun}, {Chontos},
  {Soares-Furtado}, {Eisner}, {Pereira}, {Komacek}, {Huber}, {Collins}, {Wang},
  {Stockdale}, {Quinn}, {Tronsgaard}, {Zhou}, {Nowak}, {Deeg}, {Ciardi},
  {Boyle}, {Rice}, {Dai}, {Blunt}, {Van Zandt}, {Beard}, {Akana Murphy},
  {Dalba}, {Lubin}, {Polanski}, {Brinkman}, {Howard}, {Buchhave}, {Angus},
  {Ricker}, {Jenkins}, {Wohler}, {Goeke}, {Levine}, {Colon}, {Huang},
  {Kunimoto}, {Shporer}, {Latham}, {Seager}, {Vanderspek}, \&
  {Winn}}]{grunblatt2022}
{Grunblatt}, S.~K., {Saunders}, N., {Sun}, M., {et~al.} 2022, \aj, 163, 120

\bibitem[{{Grunblatt} {et~al.}(2023){Grunblatt}, {Saunders}, {Chontos},
  {Hattori}, {Veras}, {Huber}, {Angus}, {Rice}, {Breivik}, {Blunt},
  {Giacalone}, {Lubin}, {Isaacson}, {Howard}, {Ciardi}, {Safonov}, {Strakhov},
  {Latham}, {Bieryla}, {Ricker}, {Jenkins}, {Tenenbaum}, {Shporer}, {Morgan},
  {Kostov}, {Osborn}, {Dragomir}, {Seager}, {Vanderspek}, \&
  {Winn}}]{grunblatt2023a}
{Grunblatt}, S.~K., {Saunders}, N., {Chontos}, A., {et~al.} 2023, \aj, 165, 44

\bibitem[{{Guerrero} {et~al.}(2021){Guerrero}, {Seager}, {Huang}, {Vanderburg},
  {Garcia Soto}, {Mireles}, {Hesse}, {Fong}, {Glidden}, {Shporer}, {Latham},
  {Collins}, {Quinn}, {Burt}, {Dragomir}, {Crossfield}, {Vanderspek},
  {Fausnaugh}, {Burke}, {Ricker}, {Daylan}, {Essack}, {G{\"u}nther}, {Osborn},
  {Pepper}, {Rowden}, {Sha}, {Villanueva}, {Yahalomi}, {Yu}, {Ballard},
  {Batalha}, {Berardo}, {Chontos}, {Dittmann}, {Esquerdo}, {Mikal-Evans},
  {Jayaraman}, {Krishnamurthy}, {Louie}, {Mehrle}, {Niraula}, {Rackham},
  {Rodriguez}, {Rowden}, {Sousa-Silva}, {Watanabe}, {Wong}, {Zhan},
  {Zivanovic}, {Christiansen}, {Ciardi}, {Swain}, {Lund}, {Mullally},
  {Fleming}, {Rodriguez}, {Boyd}, {Quintana}, {Barclay}, {Col{\'o}n},
  {Rinehart}, {Schlieder}, {Clampin}, {Jenkins}, {Twicken}, {Caldwell},
  {Coughlin}, {Henze}, {Lissauer}, {Morris}, {Rose}, {Smith}, {Tenenbaum},
  {Ting}, {Wohler}, {Bakos}, {Bean}, {Berta-Thompson}, {Bieryla}, {Bouma},
  {Buchhave}, {Butler}, {Charbonneau}, {Doty}, {Ge}, {Holman}, {Howard},
  {Kaltenegger}, {Kane}, {Kjeldsen}, {Kreidberg}, {Lin}, {Minsky}, {Narita},
  {Paegert}, {P{\'a}l}, {Palle}, {Sasselov}, {Spencer}, {Sozzetti}, {Stassun},
  {Torres}, {Udry}, \& {Winn}}]{guerrero2021}
{Guerrero}, N.~M., {Seager}, S., {Huang}, C.~X., {et~al.} 2021, arXiv e-prints,
  arXiv:2103.12538

\bibitem[{{Hamer} \& {Schlaufman}(2019)}]{hamer2019}
{Hamer}, J.~H., \& {Schlaufman}, K.~C. 2019, \aj, 158, 190

\bibitem[{{Hon} {et~al.}(2023){Hon}, {Huber}, {Rui}, {Fuller}, {Veras},
  {Kuszlewicz}, {Kochukhov}, {Stokholm}, {R{\o}rsted}, {Y{\i}ld{\i}z}, {Orhan},
  {{\"O}rtel}, {Jiang}, {Hey}, {Isaacson}, {Zhang}, {Vrard}, {Stassun},
  {Shappee}, {Tayar}, {Claytor}, {Beard}, {Bedding}, {Brinkman}, {Campante},
  {Chaplin}, {Chontos}, {Giacalone}, {Holcomb}, {Howard}, {Lubin},
  {MacDougall}, {Montet}, {Murphy}, {Ong}, {Pidhorodetska}, {Polanski}, {Rice},
  {Stello}, {Tyler}, {Van Zandt}, \& {Weiss}}]{hon2023}
{Hon}, M., {Huber}, D., {Rui}, N.~Z., {et~al.} 2023, \nat, 618, 917

\bibitem[{{Howard} {et~al.}(2010){Howard}, {Johnson}, {Marcy}, {Fischer},
  {Wright}, {Bernat}, {Henry}, {Peek}, {Isaacson}, {Apps}, {Endl}, {Cochran},
  {Valenti}, {Anderson}, \& {Piskunov}}]{howard2010}
{Howard}, A.~W., {Johnson}, J.~A., {Marcy}, G.~W., {et~al.} 2010, \apj, 721,
  1467

\bibitem[{{Huang} {et~al.}(2020{\natexlab{a}}){Huang}, {Vanderburg}, {P{\'a}l},
  {Sha}, {Yu}, {Fong}, {Fausnaugh}, {Shporer}, {Guerrero}, {Vanderspek}, \&
  {Ricker}}]{huang2020a}
{Huang}, C.~X., {Vanderburg}, A., {P{\'a}l}, A., {et~al.} 2020{\natexlab{a}},
  Research Notes of the American Astronomical Society, 4, 204

\bibitem[{{Huang} {et~al.}(2020{\natexlab{b}}){Huang}, {Vanderburg}, {P{\'a}l},
  {Sha}, {Yu}, {Fong}, {Fausnaugh}, {Shporer}, {Guerrero}, {Vanderspek}, \&
  {Ricker}}]{huang2020b}
---. 2020{\natexlab{b}}, Research Notes of the American Astronomical Society,
  4, 206

\bibitem[{{Huber} {et~al.}(2017){Huber}, {Zinn}, {Bojsen-Hansen},
  {Pinsonneault}, {Sahlholdt}, {Serenelli}, {Silva Aguirre}, {Stassun},
  {Stello}, {Tayar}, {Bastien}, {Bedding}, {Buchhave}, {Chaplin}, {Davies},
  {Garc{\'\i}a}, {Latham}, {Mathur}, {Mosser}, \& {Sharma}}]{huber2017}
{Huber}, D., {Zinn}, J., {Bojsen-Hansen}, M., {et~al.} 2017, \apj, 844, 102

\bibitem[{{Huber} {et~al.}(2019){Huber}, {Chaplin}, {Chontos}, {Kjeldsen},
  {Christensen-Dalsgaard}, {Bedding}, {Ball}, {Brahm}, {Espinoza}, {Henning},
  {Jord{\'a}n}, {Sarkis}, {Knudstrup}, {Albrecht}, {Grundahl}, {Fredslund
  Andersen}, {Pall{\'e}}, {Crossfield}, {Fulton}, {Howard}, {Isaacson},
  {Weiss}, {Handberg}, {Lund}, {Serenelli}, {R{\o}rsted Mosumgaard},
  {Stokholm}, {Bieryla}, {Buchhave}, {Latham}, {Quinn}, {Gaidos}, {Hirano},
  {Ricker}, {Vanderspek}, {Seager}, {Jenkins}, {Winn}, {Antia}, {Appourchaux},
  {Basu}, {Bell}, {Benomar}, {Bonanno}, {Buzasi}, {Campante}, {{\c{C}}elik
  Orhan}, {Corsaro}, {Cunha}, {Davies}, {Deheuvels}, {Grunblatt}, {Hasanzadeh},
  {Di Mauro}, {Garc{\'\i}a}, {Gaulme}, {Girardi}, {Guzik}, {Hon}, {Jiang},
  {Kallinger}, {Kawaler}, {Kuszlewicz}, {Lebreton}, {Li}, {Lucas}, {Lundkvist},
  {Mann}, {Mathis}, {Mathur}, {Mazumdar}, {Metcalfe}, {Miglio}, {Monteiro},
  {Mosser}, {Noll}, {Nsamba}, {Ong}, {{\"O}rtel}, {Pereira}, {Ranadive},
  {R{\'e}gulo}, {Rodrigues}, {Roxburgh}, {Silva Aguirre}, {Smalley},
  {Schofield}, {Sousa}, {Stassun}, {Stello}, {Tayar}, {White}, {Verma},
  {Vrard}, {Y{\i}ld{\i}z}, {Baker}, {Bazot}, {Beichmann}, {Bergmann}, {Bugnet},
  {Cale}, {Carlino}, {Cartwright}, {Christiansen}, {Ciardi}, {Creevey},
  {Dittmann}, {Do Nascimento}, {Van Eylen}, {F{\"u}r{\'e}sz}, {Gagn{\'e}},
  {Gao}, {Gazeas}, {Giddens}, {Hall}, {Hekker}, {Ireland}, {Latouf}, {LeBrun},
  {Levine}, {Matzko}, {Natinsky}, {Page}, {Plavchan}, {Mansouri-Samani},
  {McCauliff}, {Mullally}, {Orenstein}, {Garcia Soto}, {Paegert}, {van Saders},
  {Schnaible}, {Soderblom}, {Szab{\'o}}, {Tanner}, {Tinney}, {Teske}, {Thomas},
  {Trampedach}, {Wright}, {Yuan}, \& {Zohrabi}}]{huber2019}
{Huber}, D., {Chaplin}, W.~J., {Chontos}, A., {et~al.} 2019, \aj, 157, 245

\bibitem[{{Hut}(1981)}]{hut1981}
{Hut}, P. 1981, \aap, 99, 126

\bibitem[{{Isaacson} \& {Fischer}(2010)}]{isaacson2010}
{Isaacson}, H., \& {Fischer}, D. 2010, \apj, 725, 875

\bibitem[{{Jenkins} {et~al.}(2016){Jenkins}, {Twicken}, {McCauliff},
  {Campbell}, {Sanderfer}, {Lung}, {Mansouri-Samani}, {Girouard}, {Tenenbaum},
  {Klaus}, {Smith}, {Caldwell}, {Chacon}, {Henze}, {Heiges}, {Latham},
  {Morgan}, {Swade}, {Rinehart}, \& {Vanderspek}}]{jenkins2016}
{Jenkins}, J.~M., {Twicken}, J.~D., {McCauliff}, S., {et~al.} 2016, Society of
  Photo-Optical Instrumentation Engineers (SPIE) Conference Series, Vol. 9913,
  {The TESS science processing operations center}, 99133E

\bibitem[{{Johnson} {et~al.}(2013){Johnson}, {Morton}, \&
  {Wright}}]{johnson2013}
{Johnson}, J.~A., {Morton}, T.~D., \& {Wright}, J.~T. 2013, \apj, 763, 53

\bibitem[{{Johnson} {et~al.}(2007){Johnson}, {Fischer}, {Marcy}, {Wright},
  {Driscoll}, {Butler}, {Hekker}, {Reffert}, \& {Vogt}}]{johnson2007}
{Johnson}, J.~A., {Fischer}, D.~A., {Marcy}, G.~W., {et~al.} 2007, \apj, 665,
  785

\bibitem[{{Kab{\'a}th} {et~al.}(2022){Kab{\'a}th}, {Chaturvedi}, {MacQueen},
  {Skarka}, {{\v{S}}ubjak}, {Esposito}, {Cochran}, {Bellomo}, {Karjalainen},
  {Guenther}, {Endl}, {Csizmadia}, {Karjalainen}, {Hatzes}, {{\v{Z}}{\'a}k},
  {Gandolfi}, {Boffin}, {Vines}, {Livingston}, {Garc{\'\i}a}, {Mathur},
  {Gonz{\'a}lez-Cuesta}, {Bla{\v{z}}ek}, {Caldwell}, {Col{\'o}n}, {Deeg},
  {Erikson}, {Van Eylen}, {Fong}, {Fridlund}, {Fukui}, {F{\H{u}}r{\'e}sz},
  {Goeke}, {Goffo}, {Howell}, {Jenkins}, {Klagyivik}, {Korth}, {Latham},
  {Luque}, {Moldovan}, {Murgas}, {Narita}, {Orell-Miquel}, {Palle},
  {Parviainen}, {Persson}, {Reed}, {Redfield}, {Ricker}, {Seager}, {Serrano},
  {Shporer}, {Smith}, {Watanabe}, {Winn}, \& {the KESPRINT team}}]{kabath2022}
{Kab{\'a}th}, P., {Chaturvedi}, P., {MacQueen}, P.~J., {et~al.} 2022, arXiv
  e-prints, arXiv:2205.01860

\bibitem[{{Kane}(2023)}]{kane2023}
{Kane}, S.~R. 2023, \apj, 958, 120

\bibitem[{{Kipping}(2013)}]{kipping2013}
{Kipping}, D.~M. 2013, \mnras, 435, 2152

\bibitem[{{Kolbl} {et~al.}(2015){Kolbl}, {Marcy}, {Isaacson}, \&
  {Howard}}]{kolbl2015}
{Kolbl}, R., {Marcy}, G.~W., {Isaacson}, H., \& {Howard}, A.~W. 2015, \aj, 149,
  18

\bibitem[{{Kraft}(1967)}]{kraft1967}
{Kraft}, R.~P. 1967, \apj, 150, 551

\bibitem[{{Li} {et~al.}(2019){Li}, {Tenenbaum}, {Twicken}, {Burke}, {Jenkins},
  {Quintana}, {Rowe}, \& {Seader}}]{li2019}
{Li}, J., {Tenenbaum}, P., {Twicken}, J.~D., {et~al.} 2019, \pasp, 131, 024506

\bibitem[{{Lopez} \& {Fortney}(2016)}]{lopez2015}
{Lopez}, E.~D., \& {Fortney}, J.~J. 2016, \apj, 818, 4

\bibitem[{{Lovis} \& {Pepe}(2007)}]{lovis2007}
{Lovis}, C., \& {Pepe}, F. 2007, \aap, 468, 1115

\bibitem[{{Lubin} {et~al.}(2021){Lubin}, {Robertson}, {Stefansson}, {Ninan},
  {Mahadevan}, {Endl}, {Ford}, {Wright}, {Beard}, {Bender}, {Cochran},
  {Diddams}, {Fredrick}, {Halverson}, {Kanodia}, {Metcalf}, {Ramsey}, {Roy},
  {Schwab}, \& {Terrien}}]{lubin2021b}
{Lubin}, J., {Robertson}, P., {Stefansson}, G., {et~al.} 2021, \aj, 162, 61

\bibitem[{{Lundkvist} {et~al.}(2016){Lundkvist}, {Kjeldsen}, {Albrecht},
  {Davies}, {Basu}, {Huber}, {Justesen}, {Karoff}, {Silva Aguirre}, {van
  Eylen}, {Vang}, {Arentoft}, {Barclay}, {Bedding}, {Campante}, {Chaplin},
  {Christensen-Dalsgaard}, {Elsworth}, {Gilliland}, {Handberg}, {Hekker},
  {Kawaler}, {Lund}, {Metcalfe}, {Miglio}, {Rowe}, {Stello}, {Tingley}, \&
  {White}}]{lundkvist2016}
{Lundkvist}, M.~S., {Kjeldsen}, H., {Albrecht}, S., {et~al.} 2016, Nature
  Communications, 7, 11201

\bibitem[{{Mandel} \& {Agol}(2002)}]{mandel2002}
{Mandel}, K., \& {Agol}, E. 2002, \apjl, 580, L171

\bibitem[{{Miyazaki} \& {Masuda}(2023)}]{miyazaski2023}
{Miyazaki}, S., \& {Masuda}, K. 2023, \aj, 166, 209

\bibitem[{{Petigura}(2015)}]{petigura2015}
{Petigura}, E.~A. 2015, PhD thesis, University of California, Berkeley

\bibitem[{{Petigura} {et~al.}(2017{\natexlab{a}}){Petigura}, {Howard}, {Marcy},
  {Johnson}, {Isaacson}, {Cargile}, {Hebb}, {Fulton}, {Weiss}, {Morton},
  {Winn}, {Rogers}, {Sinukoff}, {Hirsch}, \& {Crossfield}}]{petigura2017a}
{Petigura}, E.~A., {Howard}, A.~W., {Marcy}, G.~W., {et~al.}
  2017{\natexlab{a}}, \aj, 154, 107

\bibitem[{{Petigura} {et~al.}(2017{\natexlab{b}}){Petigura}, {Sinukoff},
  {Lopez}, {Crossfield}, {Howard}, {Brewer}, {Fulton}, {Isaacson}, {Ciardi},
  {Howell}, {Everett}, {Horch}, {Hirsch}, {Weiss}, \&
  {Schlieder}}]{petigura2017b}
{Petigura}, E.~A., {Sinukoff}, E., {Lopez}, E.~D., {et~al.} 2017{\natexlab{b}},
  \aj, 153, 142

\bibitem[{{Rodriguez} {et~al.}(2021){Rodriguez}, {Quinn}, {Zhou}, {Vanderburg},
  {Nielsen}, {Wittenmyer}, {Brahm}, {Reed}, {Huang}, {Vach}, {Ciardi},
  {Oelkers}, {Stassun}, {Hellier}, {Gaudi}, {Eastman}, {Collins}, {Bieryla},
  {Christian}, {Latham}, {Carleo}, {Wright}, {Matthews}, {Gonzales}, {Ziegler},
  {Dressing}, {Howell}, {Tan}, {Wittrock}, {Plavchan}, {McLeod}, {Baker},
  {Wang}, {Radford}, {Schwarz}, {Esposito}, {Ricker}, {Vanderspek}, {Seager},
  {Winn}, {Jenkins}, {Addison}, {Anderson}, {Barclay}, {Beatty}, {Berlind},
  {Bouchy}, {Bowen}, {Bowler}, {Brasseur}, {Brice{\~n}o}, {Caldwell},
  {Calkins}, {Cartwright}, {Chaturvedi}, {Chaverot}, {Chimaladinne},
  {Christiansen}, {Collins}, {Crossfield}, {Eastridge}, {Espinoza}, {Esquerdo},
  {Feliz}, {Fenske}, {Fong}, {Gan}, {Giacalone}, {Gill}, {Gordon}, {Granados},
  {Grieves}, {Guenther}, {Guerrero}, {Henning}, {Henze}, {Hesse}, {Hobson},
  {Horner}, {James}, {Jensen}, {Jimenez}, {Jord{\'a}n}, {Kane}, {Kielkopf},
  {Kim}, {Kuhn}, {Latouf}, {Law}, {Levine}, {Lund}, {Mann}, {Mao}, {Matson},
  {Mengel}, {Mink}, {Newman}, {O'Dwyer}, {Okumura}, {Palle}, {Pepper},
  {Quintana}, {Sarkis}, {Savel}, {Schlieder}, {Schnaible}, {Shporer}, {Sefako},
  {Seidel}, {Siverd}, {Skinner}, {Stalport}, {Stevens}, {Stibbards}, {Tinney},
  {West}, {Yahalomi}, \& {Zhang}}]{rodriguez2021}
{Rodriguez}, J.~E., {Quinn}, S.~N., {Zhou}, G., {et~al.} 2021, \aj, 161, 194

\bibitem[{{Rodriguez} {et~al.}(2023){Rodriguez}, {Quinn}, {Vanderburg}, {Zhou},
  {Eastman}, {Thygesen}, {Cale}, {Ciardi}, {Reed}, {Oelkers}, {Collins},
  {Bieryla}, {Latham}, {Gonzales}, {Scott Gaudi}, {Hellier}, {Jones}, {Brahm},
  {Sokolovsky}, {Schulte}, {Srdoc}, {Kielkopf}, {Grau Horta}, {Massey},
  {Evans}, {Stephens}, {McLeod}, {Chazov}, {Krushinsky}, {Ghachoui}, {Safonov},
  {Dedrick}, {Conti}, {Laloum}, {Giacalone}, {Ziegler}, {Guerra Serra}, {Naves
  Nogues}, {Murgas}, {Michaels}, {Ricker}, {Vanderspek}, {Seager}, {Winn},
  {Jenkins}, {Addison}, {Alfaro}, {Anderson}, {Aydi}, {Beatty}, {Bedding},
  {Belinski}, {Benkhaldoun}, {Berlind}, {Blake}, {Bowen}, {Bowler}, {Boyle},
  {Branson}, {Brice{\~n}o}, {Calkins}, {Campbell}, {Christiansen}, {Chomiuk},
  {Collins}, {Cornachione}, {Daassou}, {Dressing}, {Esquerdo}, {Feliz}, {Fong},
  {Fukui}, {Gan}, {Gill}, {Goliguzova}, {Hansen}, {Henning}, {Hintz}, {Hobson},
  {Horner}, {Huang}, {James}, {Jensen}, {Johnson}, {Jord{\'a}n}, {Kane},
  {Barkaoui}, {Kim}, {Kim}, {Kuhn}, {Law}, {Lewin}, {Liu}, {Lund}, {Mann},
  {McCrady}, {Mengel}, {Mink}, {Murphy}, {Narita}, {Newman}, {Okumura},
  {Osborn}, {Paegert}, {Palle}, {Pepper}, {Plavchan}, {Popov}, {Rabus},
  {Ranshaw}, {Rodriguez}, {Roh}, {Reefe}, {Savel}, {Schwarz}, {Shporer},
  {Siverd}, {Sliski}, {Stassun}, {Stevens}, {Soubkiou}, {Ting}, {Tinney},
  {Vowell}, {Walton}, {West}, {Wilson}, {Wittenmyer}, {Wittrock}, {Wolf},
  {Wright}, {Zhang}, \& {Zobel}}]{rodriguez2023}
{Rodriguez}, J.~E., {Quinn}, S.~N., {Vanderburg}, A., {et~al.} 2023, \mnras,
  521, 2765

\bibitem[{{Rosenthal} {et~al.}(2021){Rosenthal}, {Fulton}, {Hirsch},
  {Isaacson}, {Howard}, {Dedrick}, {Sherstyuk}, {Blunt}, {Petigura}, {Knutson},
  {Behmard}, {Chontos}, {Crepp}, {Crossfield}, {Dalba}, {Fischer}, {Henry},
  {Kane}, {Kosiarek}, {Marcy}, {Rubenzahl}, {Weiss}, \&
  {Wright}}]{rosenthal2021}
{Rosenthal}, L.~J., {Fulton}, B.~J., {Hirsch}, L.~A., {et~al.} 2021, \apjs,
  255, 8

\bibitem[{{Saunders} {et~al.}(2022){Saunders}, {Grunblatt}, {Huber}, {Collins},
  {Jensen}, {Vanderburg}, {Brahm}, {Jord{\'a}n}, {Espinoza}, {Henning},
  {Hobson}, {Quinn}, {Zhou}, {Butler}, {Crause}, {Kuhn}, {Moses Mogotsi},
  {Hellier}, {Angus}, {Hattori}, {Chontos}, {Ricker}, {Jenkins}, {Tenenbaum},
  {Latham}, {Seager}, {Vanderspek}, {Winn}, {Stockdale}, \&
  {Cloutier}}]{saunders2022}
{Saunders}, N., {Grunblatt}, S.~K., {Huber}, D., {et~al.} 2022, \aj, 163, 53

\bibitem[{{Schlaufman} \& {Winn}(2013)}]{schlaufman2013}
{Schlaufman}, K.~C., \& {Winn}, J.~N. 2013, \apj, 772, 143

\bibitem[{{Sha} {et~al.}(2021){Sha}, {Huang}, {Shporer}, {Rodriguez},
  {Vanderburg}, {Brahm}, {Hagelberg}, {Matthews}, {Ziegler}, {Livingston},
  {Stassun}, {Wright}, {Crane}, {Espinoza}, {Bouchy}, {Bakos}, {Collins},
  {Zhou}, {Bieryla}, {Hartman}, {Wittenmyer}, {Nielsen}, {Plavchan}, {Bayliss},
  {Sarkis}, {Tan}, {Cloutier}, {Mancini}, {Jord{\'a}n}, {Wang}, {Henning},
  {Narita}, {Penev}, {Teske}, {Kane}, {Mann}, {Addison}, {Tamura}, {Horner},
  {Barbieri}, {Burt}, {D{\'\i}az}, {Crossfield}, {Dragomir}, {Drass},
  {Feinstein}, {Zhang}, {Hart}, {Kielkopf}, {Jensen}, {Montet}, {Ottoni},
  {Schwarz}, {Rojas}, {Nespral}, {Torres}, {Mengel}, {Udry}, {Zapata},
  {Snoddy}, {Okumura}, {Ricker}, {Vanderspek}, {Latham}, {Winn}, {Seager},
  {Jenkins}, {Col{\'o}n}, {Henze}, {Krishnamurthy}, {Ting}, {Vezie}, \&
  {Villanueva}}]{sha2021}
{Sha}, L., {Huang}, C.~X., {Shporer}, A., {et~al.} 2021, \aj, 161, 82

\bibitem[{{Skrutskie} {et~al.}(2006){Skrutskie}, {Cutri}, {Stiening},
  {Weinberg}, {Schneider}, {Carpenter}, {Beichman}, {Capps}, {Chester},
  {Elias}, {Huchra}, {Liebert}, {Lonsdale}, {Monet}, {Price}, {Seitzer},
  {Jarrett}, {Kirkpatrick}, {Gizis}, {Howard}, {Evans}, {Fowler}, {Fullmer},
  {Hurt}, {Light}, {Kopan}, {Marsh}, {McCallon}, {Tam}, {Van Dyk}, \&
  {Wheelock}}]{skrutskie2006}
{Skrutskie}, M.~F., {Cutri}, R.~M., {Stiening}, R., {et~al.} 2006, \aj, 131,
  1163

\bibitem[{{Smith} {et~al.}(2012){Smith}, {Stumpe}, {Van Cleve}, {Jenkins},
  {Barclay}, {Fanelli}, {Girouard}, {Kolodziejczak}, {McCauliff}, {Morris}, \&
  {Twicken}}]{smith2012}
{Smith}, J.~C., {Stumpe}, M.~C., {Van Cleve}, J.~E., {et~al.} 2012, \pasp, 124,
  1000

\bibitem[{{Stassun} {et~al.}(2019){Stassun}, {Oelkers}, {Paegert}, {Torres},
  {Pepper}, {De Lee}, {Collins}, {Latham}, {Muirhead}, {Chittidi},
  {Rojas-Ayala}, {Fleming}, {Rose}, {Tenenbaum}, {Ting}, {Kane}, {Barclay},
  {Bean}, {Brassuer}, {Charbonneau}, {Ge}, {Lissauer}, {Mann}, {McLean},
  {Mullally}, {Narita}, {Plavchan}, {Ricker}, {Sasselov}, {Seager}, {Sharma},
  {Shiao}, {Sozzetti}, {Stello}, {Vanderspek}, {Wallace}, \&
  {Winn}}]{stassun2019}
{Stassun}, K.~G., {Oelkers}, R.~J., {Paegert}, M., {et~al.} 2019, \aj, 158, 138

\bibitem[{{Stumpe} {et~al.}(2014){Stumpe}, {Smith}, {Catanzarite}, {Van Cleve},
  {Jenkins}, {Twicken}, \& {Girouard}}]{stumpe2014}
{Stumpe}, M.~C., {Smith}, J.~C., {Catanzarite}, J.~H., {et~al.} 2014, \pasp,
  126, 100

\bibitem[{{Stumpe} {et~al.}(2012){Stumpe}, {Smith}, {Van Cleve}, {Twicken},
  {Barclay}, {Fanelli}, {Girouard}, {Jenkins}, {Kolodziejczak}, {McCauliff}, \&
  {Morris}}]{stumpe2012}
{Stumpe}, M.~C., {Smith}, J.~C., {Van Cleve}, J.~E., {et~al.} 2012, \pasp, 124,
  985

\bibitem[{{Tayar} {et~al.}(2022){Tayar}, {Claytor}, {Huber}, \& {van
  Saders}}]{tayar22}
{Tayar}, J., {Claytor}, Z.~R., {Huber}, D., \& {van Saders}, J. 2022, \apj,
  927, 31

\bibitem[{{Twicken} {et~al.}(2018){Twicken}, {Catanzarite}, {Clarke},
  {Girouard}, {Jenkins}, {Klaus}, {Li}, {McCauliff}, {Seader}, {Tenenbaum},
  {Wohler}, {Bryson}, {Burke}, {Caldwell}, {Haas}, {Henze}, \&
  {Sanderfer}}]{twicken2018}
{Twicken}, J.~D., {Catanzarite}, J.~H., {Clarke}, B.~D., {et~al.} 2018, \pasp,
  130, 064502

\bibitem[{{Van Eylen} {et~al.}(2016){Van Eylen}, {Albrecht}, {Gandolfi}, {Dai},
  {Winn}, {Hirano}, {Narita}, {Bruntt}, {Prieto-Arranz}, {B{\'e}jar}, {Nowak},
  {Lund}, {Palle}, {Ribas}, {Sanchis-Ojeda}, {Yu}, {Arriagada}, {Butler},
  {Crane}, {Handberg}, {Deeg}, {Jessen-Hansen}, {Johnson}, {Nespral}, {Rogers},
  {Ryu}, {Shectman}, {Shrotriya}, {Slumstrup}, {Takeda}, {Teske}, {Thompson},
  {Vanderburg}, \& {Wittenmyer}}]{vaneylen2016}
{Van Eylen}, V., {Albrecht}, S., {Gandolfi}, D., {et~al.} 2016, \aj, 152, 143

\bibitem[{{Van Eylen} {et~al.}(2018){Van Eylen}, {Dai}, {Mathur}, {Gandolfi},
  {Albrecht}, {Fridlund}, {Garc{\'{\i}}a}, {Guenther}, {Hjorth}, {Justesen},
  {Livingston}, {Lund}, {P{\'e}rez Hern{\'a}ndez}, {Prieto-Arranz}, {Regulo},
  {Bugnet}, {Everett}, {Hirano}, {Nespral}, {Nowak}, {Palle}, {Silva Aguirre},
  {Trifonov}, {Winn}, {Barrag{\'a}n}, {Beck}, {Chaplin}, {Cochran},
  {Csizmadia}, {Deeg}, {Endl}, {Heeren}, {Grziwa}, {Hatzes}, {Hidalgo},
  {Korth}, {Mathis}, {Monta{\~n}es Rodriguez}, {Narita}, {Patzold}, {Persson},
  {Rodler}, \& {Smith}}]{vaneylen2018b}
{Van Eylen}, V., {Dai}, F., {Mathur}, S., {et~al.} 2018, \mnras, 478, 4866

\bibitem[{{Vanderburg} {et~al.}(2020){Vanderburg}, {Rappaport}, {Xu},
  {Crossfield}, {Becker}, {Gary}, {Murgas}, {Blouin}, {Kaye}, {Palle}, {Melis},
  {Morris}, {Kreidberg}, {Gorjian}, {Morley}, {Mann}, {Parviainen}, {Pearce},
  {Newton}, {Carrillo}, {Zuckerman}, {Nelson}, {Zeimann}, {Brown},
  {Tronsgaard}, {Klein}, {Ricker}, {Vanderspek}, {Latham}, {Seager}, {Winn},
  {Jenkins}, {Adams}, {Benneke}, {Berardo}, {Buchhave}, {Caldwell},
  {Christiansen}, {Collins}, {Col{\'o}n}, {Daylan}, {Doty}, {Doyle},
  {Dragomir}, {Dressing}, {Dufour}, {Fukui}, {Glidden}, {Guerrero}, {Guo},
  {Heng}, {Henriksen}, {Huang}, {Kaltenegger}, {Kane}, {Lewis}, {Lissauer},
  {Morales}, {Narita}, {Pepper}, {Rose}, {Smith}, {Stassun}, \&
  {Yu}}]{vanderburg2020}
{Vanderburg}, A., {Rappaport}, S.~A., {Xu}, S., {et~al.} 2020, \nat, 585, 363

\bibitem[{{Villaver} \& {Livio}(2009)}]{villaver2009}
{Villaver}, E., \& {Livio}, M. 2009, \apjl, 705, L81

\bibitem[{{Vissapragada} {et~al.}(2022){Vissapragada}, {Chontos},
  {Greklek-McKeon}, {Knutson}, {Dai}, {P{\'e}rez Gonz{\'a}lez}, {Grunblatt},
  {Huber}, \& {Saunders}}]{vissapragada2022}
{Vissapragada}, S., {Chontos}, A., {Greklek-McKeon}, M., {et~al.} 2022, \apjl,
  941, L31

\bibitem[{{Vogt} {et~al.}(1994){Vogt}, {Allen}, {Bigelow}, {Bresee}, {Brown},
  {Cantrall}, {Conrad}, {Couture}, {Delaney}, {Epps}, {Hilyard}, {Hilyard},
  {Horn}, {Jern}, {Kanto}, {Keane}, {Kibrick}, {Lewis}, {Osborne},
  {Pardeilhan}, {Pfister}, {Ricketts}, {Robinson}, {Stover}, {Tucker}, {Ward},
  \& {Wei}}]{vogt1994}
{Vogt}, S.~S., {Allen}, S.~L., {Bigelow}, B.~C., {et~al.} 1994, in Society of
  Photo-Optical Instrumentation Engineers (SPIE) Conference Series, Vol. 2198,
  Society of Photo-Optical Instrumentation Engineers (SPIE) Conference Series,
  ed. D.~L. {Crawford} \& E.~R. {Craine}, 362

\bibitem[{{Weiss} {et~al.}(2013){Weiss}, {Marcy}, {Rowe}, {Howard}, {Isaacson},
  {Fortney}, {Miller}, {Demory}, {Fischer}, {Adams}, {Dupree}, {Howell},
  {Kolbl}, {Johnson}, {Horch}, {Everett}, {Fabrycky}, \& {Seager}}]{weiss2013}
{Weiss}, L.~M., {Marcy}, G.~W., {Rowe}, J.~F., {et~al.} 2013, \apj, 768, 14

\bibitem[{{Wittenmyer} {et~al.}(2022){Wittenmyer}, {Clark}, {Trifonov},
  {Addison}, {Wright}, {Stassun}, {Horner}, {Lowson}, {Kielkopf}, {Kane},
  {Plavchan}, {Shporer}, {Zhang}, {Bowler}, {Mengel}, {Okumura}, {Rabus},
  {Johnson}, {Harbeck}, {Tronsgaard}, {Buchhave}, {Collins}, {Collins}, {Gan},
  {Jensen}, {Howell}, {Furlan}, {Gnilka}, {Lester}, {Matson}, {Scott},
  {Ricker}, {Vanderspek}, {Latham}, {Seager}, {Winn}, {Jenkins}, {Rudat},
  {Quintana}, {Rodriguez}, {Caldwell}, {Quinn}, {Essack}, \&
  {Bouma}}]{wittenmyer2022}
{Wittenmyer}, R.~A., {Clark}, J.~T., {Trifonov}, T., {et~al.} 2022, \aj, 163,
  82

\bibitem[{{Wolszczan} \& {Frail}(1992)}]{wolszczan1992}
{Wolszczan}, A., \& {Frail}, D.~A. 1992, \nat, 355, 145

\bibitem[{{Wolthoff} {et~al.}(2022){Wolthoff}, {Reffert}, {Quirrenbach},
  {Jones}, {Wittenmyer}, \& {Jenkins}}]{wolfhoff2022}
{Wolthoff}, V., {Reffert}, S., {Quirrenbach}, A., {et~al.} 2022, \aap, 661, A63

\bibitem[{{Yee} {et~al.}(2017){Yee}, {Petigura}, \& {von Braun}}]{yee2017}
{Yee}, S.~W., {Petigura}, E.~A., \& {von Braun}, K. 2017, \apj, 836, 77

\bibitem[{{Yu} {et~al.}(2018){Yu}, {Huber}, {Bedding}, \& {Stello}}]{yu2018}
{Yu}, J., {Huber}, D., {Bedding}, T.~R., \& {Stello}, D. 2018, \mnras, 480, L48

\bibitem[{{Zahn}(1977)}]{zahn1977}
{Zahn}, J.-P. 1977, \aap, 57, 383

\bibitem[{{Zahn}(1989)}]{zahn1989}
---. 1989, \aap, 220, 112

\end{thebibliography}

\end{document}